\documentclass[11pt]{article} 

\makeatletter
\def\@saveprimitive#1#2{%
  \let#2#1%
}
\makeatother

\usepackage{mathtools} 

\usepackage[margin=1in]{geometry}
\usepackage{microtype}          
\usepackage[sc]{mathpazo}       
\usepackage{setspace}           

\usepackage{amsfonts} 
\newcommand{\defeq}{\vcentcolon=}

\usepackage{adjustbox}
\usepackage{longtable}
\usepackage{booktabs}
\usepackage{placeins}
\usepackage{algorithm}
\usepackage{algorithmic}
\usepackage{graphicx}
\usepackage{enumerate}
\usepackage{caption}
\usepackage{subcaption}
\usepackage{multirow}
\usepackage{enumitem}
\usepackage{siunitx}[=2020-02-20]
\usepackage{color}
\usepackage{float}
\usepackage{bbm}
\usepackage{xcolor}
\usepackage{tikz}
\usepackage{lipsum}
\usepackage{tcolorbox}
\tcbuselibrary{skins}
\usepackage{tabularx}
\usepackage{listings}
\usepackage{titlesec}
\usepackage{rotating}
\usepackage{pdflscape}
\usepackage{natbib}

\definecolor{linkblue}{RGB}{30,70,160}
\definecolor{citeblue}{RGB}{20,100,120}
\usepackage[
  colorlinks=true,
  linkcolor=linkblue,
  citecolor=citeblue,
  urlcolor=linkblue
]{hyperref}

\definecolor{colour1}{RGB}{166,206,227}
\definecolor{colour2}{RGB}{31,120,180}
\definecolor{colour3}{RGB}{178,55,250}
\definecolor{colour4}{RGB}{51,160,44}
\definecolor{darkgreen}{rgb}{0.0, 0.2, 0.13}
\definecolor{darkslateblue}{rgb}{0.28, 0.24, 0.55}
\definecolor{darkviolet}{rgb}{0.58, 0.0, 0.83}
\definecolor{Goldenrod}{RGB}{255, 223, 0}

\graphicspath{{graphics/}}

\newcommand{\ja}[1]{#1}

\newcommand{\as}[1]{#1}

\newenvironment{keywords}{%
  \vspace{0.75em}\noindent\textbf{Keywords:}\ }{\par}
\newenvironment{jelcodes}{%
  \vspace{0.25em}\noindent\textbf{JEL classification:}\ }{\par}

\titleformat{\section}
  {\large\bfseries}
  {\thesection}
  {0.75em}
  {}

\titleformat{\subsection}
  {\normalsize\bfseries}
  {\thesubsection}
  {0.6em}
  {}

\titleformat{\subsubsection}
  {\normalsize\itshape}
  {\thesubsubsection}
  {0.5em}
  {}

\titlespacing*{\section}
  {0pt}{2.0ex plus 0.8ex minus 0.3ex}{1.2ex plus 0.3ex}

\titlespacing*{\subsection}
  {0pt}{1.8ex plus 0.6ex minus 0.2ex}{0.8ex plus 0.2ex}

\titlespacing*{\subsubsection}
  {0pt}{1.5ex plus 0.5ex minus 0.2ex}{0.6ex plus 0.2ex}

\title{The Market Maker's Dilemma:\\
  Navigating the Fill Probability vs.\ Post-Fill Returns Trade-Off}

\newcommand{\thedate}{\today} 

\begin{document}

\begin{center}
  {\LARGE\bfseries The Market Maker's Dilemma:\\
  Navigating the Fill Probability vs.\ Post-Fill Returns Trade-Off \par}
  \vspace{1.0em}
  {\normalsize
    Jakob Albers\textsuperscript{1}\quad
    Mihai Cucuringu\textsuperscript{1,2,3}\quad
    Sam Howison\textsuperscript{4}\quad
    Alexander Y.\ Shestopaloff\textsuperscript{5,6}\par}
  \vspace{0.4em}
  {\footnotesize
    \textsuperscript{1}Department of Statistics, University of Oxford, Oxford, UK\\
    \textsuperscript{2}Oxford-Man Institute of Quantitative Finance, University of Oxford, UK\\
    \textsuperscript{3}Department of Mathematics, University of California Los Angeles, Los Angeles, US\\
    \textsuperscript{4}Mathematical Institute, University of Oxford, Oxford, UK\\
    \textsuperscript{5}School of Mathematical Sciences, Queen Mary University of London, London, UK\\
    \textsuperscript{6}Department of Mathematics and Statistics, Memorial University of Newfoundland, St.\ John's, NL, Canada
  \par}
  \vspace{0.6em}
  {\footnotesize
    Corresponding author: \texttt{jakob.albers@merton.ox.ac.uk}\par}
  \vspace{0.6em}
  {\footnotesize \thedate\par}
\end{center}
\vspace{1em}

\begin{abstract}
Using data from a live trading experiment on the Binance Bitcoin perpetual, we examine the effects of (i) basic order book mechanics and (ii) the persistence of price changes from immediate to short timescales, revealing the interplay between returns, queue sizes, and orders' queue positions. We document a fundamental trade-off: a negative correlation between maker fill likelihood and post-fill returns. This dictates that viable maker strategies often require a contrarian approach, counter-trading the prevailing order book imbalance. These dynamics render commonly-cited strategies highly unprofitable, leading us to model `Reversals': situations where a contrarian maker strategy at the touch proves effective.
\end{abstract}



\section{Introduction}
\label{sec:introduction}

Traders in a limit order book (LOB) market have three immediate objectives in mind for their orders: to have a high probability that an order fills; to achieve a good fill price (without `walking the book'); and to achieve a good post-fill return.
These objectives are generally conflicting, rendering ``optimal'' trading a difficult task. We explore aspects of this task in the context of high-frequency trading (HFT).

It is conceptually helpful here to have in mind  two perspectives. One is that of Optimal Execution (OE), in which a trader makes a net one-way trade or set of trades to buy or sell a position, which may be relatively large and may be traded on behalf of a client. The other is   that of `Balanced Inventory' (BI) trading, aiming at small profits from a large number of  trades, in which the trader must keep their inventory broadly  in balance, eventually unwinding all positions they enter into. Naturally, the perspectives are not comprehensive, and they  may overlap (indeed, BI trading is a sequence of small OE trades);  the objectives may have different priorities, and the priorities may differ between  short and longer timescales.

Market participants in LOBs  can execute trades as (liquidity) takers, makers (liquidity providers), or as a combination of both.
Maker orders (that is, passive limit orders posted at the top bid/ask price, or deeper into the book)\footnote{In the academic literature, `limit order' often refers exclusively to same-side maker orders (passive buy orders on the bid side, sell orders on the ask side). In practice, limit orders can also be posted on the `other' side of the book, as \emph{marketable limit orders}, for example a buy limit order posted at the top ask price. Such an order executes immediately as a taker order (paying the taker fee) against any passive orders from the top of the book to the limit price, with the remainder forming a new queue of one maker order at the limit price.
We adopt this more precise terminology---where a limit order can act as a maker, a taker, or a combination---and encourage other academics to do the same.}   offer a better price (by at least the spread) and significantly lower trading fees, sometimes even earning a rebate.
These    advantages may suggest that maker orders are a more cost-effective way to execute trades than taker orders (market orders or marketable limit orders). However, they also come with two notable disadvantages.

First, maker orders face \emph{execution uncertainty}: they do not fill immediately and may not fill at all, resulting in a fill probability less than 1; market orders have a fill probability of 1.\footnote{Marketable limit orders have a fill probability slightly below 1, see ~\cite{cartea2021latencyrisk}.}
Second, there is a negative correlation between a maker order’s fill probability and its subsequent return conditioned on filling.
This effect, termed the \emph{negative drift} of maker orders by~\cite{DeLise:2024aa}, is especially pronounced at the short timescales which are the primary focus of our paper.
A simple mechanical fact underpins this observation. If the next price move is against a maker order in the top-of-book queue, that order automatically fills, with probability 1, a phenomenon often
referred to as \emph{adverse selection} \cite{Menkveld2013}.
\ja{Such price moves are often driven by short-lived opportunities, perhaps even amounting to arbitrage, arising as prices adjust with a lag to external signals,}\ja{ exposing makers to the risk of trading at stale quotes (termed \emph{toxic arbitrages} in~\cite{foucault_2016}).}
If, in contrast, the next price change is both in the same direction as that of the maker order and occurs before the order is filled, then the order may no longer be at the top of the book, thus not profiting from the favorable price change.
In summary, the fundamental challenge for maker strategies is to achieve both a high fill probability and a positive return if filled: the trade-off between those two aspects  renders profitable maker trading inherently difficult. Few academic studies examine it in detail or propose ways to walk that tightrope;
we see below that this may necessitate submitting orders under apparently adverse conditions using an additional signal (without which the orders would be money-losing).

Market orders, in contrast,  fill with probability 1 (although a large order may still walk the book, and thereby cause an immediate price change as above). Moreover, simple strategies such as trading in the direction of the order imbalance between the top bid and ask queues (formally defined in~\eqref{eq:imb} below) have been shown to yield positive short-term returns~\cite{Lipton:2013aa,Stoikov_2017}. The challenge to the liquidity-taking trader is to achieve positive returns exceeding the exchange's taker fee, which can be viewed as the profitability threshold.

We investigate these trade-offs using an experimental approach on the most liquid crypto market worldwide, the Binance USDT-margined bitcoin perpetual.\footnote{If the Binance Bitcoin perpetual were a stock traded on the NASDAQ, it would rank first in terms of average daily trading volume and fifth in terms of market capitalization at the time of writing (August 9th, 2024). For more details, see https://www.nasdaq.com/market-activity/daily-market-statistics.} We have assembled a data set comprising 232,897 minimum-sized maker orders submitted over a one-week period, agnostic to any trading signal.
If not filled by a taker order arrival, the orders remained in the queue and were canceled if and only if a price change left them posted at a deeper level in the book, i.e., no longer at the touch.
This data set allows us to address the following questions:
\begin{enumerate}
    \item How likely is a maker order to be filled, \ja{and what can we infer about the probability of larger orders filling based on our data for minimum-sized orders}? We investigate this in relation to the already-posted quantities at the top bid and top ask.
    \item How can we update our estimate of an order’s fill probability when those quantities change after the order’s submission (i.e., while it is in the queue), and how does this inform our understanding of larger order fills?
    \item What distinguishes filled orders yielding positive returns from those achieving negative returns?
    \item \ja{What is the relationship between an order’s queue position and its subsequent return upon filling?}
    \item What strategy can traders employ to post maker orders that have a high probability of filling and concurrently a positive expected return conditioned on filling?
\end{enumerate}
We address the questions of fill probability and price drift of maker orders posted at the best prices (top bid and top ask);
these price levels are of paramount importance for practitioners because the majority of trading volume (both maker and taker) occurs at them.
We also note that, on Binance, the spread is virtually always one tick wide, except fleetingly after price changes, whereupon there is fierce competition to close the spread immediately.

\subsection{Related Literature and Context}

At first sight, we have posed relatively specific microstructural questions. However, being generic, they sit in, and contribute to, a far broader context than the Bitcoin market that we study.
We therefore begin with a high-level overview of relevant topics and related literature, before proceeding to a more detailed analysis in the sections that follow.

\subsubsection{Maker Order Execution}
\label{subsubsec:maker_order_execution}

The study of fill probabilities for maker orders has garnered significant interest in the literature. One should  distinguish between estimating (experimentally) the fill probability of randomly placed orders, agnostic to any trading or cancellation signal, and inferring (from historical LOB data) the fill probability of actual orders placed by market participants.
The former is a more fundamental question: if one were to post an order, perhaps conditioned on the order book state or other observable information, how likely is that order to fill?
The latter question examines the behavior of market participants, determining the fill frequency of the orders they chose to place.
While both are of interest, the practical applicability of the latter is subject to systematic biases, which we will refer to as \textit{signal bias}: the estimated fill probabilities are implicitly conditioned on the trader's decision \emph{not} to cancel the order, with the trader following a cancellation policy that is a priori unknown to us---and the importance of this can hardly be overstated given that $>$95\% of submitted maker orders are cancelled; see \cite{Gould:2010aa} and \cite{Brogaard2014}. Traders generally also follow private, a priori unknown order submission policies, e.g., submitting only when they deem adverse selection to be unlikely.

Two recent works, \cite{Fabre_2023} and \cite{Lokin:2024aa}, use the latter approach concerning the behavior of market participants, albeit without explicitly highlighting the important distinction explained above.
\cite{Fabre_2023} conclude that fill probabilities are significantly influenced by market conditions such as bid-ask spread, order size, and liquidity imbalance.
Meanwhile, \cite{Lokin:2024aa} develop a stochastic model of LOB dynamics as a state-dependent queueing system and estimate fill probabilities of maker orders within that framework, concluding that this can be done to a reasonable degree of accuracy.
\ja{The estimated fill probabilities can differ significantly depending on which of the two aforementioned questions one seeks to answer; for instance, \cite{Lokin:2024aa} reports a fill probability of less than 2\% for orders placed by other market participants when the spread is one tick wide and bid and ask queues are balanced, compared to the 50\% fill probability we find in the same situation. This is largely a cancellation effect: real makers cancel orders, biasing historical estimates low, while our no-cancellation policy yields a baseline unaffected by private signals (as discussed further in Section~\ref{sec:new_sec}).}

The primary motivation outlined in the strand of literature addressing the more fundamental question is the trade-off between maker and taker orders: maker orders offer better prices and fees\footnote{Though, as far as we are aware, the fee improvement is left unmentioned in the existing literature.} but do not execute immediately and may fail to execute at all.
A number of recent papers have explored this topic.
In \cite{Maglaras_2021}, the authors simulate synthetic top-of-book maker orders\footnote{That is, fictitious orders `inserted' at the back of the relevant queue, whose subsequent queue place and execution (or not) can be tracked using subsequent LOB-evolution data. Note, however, that, unlike real orders,  they have no effect on the LOB dynamics.} on the NASDAQ exchange and record whether or not they fill, as well as their time-to-fill, given actual market activity after the submission of the synthetic maker order. They then construct a deep learning-based model to predict fill probabilities and the conditional expected fill times, which outperforms benchmarks.
The more recent work by \cite{Arroyo_2023} is similar in spirit but differs in methodology: they employ convolutional neural networks and transformers to perform survival analysis of maker orders, including ones posted at deeper levels in the order book.
However, they do not distinguish the side of the book in which the order is posted relative to the imbalance, leading to inaccurate estimates when comparing the fill probabilities of two opposite-side orders posted at the same depth in a skewed order book imbalance (one which is bid-dominated or ask-dominated).

We add to this strand of literature as follows.
Our data set of actual orders and their outcomes is free from signal-bias, and unlike a synthetic-order approach, not reliant on a posteriori inference of when such a synthetic order might have filled.\footnote{Such inferences may be possible with a reasonable degree of accuracy, but they are still open to error, in some cases due to factors such as latency.}
We leverage our data to analyze fill probabilities of maker orders,
and demonstrate that they can be described in simple terms, and indeed be predicted with a high degree of accuracy, as a function of top-of-book quantities and an order’s queue position,
calling into question the need for black-box machinery lacking interpretability and ease of replicability to solve what turns out to be a relatively simple problem.

\subsubsection{Trade-offs and Trading Decisions}
\label{subsubsec:trade_offs}

As well as using our data set to estimate fill probabilities, we use it to illustrate the difficulty of leveraging that knowledge (or predictability)\textbf{} into trading edge.
In particular, in addition to the trade-off  between price and immediacy of execution, we highlight the trade-off between fill probability itself and the subsequent returns of filled orders.
We support our earlier assertions: first, that it is not at all easy to combine a high fill probability for maker orders with a positive subsequent return, due to one or other of adverse selection if filled, and low fill probability when subsequent returns are good; and second,  for taker orders, to achieve short-term returns in excess of the taker fee. One can see this as an instance of an `Unprofitability Principle' stated in cartoon form as
\[
\text{Ease of prediction}\times \text{ease of exploitation} < c
\]
for some positive constant $c$.

Adverse selection of maker orders has been studied in early works such as ~\cite{foucault_2002} which discusses the need for market makers to monitor information sources to cancel stale quotes, while ~\cite{foucault_2015} and ~\cite{foucault_2016} examine the adverse selection risks market makers face from informed HFT takers reacting to news or arbitrageurs exploiting short-lived price adjustments.
More recently, ~\cite{DeLise:2024aa} provided a more granular examination of adverse selection by analyzing post-fill price trajectories of maker orders.
Our contribution extending these works lies in a precise description of the relationship between fill probability and post-fill returns at small timescales, and in demonstrating how this relationship shapes the way in which market makers should go about constructing their strategies.

To illustrate the role of trade-offs, suppose a trader decides whether or not to employ maker orders on the basis of their estimated fill probability (however obtained), posting an order if this is deemed sufficiently high.
The fill probability/subsequent return trade-off means that the trader may
inadvertently submit orders to the shorter of the bid and ask queues which, as we suggest above, makes them likely to suffer  significant adverse immediate price movement.
This is clearly detrimental for market makers and BI traders more generally, as an adverse price movement after a fill yields a trading loss if the filled order is later unwound  at a worse price.
In the OE setting, it is equally problematic: although no `loss' resulting from the adverse price movement needs to be realized, there is an opportunity cost in that the position might have been entered later at a better price.
The outcome may be a disappointed client and a loss of repeat business for the trader.

\subsubsection{Trading with Signals}

We have noted (and analyze in more detail below) that some apparently natural strategies drawing on basic publicly available information (for example, posting a taker order in the direction of the imbalance or a maker order with high fill probability) may not yield a trading advantage. A trader may, however, take a more positive view if they have extra information in the form of a private signal, for example algorithmically generated.

Suppose, for example, that  the top bid queue is large and the top ask queue is small, \ja{a situation we describe as a price-positive imbalance.}
\footnote{In markets with large relative tick sizes (in basis points) or substantial maker rebates such as ours, liquidity typically concentrates at the top bid and ask levels, and a high fraction of overall trading volume executes against them \cite{Menkveld2013}.}
Intuitively, it is then likely that the shorter (ask) queue is the first to empty, in which case  the spread widens and may be refilled with a new queue on the bid side, giving a price increase.
This argument is corroborated in practice: it is well known that imbalance  is positively correlated with the direction
the next price movement.
(For further insights, refer to ~\cite{Gould:2015aa}.) Now suppose that a trader supplements this modestly return-positive signal with a convincing private signal, suggesting an enhanced short-term return.  How best to capitalize on this opportunity?
Submitting a maker order at the end of the large bid queue is unlikely to result in a fill, and thus may not profit from the prediction (in the less frequent instances where it does fill, the short-term return is likely to be negative, as we discuss in more detail in Section~\ref{sec:fill_outcome}).
Alternatively, a taker order is certain to fill and benefit from any favorable price change; but they must pay the taker fee.

Now, suppose the queues are as before (the imbalance is price-positive), but the trader's private signal, in which they have high confidence, predicts a price decline.
A sell maker order (ask side), counter-trading the imbalance, has a high fill probability due to the small queue and, if the private signal is correct, results in a profitable trade. A sell taker order, however,  incurs a higher fee and crosses the spread.

These arguments illustrate the important point that the choice of order type and execution mode depends on the kind of signal in response to which the trader makes decisions.
The literature contains numerous discussions of trading with signals, often under the umbrella term \textit{`optimal market making'} when they involve maker orders.
For instance,~\cite{Cartea2023} conclude that predictive signals, referred to as alpha signals, can boost market maker profitability.
Their conclusions are arrived at under the assumption that maker orders fill upon the arrival of the next opposite-direction taker order; that is,  the maker order is always at the front of  the queue. As just argued, and discussed in more detail later, this is not generally tenable.
Other papers such as ~\cite{Fushimi2018OptimalHM} and ~\cite{Ait_Sahalia_2013} rely on the same assumption, while others still make different assumptions which are also problematic: for example,~\cite{spooner_rl} assumes that cancellations remove liquidity uniformly from the queue.

Our contribution in this context also addresses the last question from Section~\ref{sec:introduction}: we develop a model to identify when the order book imbalance falsely predicts the next price change. We term this a  \textit{reversal} and discuss it further in Section~\ref{sec:reversal}.
The ease with which those situations are parlayed into a trading advantage (by placing a maker order counter to the imbalance) suggests, in line with  the Unprofitability Principle, that making accurate predictions is hard.
Indeed, to make economically significant predictions, we found it necessary to carefully develop a set of features that capture relevant microstructural patterns in past order flow and price movements.

Our examples above further illustrate the influence of queue position—whether submitting at the end of a long queue or a short one—on an order's fill probability and the effectiveness of using maker orders to act on a trading signal.
For example, posting maker orders in the direction of the imbalance implies posting at the back of the longer of the bid- and ask queues, unless one is the first to close the spread after a price change, which is extremely competitive and typically involves latency optimization and/or special order types.
Conversely, one can obtain a good queue position (in the shorter of the two queues) by posting a maker order opposite to the imbalance-implied movement, but this entails trading against the expected return.
Although queue position is rather sparsely addressed in the academic literature,~\cite{Moallemi_2016} find that bad queue positions are associated with lower fill probabilities and increased adverse selection, findings that our results corroborate; \ja{additionally,~\cite{Donnelly_2018} explore the interplay of queue position, adverse selection, and cancellation behavior, showing that within their modeling framework, the optimal strategy sometimes involves submitting an order counter to the imbalance to benefit from a queue position close to the front-of-queue, a result that is in agreement with our findings.}

\subsection{Notation and Definitions}

Our analysis is set in a standard LOB, a data structure that records queues of maker orders waiting to be executed or canceled, prioritized by price and then by time of arrival, and quantized by the minimum order size,
at price levels incremented by the tick size.
The state of the order book at time \( t \) is defined by the bid and ask sides:
\[
 \mathcal{B}_t := \left( p^b_{t, i}, Q^b_{t, i}, \left( q^b_{t, i, j} \right)_{j} \right)_{i=1,2,3,\cdots}, \qquad \mathcal{A}_t := \left( p^a_{t, i}, Q^a_{t, i}, \left( q^a_{t, i, j} \right)_{j} \right)_{i=1,2,3,\cdots}
 \]
where \( p^b_{t, i} \) and \( p^a_{t, i} \) represent bid and ask prices, \( Q^b_{t, i} \) and \( Q^a_{t, i} \) the total quantities available, and \( \left( q^b_{t, i, j} \right)_{j} \) and \( \left( q^a_{t, i, j} \right)_{j} \) the sequences of individual order quantities, ordered by arrival time, at those prices.
\ja{We use the notational convention \( B_{t} := Q^b_{t, 1} \) and \( A_{t} := Q^a_{t, 1} \) to represent the top-of-book quantities.}

We further introduce the \textit{order book imbalance},
\begin{eqnarray}
\label{eq:imb}
	\operatorname{imb}_t := \frac{B_{t} - A_{t}}{B_{t} + A_{t}} \in (-1,1).
\end{eqnarray}
Its value is close to $+1$ when the liquidity is significantly greater at the top bid level than at the top ask level, and conversely for values near $-1$. As noted above, these extremes are associated with price increases and decreases respectively.

We focus exclusively on those maker orders posted at the top level (and taker orders filling them), meaning sell orders at the top ask price \( p^a_{t, 1} \) and buy orders at the top bid price \( p^b_{t, 1} \).
These two order book levels hold a special significance as a large fraction of overall trading volume is executed against orders posted at the touch, making them especially important for market makers.
For brevity, we shall use the term \emph{maker order} to mean exclusively a maker order posted at one of the two top levels.
The analysis of maker orders posted at deeper levels in the LOB requires different tools and is beyond the scope of this paper.

We now introduce some conventions and define key properties of maker orders posted at the touch. We designate the queue (resp.\/ side) in which they are posted (bid for a buy order and ask for a sell) as the \emph{near-side} queue (resp.\/ side), and the other queue/side (ask for a buy order, bid for a sell) as the \emph{opposite-side} queue/side.
The respective prices are
\( p^{\text{near}}_{t, 1} \)
 and \( p^{\text{opp}}_{t, 1} \), while the respective queue sizes are
 \(Q_t^{\text{near}}\)
 and \(Q_t^{\text{opp}}\). In our later discussions of order book mechanics, the reader may find it helpful to visualize a buy maker order submitted to the bid (near) side, with the ask side being opposite.

We say that the order book imbalance \(\operatorname{imb}_t\) is \emph{favorable} for an order if it is associated with a subsequent positive return on the order, subject to the thresholds \(\operatorname{imb}_t > 0.5\) for a buy order and \(\operatorname{imb}_t < -0.5\) for a sell order.
Conversely, we say it is \emph{adverse}.

\begin{figure}[htbp]
\centering
\captionsetup{width=\textwidth}
\includegraphics[scale=0.25]{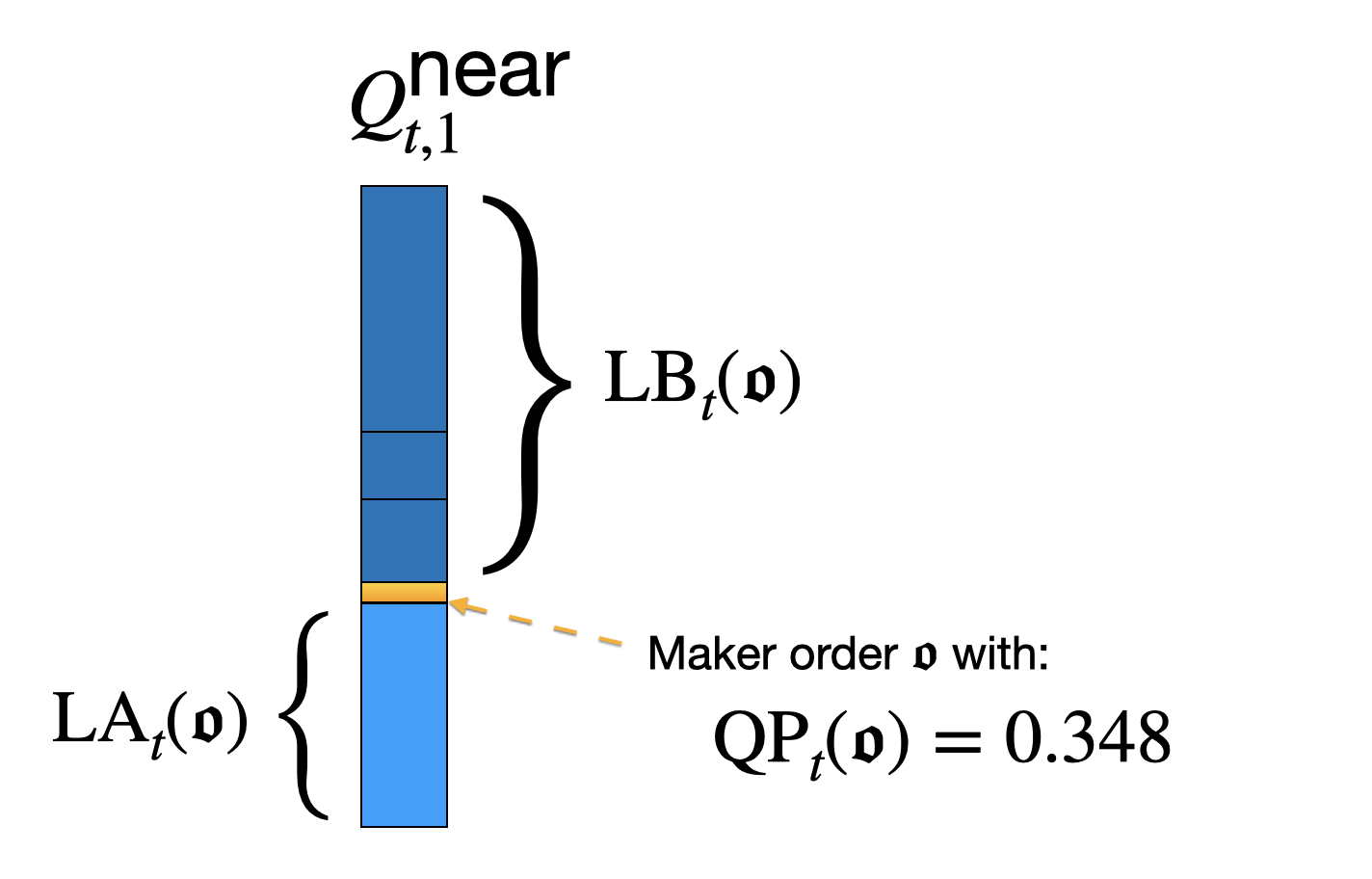}
\caption{Visualization of the order properties introduced above.}
\label{fig:order_notation_pic}
\end{figure}

Consider a buy order (and analogously for sell orders)  $\mathfrak o$ posted at the top bid price \( p^b_{t, 1} \) at time $t$, indexed by $j_0$ in the sequence \( \left( q^b_{t, 1, j} \right)_{j} \) of individual order quantities.
As illustrated in  Figure~\ref{fig:order_notation_pic}, we define the \emph{liquidity ahead} of the order as the total quantity of all preceding orders in the queue:
\begin{equation*}
	\operatorname{LA}_t(\mathfrak o) := \sum_{j < j_0} q^b_{t, 1, j}.
\end{equation*}
Similarly, the \emph{liquidity behind} of the order is the total quantity of subsequent orders in the queue:
\begin{equation*}
	\operatorname{LB}_t(\mathfrak o) := \sum_{j > j_0} q^b_{t, 1, j}.
\end{equation*}

Finally, the order's \emph{queue position} in the near-side queue is defined as
\begin{equation*}
	\operatorname{QP}_t(\mathfrak o) := \frac{\operatorname{LA}_t}{B_t - q^b_{t, 1, j_0}} =
 \frac{\operatorname{LA}_t}{\operatorname{LA}_t + \operatorname{LB}_t}, \quad \max(\operatorname{LA}_t, \operatorname{LB}_t) >0,
\end{equation*}
and equal to $0$ when there are no other orders in the queue ($\operatorname{LA}_t=\operatorname{LB}_t=0$). An order whose queue position is $0$ is at the front of the queue; if it is $1$, the order is at the back.

Note that  \(\operatorname{LA}_t(\mathfrak o)\), \(\operatorname{LB}_t(\mathfrak o)\), and \(\operatorname{QP}_t(\mathfrak o)\)
—are time series over the time interval from the moment the order is inserted into the queue until it is removed by being filled or canceled.
Every book update spawns a new value in the time series of each extant order.
Additionally, note that the liquidity ahead \(\operatorname{LA}_t(\mathfrak o)\) decreases over time, since new orders cannot be inserted into the queue at an earlier position than any existing order.

An order \(\mathfrak{o}\) has a  \emph{post time} \(t_0(\mathfrak{o})\) and a \emph{terminal time}
\(T(\mathfrak{o})\), at which the order is either filled or canceled;
that, in our experimental design, an order is canceled if and only if the price level at which it is posted ceases to be; the top-of-book on its respective side.

We describe \textit{price drift} for maker and taker orders using the microprice at time \( t \); this is calculated as the quantity-weighted average of the top bid and ask prices:
\[
p^\mu_t :=
 \frac{B_t p^a_{t, 1} + A_t p^b_{t, 1}}{B_t + A_t}
=\frac{Q^{\text{near}}_t p^{\text{opp}}_{t, 1} + Q^{\text{opp}}_t p^{\text{near}}_{t, 1}}{Q^{\text{near}}_t + Q^{\text{opp}}_t},
\]
while the midprice is calculated as the simple average of the top bid and ask prices:
\[
p^\text{mid}_t := \frac{p^a_{t, 1} + p^b_{t, 1}}{2} = \frac{p^{\text{near}}_{t, 1} + p^{\text{opp}}_{t, 1}}{2}.
\]

Consider an order \(\mathfrak{o}\) (either maker or taker) posted at time \( t_0 \) when the top-of-book price is $p_{t_0}$.
\footnote{Conventionally, for a maker order, \( p_{t_0} \) is the top price on the near-side, while for a taker order, it is the top price on the opposite-side.}
We define a sign function, \(\operatorname{sgn}(\mathfrak{o})\), equal to \( +1 \) for buy orders and \(-1\) for sell orders, and multiply the drift values by this sign to obtain a direction-independent measure, with positive (negative) values indicating favorable (adverse) drift.

Supposing \(\mathfrak{o}\) fills, we write \( t_0 + \tau_0 \) for its fill time.\footnote{This denotes the time immediately after (rather than before or during) its processing in the exchange's matching engine: this is when a maker order has been removed from the queue, or a taker order has removed other liquidity from the queue. For taker orders, \(\tau_0\) is typically very small, as these orders are processed quickly.}
The order's (instantaneous) \textit{price drift}, measured in bp, is defined as the signed microprice return when the midprice remains unchanged; otherwise, it is the return:
\begin{eqnarray}
\label{eq:drift_immediate}
\operatorname{dr}(\mathfrak o) :=
\begin{cases}
\operatorname{sgn}(\mathfrak o)\left(\frac{p^\mu_{t_0+\tau_0}}{p_{t_0}} - 1\right) \cdot 10000, & \text{if } {p^\text{mid}_{t_0}} = {p^\text{mid}_{t_0+\tau_0}}, \\
\operatorname{sgn}(\mathfrak o)\left(\frac{p^\text{mid}_{t_0+\tau_0}}{p_{t_0}} - 1\right) \cdot 10000, & \text{else}.
\end{cases}
\end{eqnarray}

We further define variants of this at longer time scales: the order's \textit{price drift after $\tau > \tau_0$ seconds} is defined as

\begin{eqnarray}
\label{eq:drift_tau}
    \operatorname{dr}_\tau(\mathfrak o) := \operatorname{sgn}(\mathfrak o)\left(\frac{p^\mu_{t_0+\tau}}{p_{t_0}} - 1\right) \cdot 10000.
\end{eqnarray}

\noindent\textbf{Contributions.} We present a large-scale live trading experiment (Section~\ref{sec:data_acquisition}) involving hundreds of thousands of randomly placed maker orders, ensuring the absence of signal-bias typical in many data sets. We analyze fill probabilities at the top of the book and show how queue sizes and order positions affect a maker order’s likelihood of execution (Section~\ref{sec:fill_prob}). We then demonstrate a trade-off between fill probability and short-term post-fill returns, since large queues and front-of-queue positions often yield the most favorable returns but suffer lower fill rates (Sections~\ref{sec:fill_outcome} and \ref{sec:new_sec}). We further illustrate that taker strategies are constrained by often-prohibitive fees, while successful maker trading requires exploiting conditions in which order book imbalance reverses, producing favorable outcomes despite fundamental signals pointing in the opposite direction (Section~\ref{sec:reversal}).

\section{Data Acquisition}
\label{sec:data_acquisition}

Our data was collected through live trading experiments carried out on Binance’s linear Bitcoin perpetual market, the most liquid crypto market globally.

\textbf{Experimental design.}
We conducted two separate experiments, each operating under a different quoting mode:
\begin{enumerate}
    \item \textbf{Continuous Quoting Mode:}
    We continuously quoted on both sides of the market: we always maintained an active buy order at the top bid price and an active sell order at the top ask price.
    If the top bid price moved up or the top ask price moved down without the respective orders filling, we canceled and reposted them at the updated top price.
    Similarly, if an order filled, we immediately posted a new one at the top-of-book price on the respective side.
    We ran our experiment for one week, from February 12 to 19, 2024, during which we submitted 232,897 minimum-sized maker orders, of which 127,051 filled, and 105,846 were canceled unfilled.

    \item \textbf{Periodic Quoting Mode:} Every ten seconds, we independently submitted a buy order at the top bid price and a sell order at the top ask price (regardless of inventory, order history, or active orders). If the top bid or ask price changed, leaving the respective order posted deeper in the book, we canceled it.
    This experiment ran from August 21 to 31, 2024, during which we submitted 172,800 minimum-sized maker orders, of which 95,998 filled, with the remainder canceled unfilled.
\end{enumerate}

The results from both modes were essentially identical and we present findings from the Continuous Quoting Mode in this paper.
The resulting dataset of orders is denoted \(\mathcal{O}\) and contains each order’s limit price, submission timestamp, and terminal state (filled or canceled).

\textbf{Alternative forms of data.}
A key contribution of this paper is its experimental methodology.
We chose to collect data via live trading experiments to obtain a signal-bias-free dataset, overcoming the inherent limitations of alternative methods, of which there are three broad categories:

\begin{enumerate}
    \item \textbf{Synthetic orders:}
    This involves the hypothetical placement of synthetic orders
    within observed market data and tracking their outcomes based on subsequent real market events.
    The synthetic order placement creates a hypothetical LOB state that includes the order; its potential market impact is ignored, an assumption that is particularly problematic if the order size is large.
    Even with full L3 data, this method cannot guarantee that a fill is realizable in the real world, and with Binance’s L2 data, accurately tracking an order’s queue position—and thus its outcome—is impossible.

    \item \textbf{Passive use of existing orders:}
    This involves observing already-placed orders by other market participants and it is subject to biases inherent in the orders of other market participants, as discussed in Section~\ref{subsubsec:maker_order_execution}.
    Moreover, this method cannot ensure that the analyzed orders originate from the same trader; in a competitive market with many active participants, the lack of trader identifiability makes this highly unlikely.

    \item \textbf{Simulation:}
    This is a broad category, ranging from simulating the entire LOB's evolution to more limited short-term forward simulations for determining an order's outcome.
    Such methods are entirely unsuitable for our analysis due to their generally poor data fidelity, as these simulators are often designed merely to replicate a set of stylized facts.

    \end{enumerate}

\begin{figure}[htbp]
\centering
\captionsetup{width=\textwidth}
\includegraphics[scale=0.39]{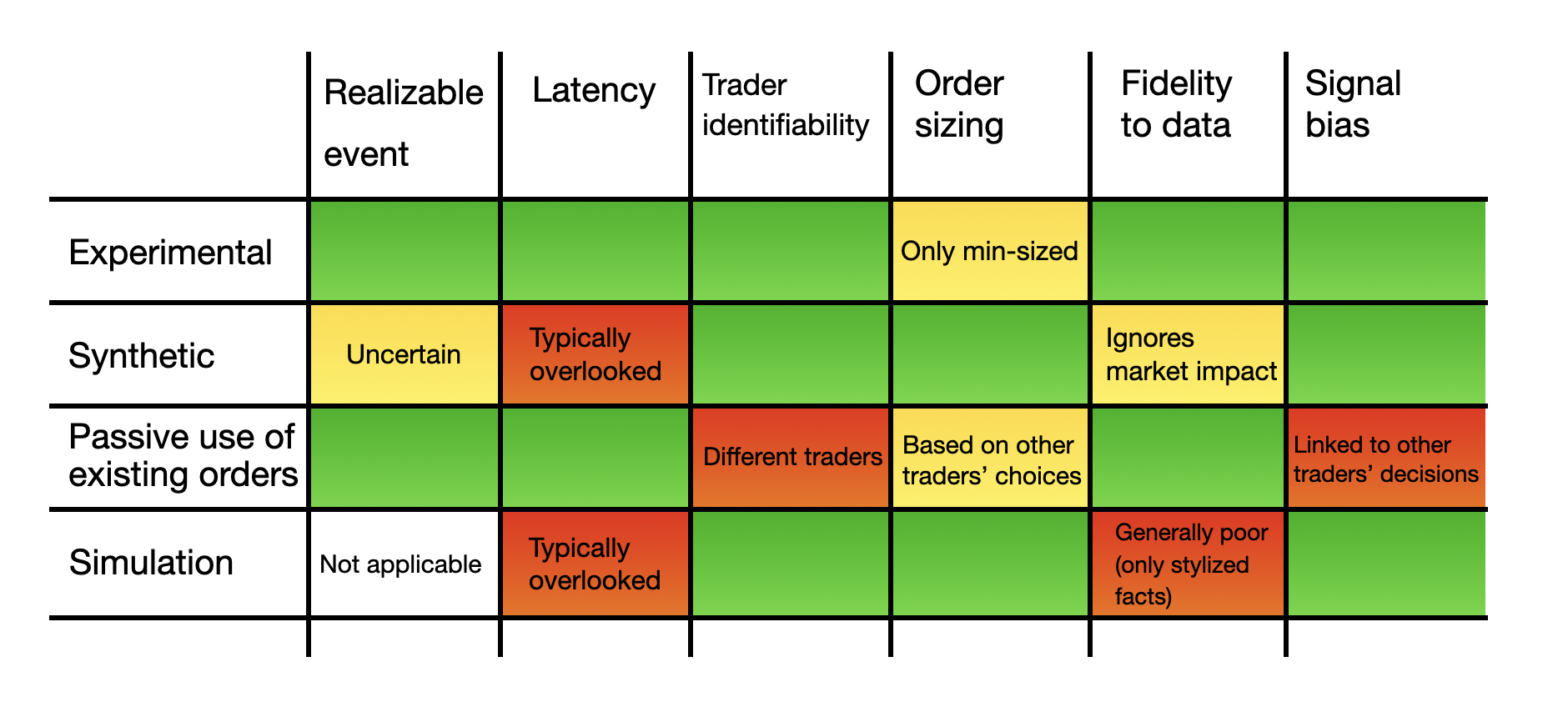}
\caption{Comparison of alternative forms of data, assessed along six dimensions: whether fills are guaranteed to be realizable in the real world (realizable event); whether latency effects are accounted for (latency);
whether the data allows for the analysis of single-trader strategies or if the orders may stem from different traders (trader identifiability);
whether analysis extends beyond minimum-sized orders (order sizing);
the fidelity of data to real-world trading (fidelity to data);
and the presence of biases in the analyzed orders (signal bias).}
\label{fig:methods_comparative_table}
\end{figure}

Table~\ref{fig:methods_comparative_table} presents a comparative overview of the different data acquisition methods, highlighting their respective advantages and disadvantages.

\textbf{Market microstructure and fees.}
Like most crypto exchanges, Binance employs a tiered fee structure where trading volume is the main factor, though not the only one; under its most favorable tier, the taker fee is 1.5 bp and the maker fee is \(-0.5\) bp (i.e., a rebate for maker orders).
The tick size is 0.1 USD, or approximately 0.03 bp at the time of the experiment, which is negligible compared to fees.
The market generally exhibits robust liquidity, with several hundred thousand USD available at the top bid and ask levels, and a spread that is virtually always one tick wide.
In the subsequent analysis, we assume the best possible trading fees; naturally, any strategy results analyzed below would only deteriorate with less favorable fees.

\section{The Two Fundamental Attributes for Maker Orders}
\label{sec:fundamental_aspects}

We now expand on  some of our earlier observations,
examining the \text{risk profile} of an order (maker or taker). This has four basic attributes: fill price, trading fee, fill probability, and price drift.
These four attributes are the most base-level factors determining the (unrealized) PnL of an individual order (maker or taker) and, by extension, the realized PnL of  a BI strategy.
Other factors such as volatility, liquidity levels or a broader information set (from which trading signals may be derived) also affect the PnLs, but they do so through the four basic attributes.

We note the caveat that, while latency is a crucial aspect of trading affecting both maker and taker orders,\footnote{For a detailed examination of its impact on taker order execution see  \cite{Albers_2024}; similar effects apply to maker orders.} to avoid lengthy technical digressions we treat latency as being negligible.
This simplifies our arguments; on the whole, a meticulous treatment of latency reinforces our conclusions.

\begin{figure}[htbp]
\centering
\captionsetup{width=\textwidth}
\includegraphics[scale=0.39]{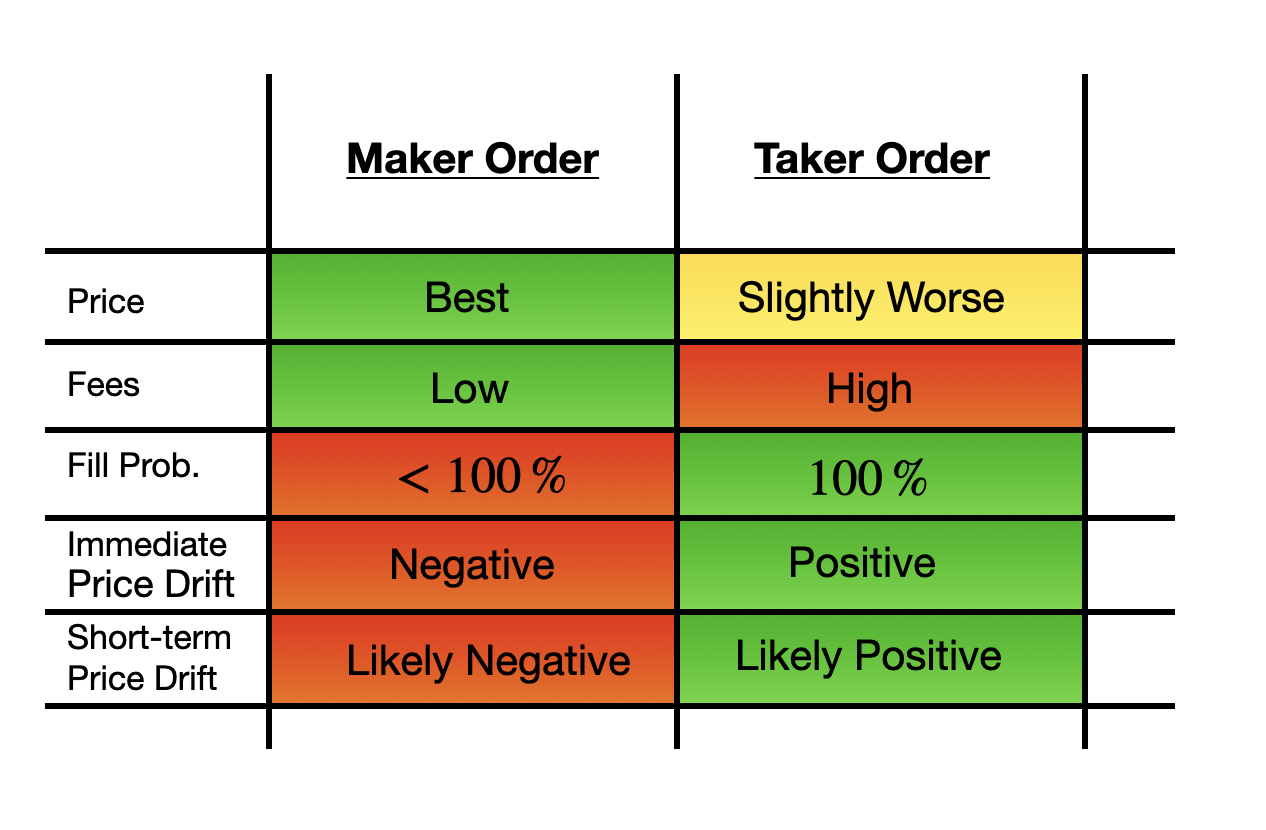}
\caption{Comparison of key attributes of maker and taker orders.}
\label{fig:maker_taker_table}
\end{figure}

Table~\ref{fig:maker_taker_table} shows a comparison of the aforementioned four attributes for a maker order posted at the touch versus a market order.
In either case, two attributes are favorable while two pose challenges.

To fix ideas, consider the case of a buy order (a sell order being the converse).
The first two rows of Table~\ref{fig:maker_taker_table} show the attributes that are
favorable for maker orders but unfavorable for taker orders.
Recalling that we neglect latency, a maker buy order  has price certainty as its price is fixed at the top bid price at submission time. A taker buy order executes at an \emph{a priori} uncertain price, which is at least as high as the top ask price, which itself is at least one tick worse than the top bid.
More significantly, maker orders enjoy a low exchange fee (or even receive a rebate), while taker orders in contrast incur a considerably higher taker fee.

The remaining rows of Table~\ref{fig:maker_taker_table} list attributes which are favorable for taker orders and unfavorable for maker orders.
\ja{While the two attributes favorable to maker orders—fill price and trading fee—are observable at the time of posting, the two unfavorable ones, namely fill probability and price drift (explored in detail below),
are unknown and effectively stochastic: they can be highly variable, making them interesting to analyze.
We term them \textit{the two fundamental attributes of maker orders}.}

Taker orders always fill immediately, whereas maker orders have a fill probability less than 1 and fill some time after submission (if at all).
Immediate price drift is favorable to taker orders. When a taker buy order is executed, it removes liquidity from the top ask price queue; where this does not consume the whole queue, it instantaneously increases both the order book imbalance and the microprice, while
an order that is large enough to clear the entire queue  directly causes the spread to widen upwards, with an increase in the mid-price. Both of these  result in an immediate favorable change in the first post-fill order book snapshot, hence the positive immediate price drift (see Equation~\ref{eq:drift_immediate}).

The situation is reversed for maker orders.
If a maker buy order, posted in the top bid queue, is filled, a sell taker order must have removed liquidity from the bid side.
If the queue remains non-empty immediately after the fill, the microprice decreases and the order book imbalance skews more price-negative; if the taker order is large enough to clear the top bid queue,
the spread widens downwards and the mid-price decreases.

It is natural to ask whether this initial movement persists for longer times.
A mid-price move  might be counteracted by  trades in the opposite direction, for example refilling  the queue hit by the taker order that first caused the spread to widen with associated immediate price drift.
However, as widely acknowledged in the literature~\cite{Lipton:2013aa,Gould:2015aa,Lehalle:2016aa,Stoikov_2017}, there is a positive correlation between order book imbalance and subsequent asset price movements, typically observed over short time scales ranging from milliseconds to seconds.
Moreover, if the taker order causes an immediate mid-price movement, the well-documented autocorrelation of high-frequency returns suggests that this trend is likely to continue in the short term.
\ja{Our empirical evidence confirms these statements, and in Section ~\ref{sec:reversal}, we explore why the  correlations involved are less than 1: the fill-producing initial movement occasionally reverts. This class of cases is of paramount importance in market making, as we shall see later.
}
In summary, maker orders experience negative instantaneous price drift and negative expected price drift over longer time scales, amounting to adverse selection.

\section{Fill Probabilities}
\label{sec:fill_prob}

Suppose \( \mathfrak{o} \) is a maker order submitted at time $t_0$ to  the near-side queue, and
consider how its fill probability depends on  $\operatorname{LA}_t(\mathfrak{o})$ on the near side and the opposite-side queue size $Q_t^{\text{opp}}$.
We present two informal arguments, the first concerning $\operatorname{LA}_t(\mathfrak{o})$ and the second concerning $Q_t^{\text{opp}}$, and validate their predictions against our empirical data.

\textbf{1: Near-side liquidity.} The order fills at time \(T(\mathfrak{o}) > t_0\) if and only if at that time a sell taker order is processed within the exchange matching engine and its quantity exceeds the order's liquidity ahead, \(
\operatorname{LA}_T(\mathfrak{o})\), consuming \(\mathfrak{o}\) and all maker orders earlier in the queue. Since the arrival probability of a taker order with size less   than a given quantity decreases as  that quantity decreases, for times $t_0<t\leq T$
we would expect less liquidity ahead 
to yield a higher fill probability.
Furthermore, since \(\operatorname{LA}_t(\mathfrak{o}) \leq \operatorname{LA}_{t_0}(\mathfrak{o})\), the order \(\mathfrak{o}\) is certain to fill if, while it remains active in the book, a taker order arrives at its price level with size exceeding \(\operatorname{LA}_{t_0}(\mathfrak{o})\). Therefore, we would generally expect orders posted on a small near-side queue to have a high likelihood of filling.

\textbf{2: Opposite-side liquidity.} The opposite-side queue size \(Q_t^{\text{opp}}\) influences the fill probability of \(\mathfrak{o}\) via a more indirect mechanism. When \(Q_t^{\text{opp}}\) is small, other things being equal,  a price increase becomes likely, as only a small taker quantity is required to cause it. Conversely, when \(Q_t^{\text{opp}}\) is large, a price increase is less likely, this increases the residence time of \(\mathfrak{o}\) in the bid queue and thereby increases the time available for a taker order exceeding \(\operatorname{LA}_t(\mathfrak{o})\) to arrive.

In summary, the combination of \((\operatorname{LA}_t(\mathfrak{o})\), \(Q_t^{\text{opp}})\) significantly influences the fill probability of \(\mathfrak{o}\), with lower values of \(\operatorname{LA}_t(\mathfrak{o})\) and higher values of \(Q_t^{\text{opp}}\) both contributing to higher fill probabilities.

\paragraph{Empirical Validation.} We now examine how fill frequencies \ja{(sometimes also referred to as empirical fill probabilities)} depend empirically on the conditions at the order's submission time \( t_0(\mathfrak{o}) \).
Specifically, we analyze the fill frequency conditioned on the near-side queue size at the time of  submission \( Q^{\text{near}}_{t_0} \)
(equivalent to the order's initial liquidity ahead) and the initial opposite-side queue size \( Q^{\text{opp}}_{t_0} \); namely, the probability
\[
\operatorname{P}\left( \mathfrak{o} \text{ fills } \, \mid \,  Q^{\text{near}}_{t_0(\mathfrak o)} = Q^{\text{near}}, \,  Q_{t_0(\mathfrak o)}^{\text{opp}} = Q^{\text{opp}} \right).
\]

To obtain our empirical estimate, we employ an equal-width binning approach for \( Q^{\text{near}} \) and \( Q^{\text{opp}} \).
Specifically, we divide the range of these variables—from their minimum observed values to their 99th percentiles during the sample period—into 20 equal-width bins, using the 99th percentile instead of the maximum to avoid skewing the analysis \ja{with very occasional large orders.}
Queue sizes exceeding the 99th percentile are included in the largest bin.
We thus obtain
a two-dimensional grid of bin pairs \((Q_i^{\text{near}}, Q_j^{\text{opp}})\), for $i, j\in \left\{ 1, \cdots, 20 \right\}$.
The bins were constructed mindful of the balance between (1) having enough data points in each bin pair to ensure reliable empirical estimates and (2) obtaining a high level of granularity in the data representation.

\begin{figure}[htbp]
\centering
\includegraphics[scale=0.6]{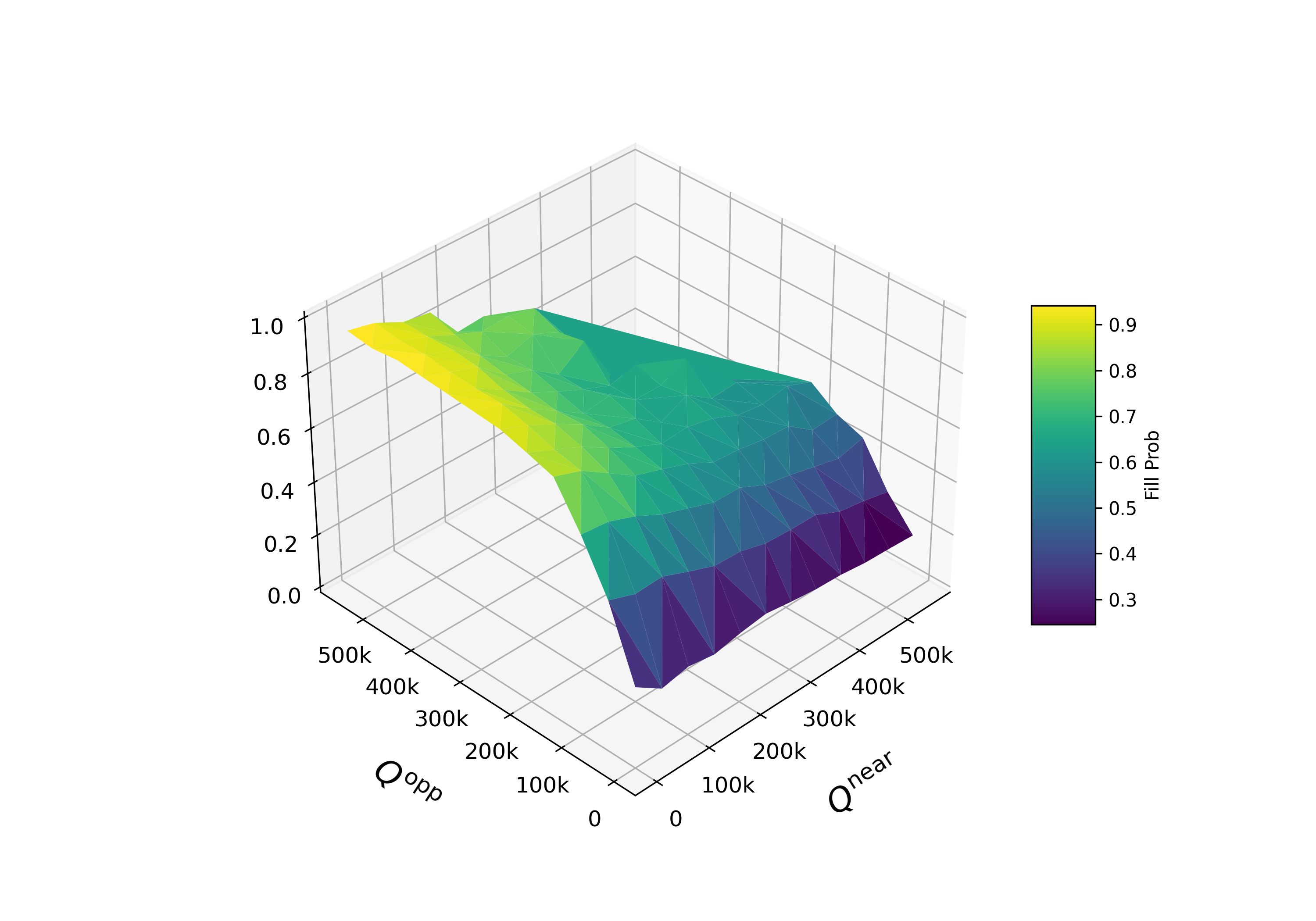}
\caption{Fill probability as a function of near and opposite-side queue sizes (in USD).}
\label{fig:mesh_buy}
\end{figure}

For each bin pair \((Q_i^{\text{near}}, Q_j^{\text{opp}})\), we calculate the empirical fill probability as the proportion of orders whose \( Q^{\text{near}}_{t_0} \) and \( Q^{\text{opp}}_{t_0} \) fall within the bins that are filled.
We denote by  \( z = \operatorname{pr}(Q^{\text{near}}, Q^{\text{opp}}) \)
a suitable smooth approximation \ja{(we used a Delaunay triangulation-based interpolation)} to the binned frequency data, shown in Figure~\ref{fig:mesh_buy}.
The plot reveals that fill probability decreases with the near-side queue size
\( Q^{\text{near}} \)
and increases with the opposite-side queue size \( Q^{\text{opp}} \).
Orders submitted when the near-side queue is large and the opposite-side queue is small have a low fill probability, as low as 30\%.
Conversely, the fill probability exceeds 90\% when the near-side queue is very small and the opposite-side queue is very large.

The highest fill probabilities occur when there is a large opposite-side queue and a nearly empty near-side queue, implying an adverse order book imbalance; however, the plot clearly shows that fill probabilities are not a function of imbalance alone, as the imbalance is constant along lines on which $Q^{\text{near}}/Q^{\text{opp}}$ is constant.

To further quantify the relationship between fill probability, queue sizes, and imbalance, we fitted an ordinary least-squares (OLS) regression model to the surface shown in Figure~\ref{fig:mesh_buy}. For this analysis, the queue size variables were normalized (by dividing by their 99th percentile values) to ensure their coefficients are on a comparable scale. The regression model takes the form:

\[
z = \beta_0 + \beta_1 Q^{\text{near}} + \beta_2 Q^{\text{opp}} + \beta_3 \text{imb} + \epsilon,
\]

where \( z \) represents the fill probability, \( Q^{\text{near}} \) and \( Q^{\text{opp}} \) denote the normalized opposite-side and near-side queue sizes (with a slight abuse of notation, we reuse these symbols for simplicity), and \(\text{imb} \defeq (Q^{\text{near}} - Q^{\text{opp}}) / (Q^{\text{near}} + Q^{\text{opp}})\) represents the imbalance.
The model achieves an \( R^2 \) of 0.946, indicating a near-perfect fit to the data, and underscoring that fill probabilities can be estimated with minimal complexity.

\ja{
The estimated coefficients are \(\beta_0 = 0.5649\), \(\beta_1 = 0.0159\), \(\beta_2 = 0.1013\), and \(\beta_3 = -0.3166\).
The p-values show that all terms are statistically significant except \(Q^{\text{opp}}\) (\(\mathrm{p} = 0.065\)), with the dominant drivers being the near-side queue size (\(\mathrm{p} = 0.000\)) and imbalance (\(\mathrm{p} = 0.000\)).
}

\paragraph{Intermediate Times}
Thus far, we have only focused on estimating an order's fill probability based on information available at the time of submission.
Can we update this estimate for an order \(\mathfrak{o}\) at an intermediate time \( t > t_0(\mathfrak{o}) \), given \(\operatorname{LA}_{t}(\mathfrak{o})\) and \(Q_{t}^{\text{opp}}\)?

A basic ``zero-intelligence" approach, ignoring   \(\operatorname{LB}_{t}(\mathfrak{o})\) and potential responses of other traders to the size of the whole near-side queue, including
\(\operatorname{LB}_{t}(\mathfrak{o})\), might suggest that the fill probability
of \(\mathfrak{o}\) at time \( t \) is the same as that of another order \(\mathfrak{o}'\) at its initial time \( t_0(\mathfrak{o'}) \) if they have the same liquidity ahead:
\(
\left( \operatorname{LA}_{t}(\mathfrak{o}), Q_{t}^{\text{opp}} \right) = \left( \operatorname{LA}_{t_0(\mathfrak{o'})}(\mathfrak{o'}), Q_{t_0(\mathfrak{o'})}^{\text{opp}} \right).
\)
The rationale is that the total quantity of liquidity removal necessary to terminate the order, either on the near-side (giving a fill) or on the opposite-side (giving a cancellation) is the same for both orders.
However, a build-up of \(\operatorname{LB}_{t}(\mathfrak{o})\) is likely to reduce the arrival rate of taker orders on the near-side, increase it on the opposite side, reduce cancellations from the near-side (particularly for front-of-queue orders), and increase the cancellation rate of opposite-side orders, as traders respond to the evolving imbalance.
All of these reactions tend to reduce the fill probability for \(\mathfrak{o}\) compared with \(\mathfrak{o}'\).
This estimate is thus an upper bound on the order's fill probability at the intermediate time. The deviation from the true fill probability is determined by the extent to which a non-zero \(\operatorname{LB}_{t}(\mathfrak{o})\) influences the decisions of other traders.

To obtain a better estimate of intermediate-time fill probability, one could account for \(\operatorname{LB}_{t}(\mathfrak{o})\) with a linear interpolation of the form \(\operatorname{QP}_t \operatorname{pr}(\operatorname{LB}_t, Q^{\text{opp}}_t) + (1-\operatorname{QP}_t) \operatorname{pr}(\operatorname{LA}_t, Q^{\text{opp}}_t)\).
We leave this for future work, as it is not central to this paper; our L2 data, in which orders are aggregated for each level,  lacks the granularity needed for a precise analysis. In particular,  the values of \(\operatorname{LA}_{t}(\mathfrak{o})\) and \(\operatorname{LB}_{t}(\mathfrak{o})\) require L3 data, which is not available on the main cryptocurrency exchanges.
\footnote{Currently, as far as we are aware, the only exchanges that make L3 data available are Bitfinex, Coinbase, Bitstamp, Kraken, and KuCoin.}

\paragraph{Larger Orders} Although we used minimum-sized orders due to cost constraints, we can use our data to find the fill (full or partial) probability of a larger order. This is not difficult: the probability that an order \( \mathfrak{o} \) of any size $q$ fills
fully,
conditioned on
\( \left( \operatorname{LA}_{t_0(\mathfrak{o})}(\mathfrak{o}), Q_{t_0(\mathfrak{o})}^{\text{opp}} \right) \)
is the same as the probability of another, minimum-sized, order \( \mathfrak{o}' \) filling, conditioned on
\( \left( \operatorname{LA}_{t_0(\mathfrak{o'})}(\mathfrak{o'}) + q, Q_{t_0(\mathfrak{o}')}^{\text{opp}} \right).
\)
One can proceed similarly for partial fills.

In conclusion, fill probabilities are relatively easy to predict based on queue sizes. However, the ease of predictability does not translate into ease of exploitation \as{for profit} due to the trade-off between fill probability and price drift, a topic we will explore in more detail in the next section.

\FloatBarrier

\vspace{-3mm}
\section{Price Drift}
\label{sec:fill_outcome}

Let us now address the second fundamental aspect of maker orders, namely the price drift, i.e. subsequent return, they realize over short time scales if filled.
As in the previous section, we begin with two informal arguments derived from basic order book mechanics, before examining our empirical data for confirmation.

\textbf{1: The near-side queue.}
An order's near-side queue size \(Q^{\text{near}}_{t}\) and queue position \(\operatorname{QP}_t(\mathfrak{o})\) both influence its post-fill price drift.
If the order is at the very end of the queue at the time $T(\mathfrak{o})$ of its execution, i.e., \(\operatorname{LB}_T(\mathfrak{o}) = 0\) (or equivalently \(\operatorname{QP}_T = 1\)), then the very taker order that yielded the fill also causes the midprice to move against the limit price of \(\mathfrak{o}\), resulting in an adverse immediate price movement.
Conversely, if the value \(\operatorname{LB}_T(\mathfrak{o})\) is very large, then the midprice is less likely to move adversely as a direct result of  the fill: that could
only happen if the taker order were so large as to consume the whole near-side queue (which is at least as large as \(\operatorname{LB}_T(\mathfrak{o})\)).
We thus expect immediate (and probably subsequent) adverse selection to increase in magnitude  the smaller \(\operatorname{LB}_T(\mathfrak{o})\) is.
\ja{An important class of examples where \(\operatorname{LB}_T(\mathfrak{o})\) is large includes orders with a front-of-queue position (\(\operatorname{QP}_T \ll 1\)) in a large near-side queue.}

\textbf{2: The opposite-side queue.}
Now consider the opposite-side queue size \(Q^{\text{opp}}_{T}\) at fill time.
If \(Q^{\text{opp}}_{T}\) is very small, then even
a small trade suffices to produce a post-fill price change favorable to the order \(\mathfrak{o}\).
Conversely, as \(Q^{\text{opp}}_{T}\) increases, this post-fill price move becomes increasingly less likely.
Altogether, we would therefore expect the degree of adverse selection\ja{—measured by the conditional expected markout return—}to worsen as the opposite-side queue size increases.

To summarize, we have argued that we should observe:
\begin{enumerate}
    \item The degree of adverse selection decreases with \(\operatorname{LB}_T(\mathfrak{o})\). Orders with \(\operatorname{QP}_T \approx 1\) are likely to generate negative returns regardless of queue sizes, while orders with \(\operatorname{QP}_T \approx 0\) may generate positive returns if the near-side queue size is sufficiently large.
    \item The degree of adverse selection is increasing in opposite-side queue size \(Q^{\text{opp}}_{T}\).
\end{enumerate}

\paragraph{Empirical Validation.}
Turning to our empirical evidence, we examine the empirical return distributions of filled orders, conditioned on queue position and the sizes of both the near-side and opposite-side queues.
The return, which we refer to as the order's \textit{markout return}, is calculated in basis points by comparing the limit price of the filled order with the midprice one second after the fill occurs—note that these returns do not include the maker rebate.
We categorize the queue sizes \(Q^{\text{opp}}_{T}\) and \(Q^{\text{near}}_{T}\) using three bins: large, medium, and small.
These bins are defined based on a quantile analysis of total queue sizes across our data set.

\ja{
To calculate the exact fill-time queue position \( \operatorname{QP}_T(\mathfrak{o}) \) for all filled orders \( \mathfrak{o} \) in our dataset, we leveraged an idiosyncrasy of Binance's public trades feed: for every taker order, the exchange publishes all corresponding maker fills ordered by execution priority, each with unique identifiers.
Makers receive these same identifiers for their posted orders at submission-time, allowing them to identify their order within a burst of maker fills after receiving a fill.
By summing the sizes of earlier fills in the burst and comparing this to the total queue size, either from the most recent LOB snapshot or inferred when the taker order changes the price, makers can compute their exact \( \operatorname{QP}_T(\mathfrak{o}) \)--we applied this method to the minimum-sized fills from our experiment.
}

\ja{Building on this, we define bins for \( \operatorname{QP}_T \) based on percentile ranges.}
For instance, 0\% - 10\% (\( \operatorname{QP}_T \in [0, 0.1) \)) represents orders within the front 10\% in the queue at the time of execution, while 75\% - 100\% (\( \operatorname{QP}_T \in [0.75, 1] \)) represents orders within the last quartile in the queue.
We then examine the empirical markout return distributions conditioned on a near-side queue size bin, an opposite-side queue size bin, and a queue position bin.

\paragraph{Large near-side queue, small opposite-side queue.} Here, the order book imbalance is favorable to the order.
We consider two extremes of queue position: the first decile (0\% to 10\%) and the upper quartile (75\% to 100\%); we use a
larger range here because there are fewer fills with late queue positions).

\begin{figure}[htbp]
    \centering
    \begin{minipage}{0.5\textwidth}
        \centering
        \captionsetup{width=.8\linewidth}
        \includegraphics[scale=0.36]{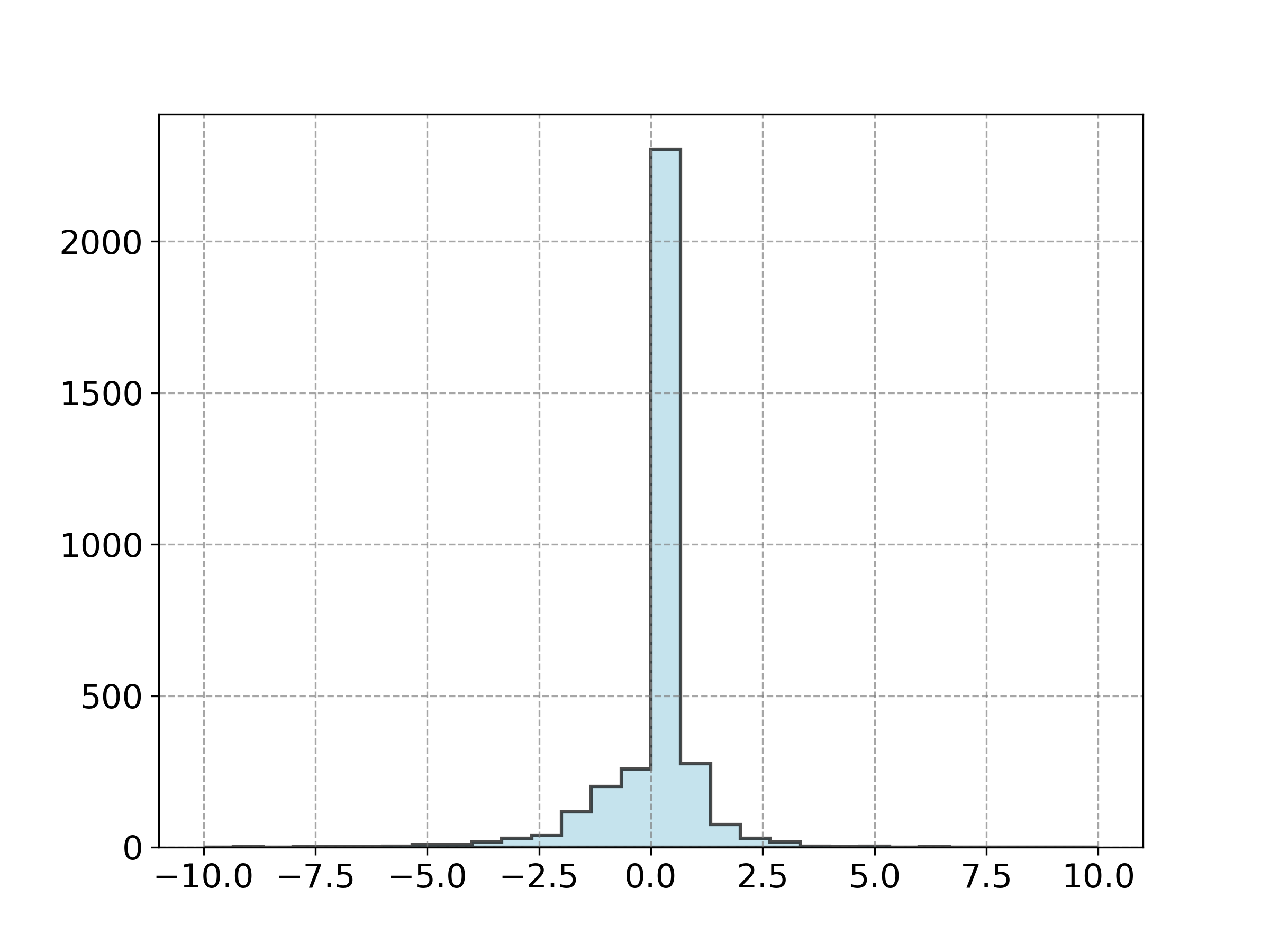}
\caption{Queue position: 0--10\% (front); large near-side queue; small opposite-side queue.}
\label{fig:early_position_large_near_small_opposite}
    \end{minipage}%
    \begin{minipage}{0.5\textwidth}
        \centering
        \captionsetup{width=.8\linewidth}
        \includegraphics[scale=0.36]{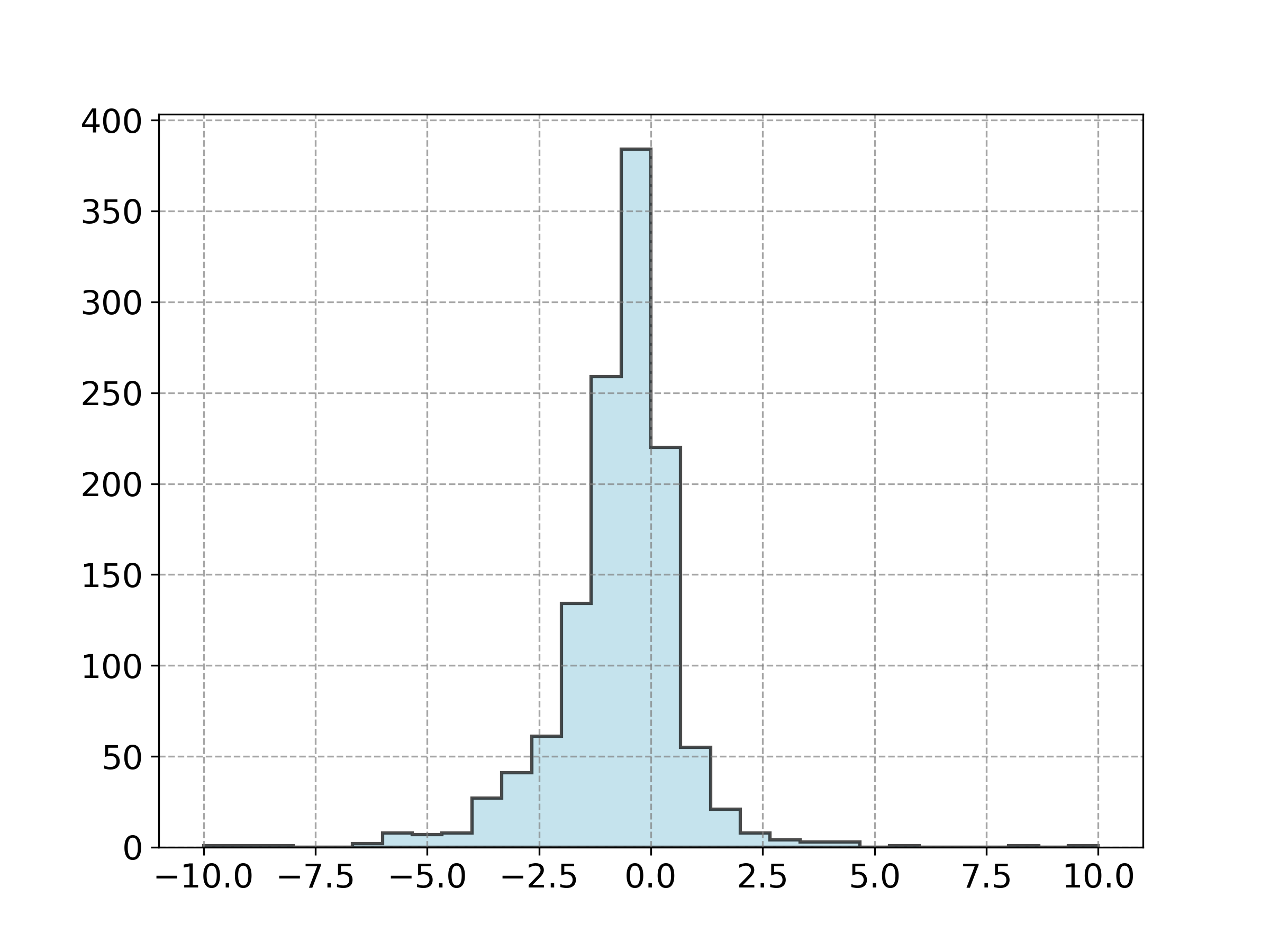}
        \caption{Queue position: 75--100\% (back); large near-side queue; small opposite-side queue.}
        \label{fig:late_position_large_near_small_opposite}
    \end{minipage}
\end{figure}

The markout return histogram for fills with a front-of-queue position is shown in Figure \ref{fig:early_position_large_near_small_opposite}, while that for back-of-queue fills is shown in Figure \ref{fig:late_position_large_near_small_opposite}.
In each histogram, a positive return  indicates a favorable subsequent price move and negative returns indicate an adverse price movement.
We further provide summary statistics of the distributions illustrated in these histograms, as well as others not shown, in Table~\ref{tab:maker_order_summary}.

The histograms reveal a striking difference in outcomes. For fills with a front-of-queue position, the distribution's mass is predominantly at positive values, with a notable spike at small positive markout returns, \ja{as reflected in the average value of $-0.058$ bp.}
Conversely, fills from maker orders at the back of the queue predominantly result in negative outcomes, with the majority of the distribution's mass being at negative values, \ja{and an average return of $-0.775$ bp.
Notably, the standard deviation of markout returns is also significantly larger for back-of-queue fills, indicating greater variability in outcomes.}
These observations suggest that trading in the direction of the order book imbalance with a maker order is only desirable if the order is filled from a front-of-queue position.\footnote{As we shall see later, trading in the direction of the imbalance with a taker order does not generally outperform the taker fee, rendering it a losing strategy overall.}

\begin{figure}[htbp]
    \centering
    \begin{minipage}{0.5\textwidth}
        \centering
        \captionsetup{width=.8\linewidth}
        \includegraphics[scale=0.36]{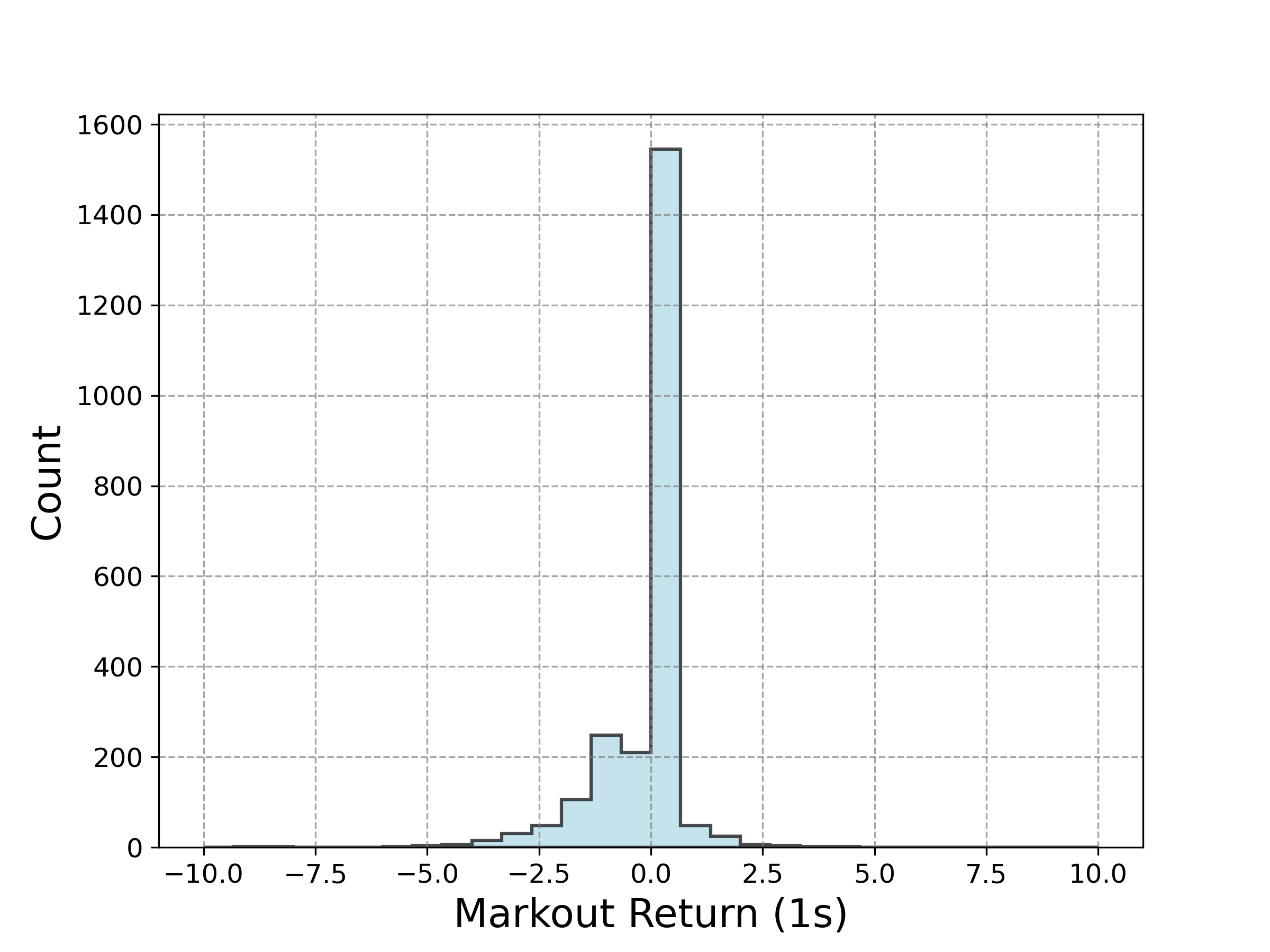}
        \caption{Queue position: 0--10\%; large near-side queue; large opposite-side queue.}
        \label{fig:early_position_large_near_large_opposite}
    \end{minipage}%
    \begin{minipage}{0.5\textwidth}
        \centering
        \captionsetup{width=.8\linewidth}
        \includegraphics[scale=0.36]{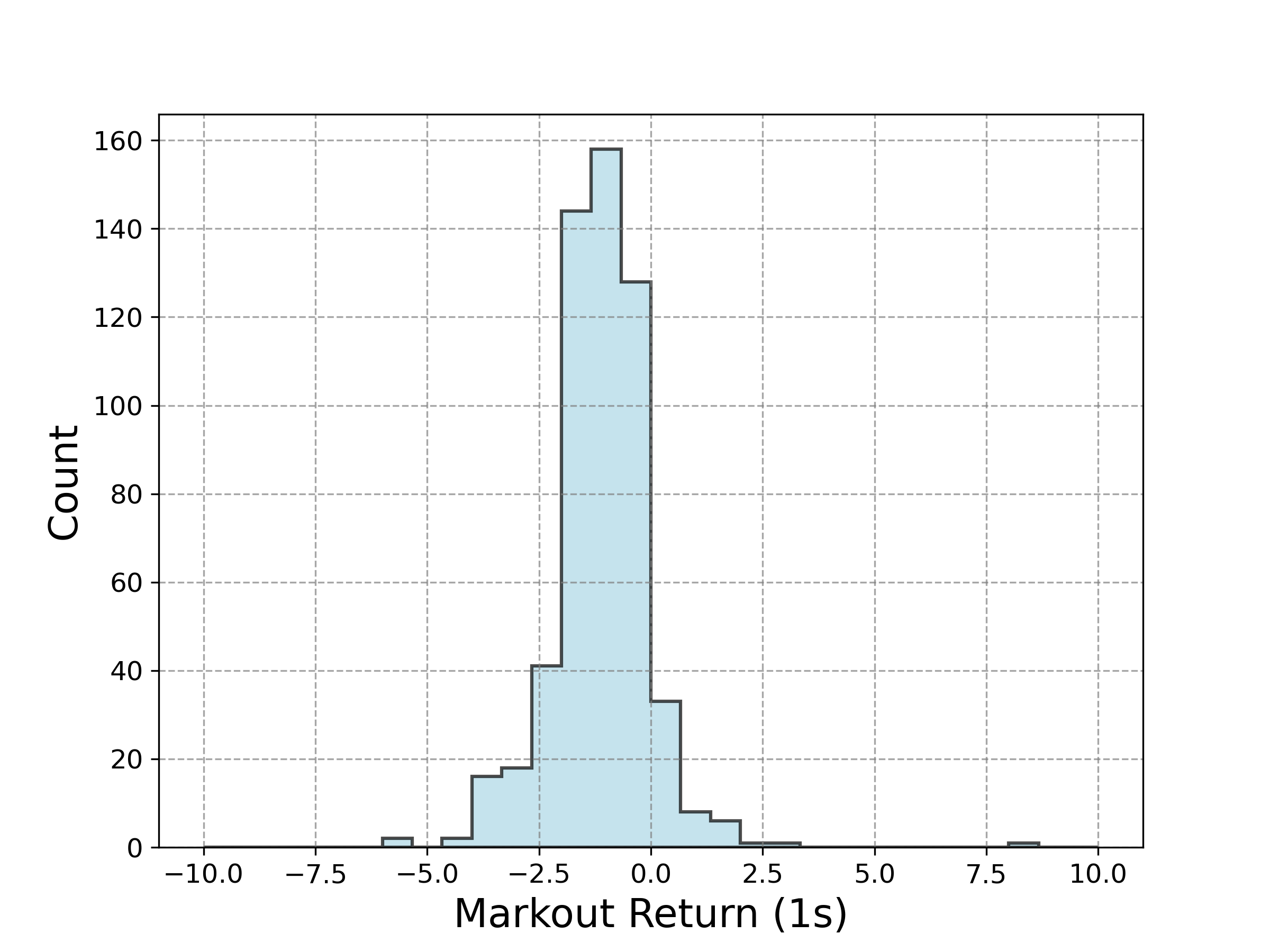}
        \caption{Queue position: 75--100\%; large near-side queue; large opposite-side queue.}
        \label{fig:late_position_large_near_large_opposite}
    \end{minipage}
\end{figure}

\paragraph{Both queues large.} Markout return  histograms for orders where
both \(Q^{\text{near}}_{T}\) and
\(Q^{\text{opp}}_{T}\) fall into the ``large" bin are shown in Figure \ref{fig:early_position_large_near_large_opposite} for
fills from orders within the 0\%-10\% front-of-queue position range, while those within the 75\%-100\% range are shown in Figure \ref{fig:late_position_large_near_large_opposite}.
Consistent with the previous pair of histograms, we see significantly better outcomes for front-of-queue orders.
\ja{The outcomes are slightly worse on average (see Table~\ref{tab:maker_order_summary}) compared with the previous case, consistent with the less favorable order book imbalance in this scenario: the average markout return is $-0.296$ bp for front-of-queue fills and $-1.157$ bp for back-of-queue fills.}
The back-of-queue histogram, in particular, shows that the vast majority of outcomes are adverse, with almost all fills resulting in negative markout returns.

\begin{figure}[htbp]
    \centering
    \begin{minipage}{0.5\textwidth}
        \centering
        \captionsetup{width=.8\linewidth}
        \includegraphics[scale=0.36]{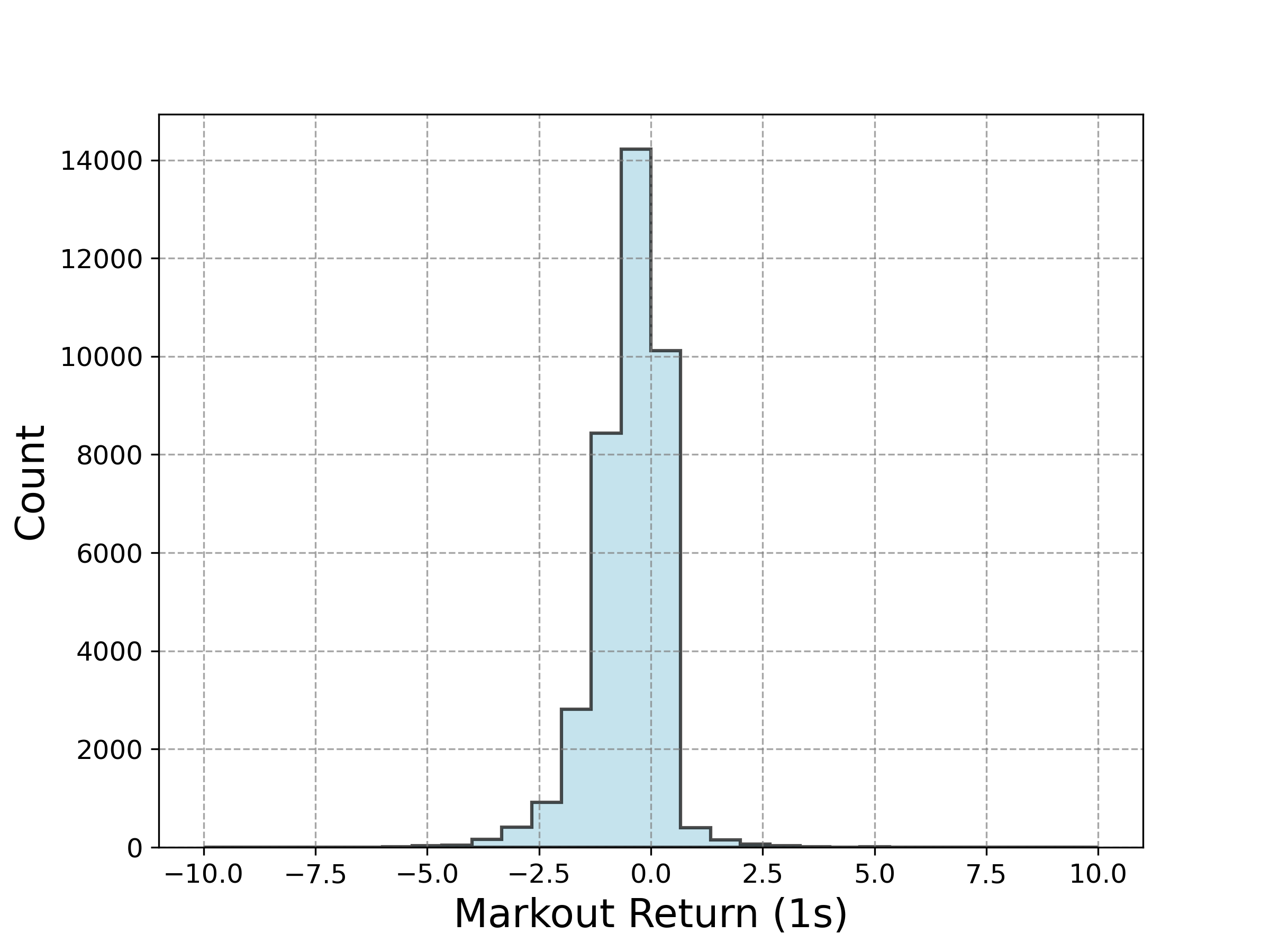}
        \caption{Queue position: 0--10\%; small near-side queue; large opposite-side queue.}
        \label{fig:early_position_small_near_large_opposite}
    \end{minipage}%
    \begin{minipage}{0.5\textwidth}
        \centering
        \captionsetup{width=.8\linewidth}
        \includegraphics[scale=0.36]{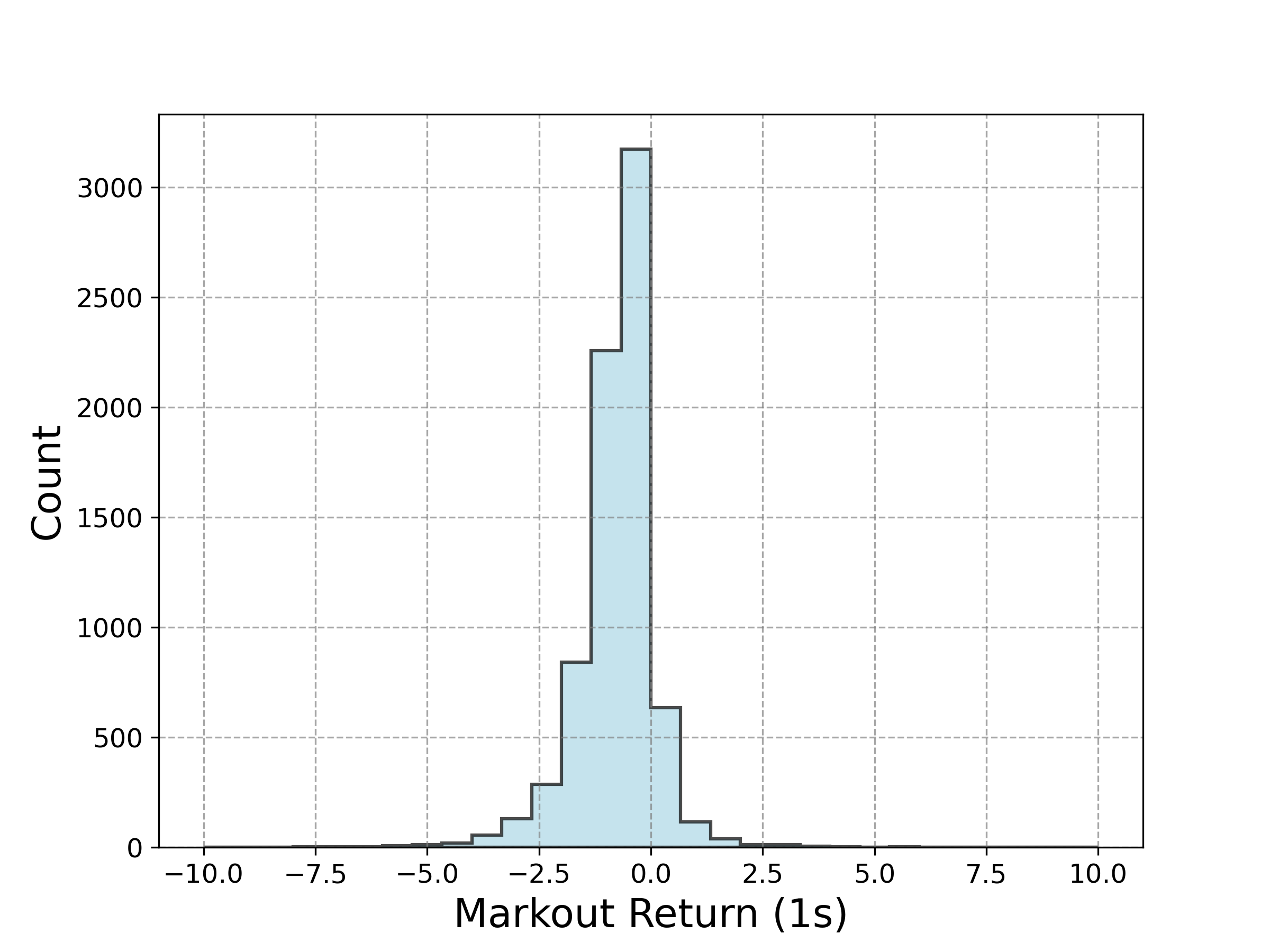}
        \caption{Queue position: 75--100\%; small near-side queue; large opposite-side queue.}
        \label{fig:late_position_small_near_large_opposite}
    \end{minipage}
\end{figure}

\paragraph{Small Near-side, Large Opposite-side Queue} The last pair of histograms in the main text (additional histograms are available in the appendix) show markout returns for orders  for which \(Q^{\text{near}}_{T}\) is small and \(Q^{\text{opp}}_{T}\) is large, indicating that the order book imbalance is unfavorable.
\ja{Outcomes for front-of-queue positions are shown in Figure \ref{fig:early_position_small_near_large_opposite}, with an average markout return of $-0.539$ bp as reflected in Table~\ref{tab:maker_order_summary}, while those for back-of-queue positions are shown in Figure \ref{fig:late_position_small_near_large_opposite}, where the average markout return is slightly worse at $-0.763$ bp.
Most outcomes in both cases are adverse, with markout returns predominantly negative.}
Interestingly, the difference between the two histograms is not significant, suggesting that when the queue is small, the queue position   does not matter much.
This aligns with our previous discussion and makes economic sense: if the queue is small, the liquidity behind an order at the beginning of the queue is not much different from that of an order at the back of the queue.

\ja{
Histograms for mid-size queues are available in Appendix~\ref{sec:more_histos}.
We omitted these here, along with histograms for intermediate queue positions, as the transitions from small to mid to large queue sizes, and from front-of-queue to middle-of-queue and to back-of-queue positions, are smooth and do not offer materially different insights from those presented above.
}

\begin{table}[htbp]
\centering
\caption{Summary statistics across different queue sizes and positions.}
\label{tab:maker_order_summary}
\begin{tabular}{llcccc}
\hline
\textbf{Queue Size} & \textbf{Queue Position} & \textbf{Avg (bp)} & \textbf{Min (bp)} & \textbf{Max (bp)} & \textbf{Std (bp)} \\
\hline
Large near, small opp. & 0 - 0.1    & $-0.058$  & $-9.2$  & $6.3$  & $1.0$ \\
                       & 0.1 - 0.4  & $-0.586$  & $-9.5$  & $7.0$  & $1.3$ \\
                       & 0.4 - 0.75 & $-0.743$  & $-8.6$  & $6.0$  & $1.3$ \\
                       & 0.75 - 1   & $-0.775$  & $-27.3$ & $9.5$  & $1.6$ \\
\hline
Large near, large opp. & 0 - 0.1    & $-0.296$  & $-8.9$  & $4.1$  & $0.9$ \\
                       & 0.1 - 0.4  & $-0.882$  & $-6.1$  & $3.0$  & $1.1$ \\
                       & 0.4 - 0.75 & $-0.967$  & $-7.2$  & $3.9$  & $1.1$ \\
                       & 0.75 - 1   & $-1.157$  & $-5.8$  & $8.0$  & $1.1$ \\
\hline
Small near, small opp. & 0 - 0.1    & $-0.562$  & $-16.7$ & $18.8$ & $1.6$ \\
                       & 0.1 - 0.4  & $-0.711$  & $-13.8$ & $8.8$  & $1.8$ \\
                       & 0.4 - 0.75 & $-0.622$  & $-14.0$ & $17.3$ & $1.9$ \\
                       & 0.75 - 1   & $-0.677$  & $-19.8$ & $20.2$ & $2.2$ \\
\hline
Small near, large opp. & 0 - 0.1    & $-0.539$  & $-19.0$ & $21.0$ & $0.8$ \\
                       & 0.1 - 0.4  & $-0.645$  & $-33.4$ & $5.2$  & $0.9$ \\
                       & 0.4 - 0.75 & $-0.686$  & $-9.8$  & $9.7$  & $0.8$ \\
                       & 0.75 - 1   & $-0.763$  & $-14.6$ & $10.7$ & $0.9$ \\
\hline
\end{tabular}
\end{table}

\FloatBarrier

\section{Fill and Be Killed?}
\label{sec:new_sec}

Many market makers report feeling a burst of anxiety immediately upon realizing they have received a fill.
That psychological distress exists for a good reason: they are painfully aware (through past loss-inflicting experiences) of the fact that it is much easier to get a fill accompanied by a subsequent adverse price change than one without.
After having examined the fundamental attributes of maker orders in isolation in the preceding sections, we now illuminate the tension between the two and the significant implications it holds for the use of maker orders broadly and market-making strategies in particular.

\paragraph{The connection between fill probability and returns for maker orders.}
We have highlighted two mechanical facts  for maker orders in our experiment: (1) If the  snapshot just after an order's terminal time (fill or cancel) shows  a mid-price move that is zero or  against the order, the order must have filled; (2)
If the snapshot shows a favorable mid-price movement the order must have been canceled. That is, the fill probability of a maker order, conditioned on the immediate mid-price change, is 0 when that change is positive, and 1 when it is zero or negative.
We now consider the relationship between an order's fill probability and its subsequent return over longer horizons.

\begin{figure}[htbp]
\centering
\captionsetup{width=\textwidth}
\includegraphics[scale=0.13]{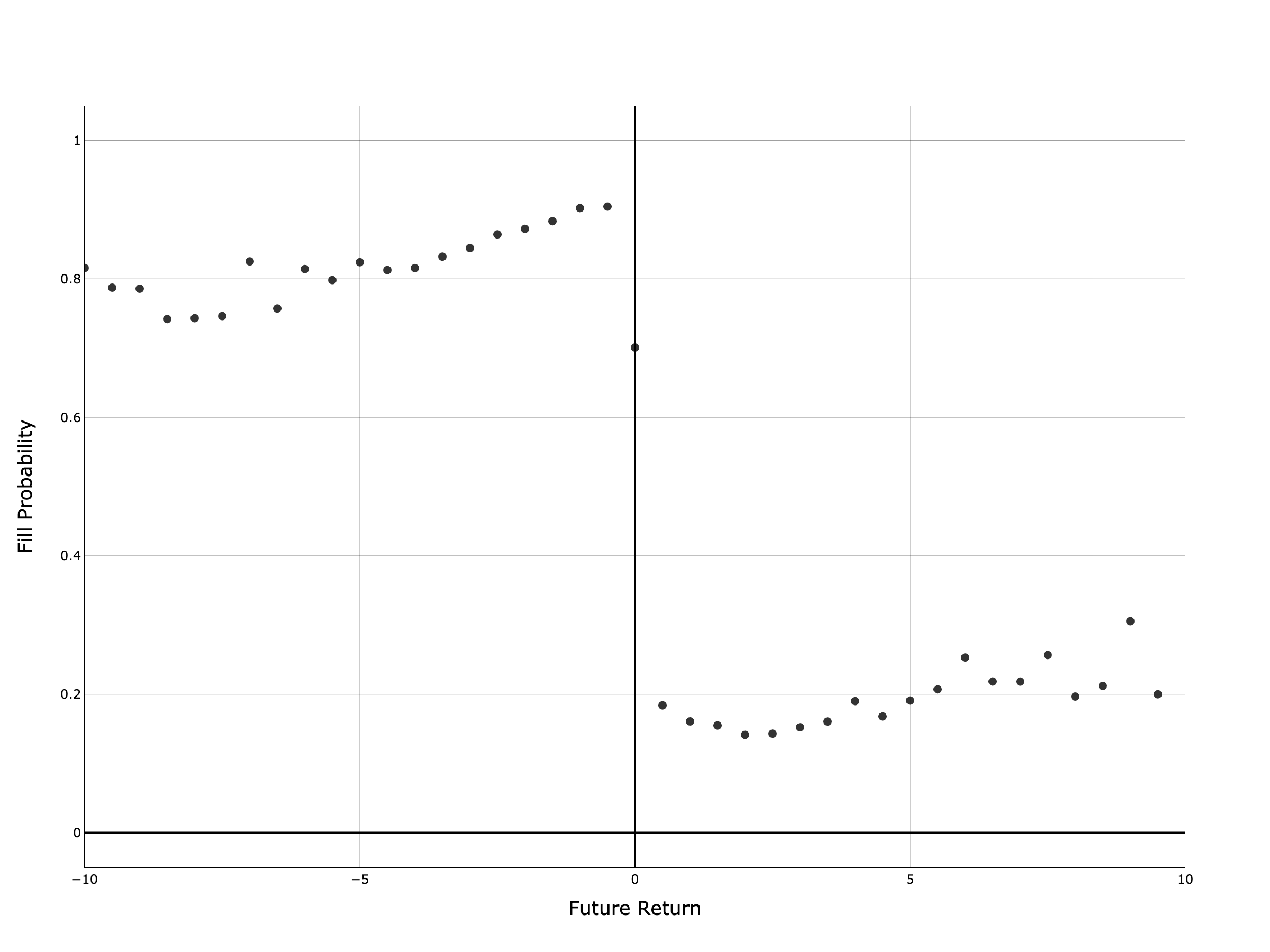}
\caption{Empirical relationship between fill probability and return measured five seconds from post-time.}
\label{fig:fill_prob_vs_fret}
\end{figure}

Figure~\ref{fig:fill_prob_vs_fret} demonstrates the relationship. The horizontal axis labels the on-paper returns of all the orders in our sample (whether filled or canceled because the top price moved away); returns are measured from the time of the order's terminal state (fill-or-cancel) five seconds into the future.
The vertical axis shows the proportion of those orders that filled; that is, an estimate of the fill probability given the subsequent return.
The plot reveals a notable anti-correlation: orders with negative subsequent five-second returns are highly likely to fill, whereas those with positive returns are much less likely to fill.
\ja{While this result alone does not guarantee an unconditional expectation of a negative return—since the negative-return orders might represent only a small portion of the entire sample, with their large negative returns being counterbalanced by small positive returns from a larger number of orders—our empirical data does show a negative expected return of approximately $-0.8$ bp, or $-0.3$ bp net of rebate.}

Our observations suggest a dilemma: maker orders are either likely to fill but with a negative short-term return (loss-making) or unlikely to fill and having a positive return (failing to realize a profit).
How does that dilemma affect the profitability of a basic market making strategy, often cited in the literature?

\paragraph{The PnL of a market making strategy.}
A common misconception vitiating the HFT and market making literature is the belief that market makers can simply post orders at the top bid and ask prices to earn the spread plus rebates.
In a number of such papers (see for instance~\cite{Cartea_Jaimungal_2016,Avellaneda_2008}), it is commonly assumed that those maker orders fill randomly at Poisson rates, with no impact on the subsequent mid-price evolution, which is often modeled as a continuous diffusion process.

The cruel reality is that such market making strategies are a quick path to bankruptcy due to negative price drift, which we now illustrate with a case study of the prototypical naive market making strategy.
We mirror the approach of~\cite{DeLise:2024aa}, who carried out a similar analysis of the same strategy in US treasury markets, and it is interesting to note that the same results hold in our completely different asset class.
That strategy can be specified as follows:
\begin{tcolorbox}[title=\textbf{Basic MM Strategy}, colback=white, coltitle=white]
\begin{enumerate}[label=\textbf{Step \arabic*:}, leftmargin=*, labelsep=2mm]
    \item Post a minimum-sized buy order at the top bid and a minimum-sized sell order at the top ask.
    \item If the position differs from zero (because one of the two initial orders has filled), post only on the side that will return the position to zero:
    \begin{itemize}
        \item Keep the unfilled order open while it is posted at the touch.
        \item Should the price change without the order being filled, so it no longer being posted at the top price, cancel the order and place a new one at the new top price.
    \end{itemize}
    \item Once a fill is received that returns the inventory back to zero, return to Step 1.
\end{enumerate}
\end{tcolorbox}

One might imagine that by buying low and selling higher (by the spread), the strategy would capture the spread and earn twice the maker rebate per pair of posts.
However, this is only true if both orders are filled without any change in the midprice. The strategy described above is "naive" primarily because it lacks a sophisticated cancellation policy; orders are left in the book even when a skilled trader might cancel them to avoid adverse selection.
In reality, it is far more common that, after one order has filled, an adverse midprice movement forces the cancellation and reposting of the unfilled order at a worse price.
We use our empirical data to assess the impact of this fact and later illustrate how a simple, signal-based cancellation policy can mitigate, but not eliminate, these losses.

In our overall experiment, we continuously quoted minimum-sized orders on both bid and ask without inventory restrictions. The orders submitted by the basic market-making  strategy are a subset of the whole. We extract these and analyze the performance of the strategy.\footnote{
Our results with minimum-sized orders are indicative of what one might expect with larger order sizes.
One might reasonably expect results to deteriorate with larger order sizes: the literature on market impact tells us that the market impact of taker orders grows with their size.
Hence, since the market impact of a taker order is the same as that of  adverse selection on the maker order on the other end of the trade, the subsequent expected return of a maker order ought to decrease in its filled amount.
}

\begin{figure}[htbp]
    \centering
    \begin{minipage}{0.5\textwidth}
        \centering
        \captionsetup{width=.8\linewidth}
        \includegraphics[scale=0.36]{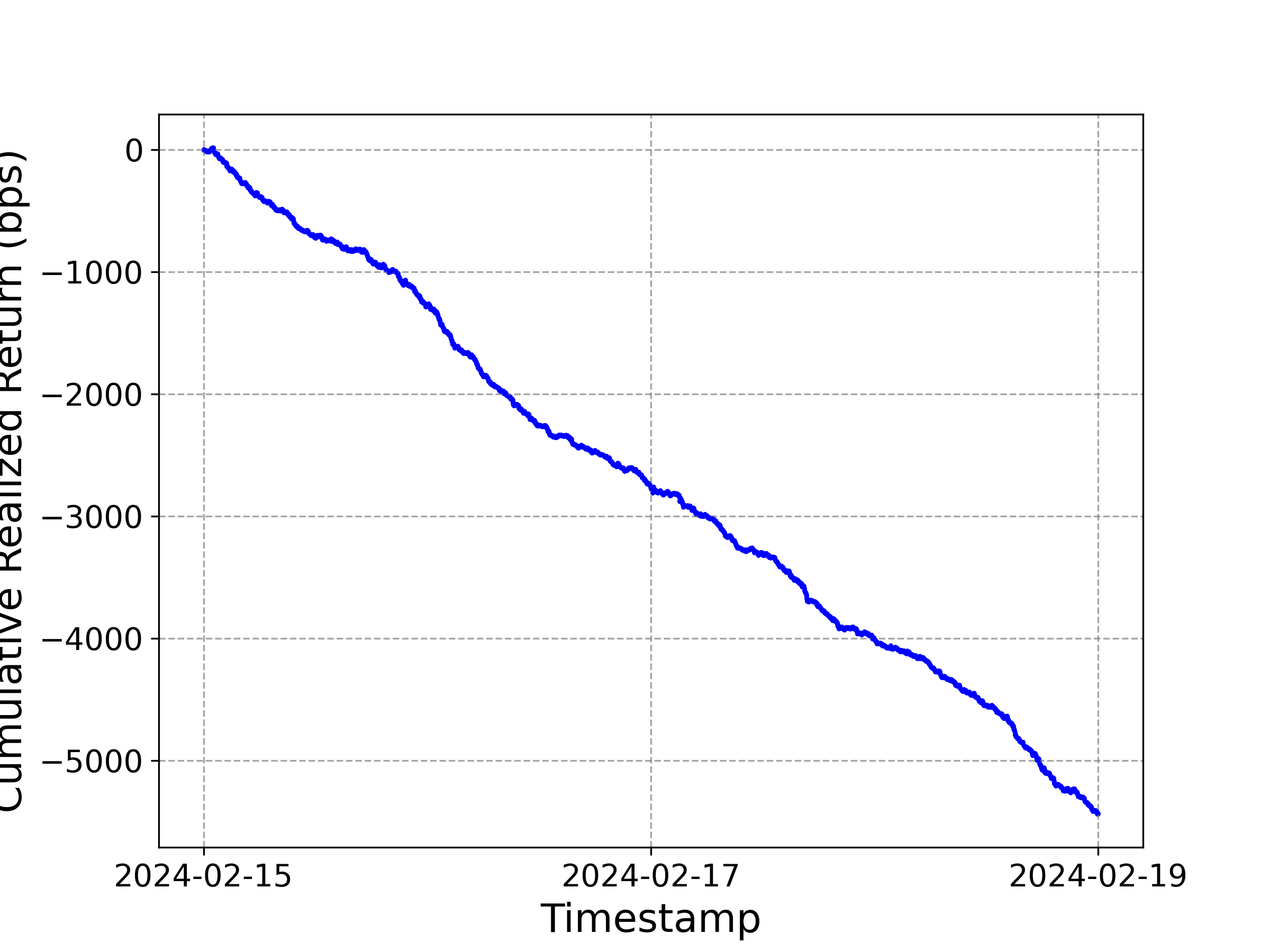}
        \caption{PnL plot of the basic MM strategy, net of fees.}
        \label{fig:timeseries_realized_bp_diff_cumsum}
    \end{minipage}%
    \begin{minipage}{0.5\textwidth}
        \centering
        \captionsetup{width=.8\linewidth}
        \includegraphics[scale=0.36]{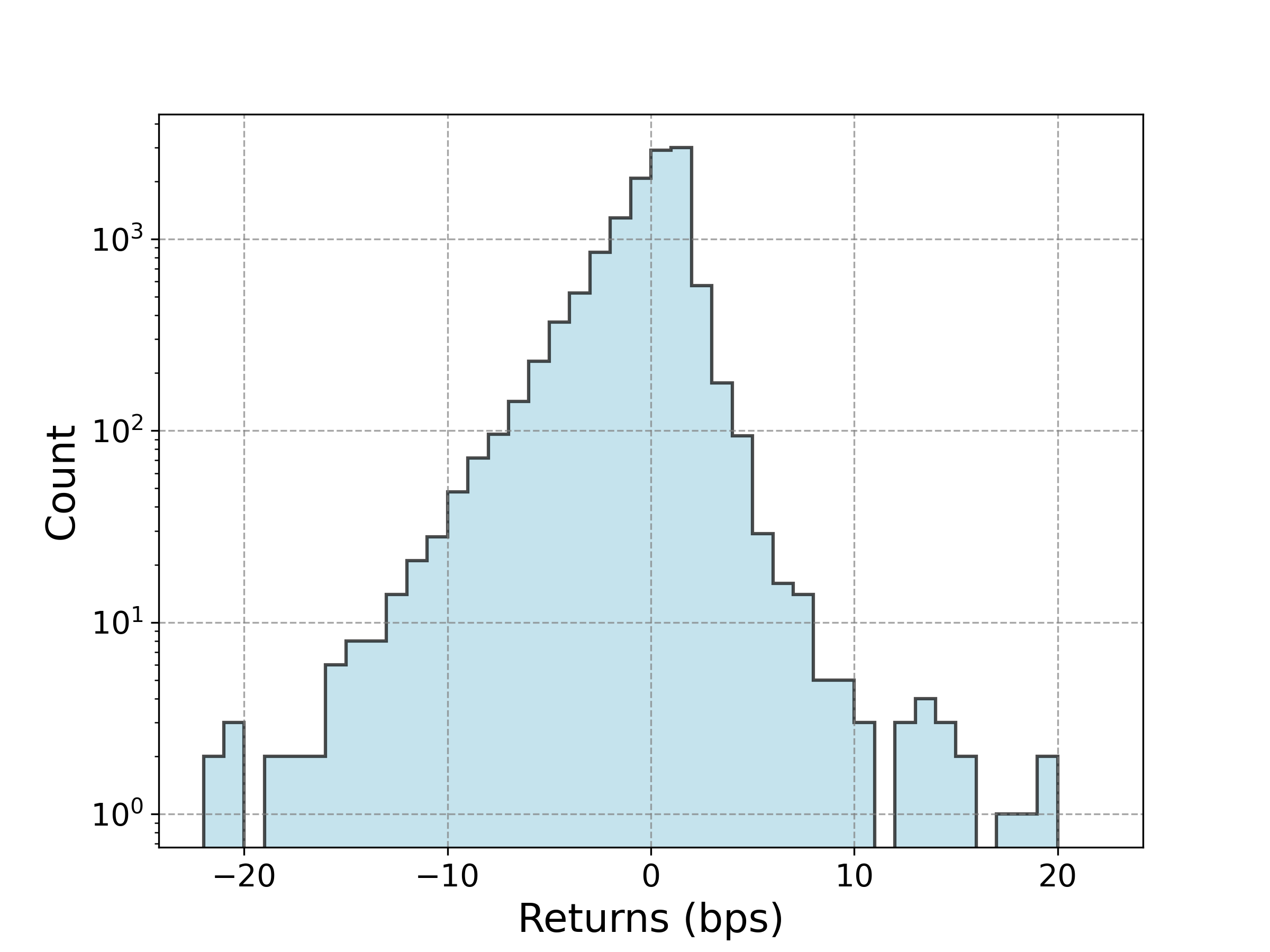}
        \caption{Histogram of pre-fee returns of roundtrip trades of the basic MM strategy.}
        \label{fig:realized_bp_diff_naive}
    \end{minipage}
\end{figure}

In Figure~\ref{fig:timeseries_realized_bp_diff_cumsum} we show the strategy’s cumulative return  (including maker rebates) in basis points over a period of approximately three days and seven hours.
By the end of this period, the strategy has lost almost  60\%; in annualized units, the Sharpe ratio is a staggering  $-109.0$.
The average holding time between opening and closing of a position is approximately 11 seconds.

Figure~\ref{fig:realized_bp_diff_naive} illustrates the distribution of realized spreads (the price difference in basis points between an opening fill that changes the position from 0 to $\pm 1$, and the subsequent rebalancing fill) with the frequency on a logarithmic scale.
While the most frequent outcome for a round-trip is indeed a modest positive return, the strategy is plagued by a high occurrence of larger losses, and occasional catastrophic ones, which overwhelm the small gains.

The naive market making strategy is thus a recipe for poverty -- heaven help the starry-eyed novice traders, fresh from academia, who, seduced by maker rebates, attempt to implement this strategy in practice.

\FloatBarrier

\paragraph{Trading with the imbalance.}
The naive strategy, which is prima facie appealing because it receives the rebate and benefits from the spread, is in fact highly unprofitable. We now ask whether  the same is true for approaches  which incorporate simple  signals into decision-making, posting orders more selectively based on a signal or canceling them when market conditions change adversely.

A canonical example of a trading signal, widely accepted in the microstructure literature, is the order book imbalance; indeed, we have referred to it repeatedly above.
We investigate a set of imbalance-based strategies commonly found in the literature (see for instance~\cite{Cartea_2018aa,Cont_2023}) which, while still relatively naive, also have prima facie appeal. We show that this, too, is a false hope.

The strategies examined include:
\begin{itemize}
    \item \textbf{Imbalance-based maker strategy:} Post maker orders in the direction of the imbalance. This strategy differs from the naive one by only submitting new orders when the imbalance is favorable to the maker order (Imbalance threshold to post: magnitude $> 0.5$).
    \item \textbf{Imbalance-based maker strategy with cancellation:} Similar to the above, but with an added rule to cancel posted orders if the imbalance shifts adversely to the order (Imbalance threshold to post: magnitude $> 0.5$; cancel threshold: 0.0).
    \item \textbf{Imbalance-based taker strategy:} Trade in the direction of the imbalance using taker orders, employing the same inventory-balancing approach as the maker strategies (Imbalance threshold to submit a taker order: magnitude \( > 0.5 \)).
    \footnote{Latency is ignored here, potentially leading to an overstatement of results, see~\cite{Albers_2024}.}
\end{itemize}

\begin{figure}[htbp]
    \centering
    \hspace{-0.06\textwidth}
    \begin{minipage}{0.4\textwidth}
        \centering
        \captionsetup{width=.8\linewidth}
        \includegraphics[scale=0.37]{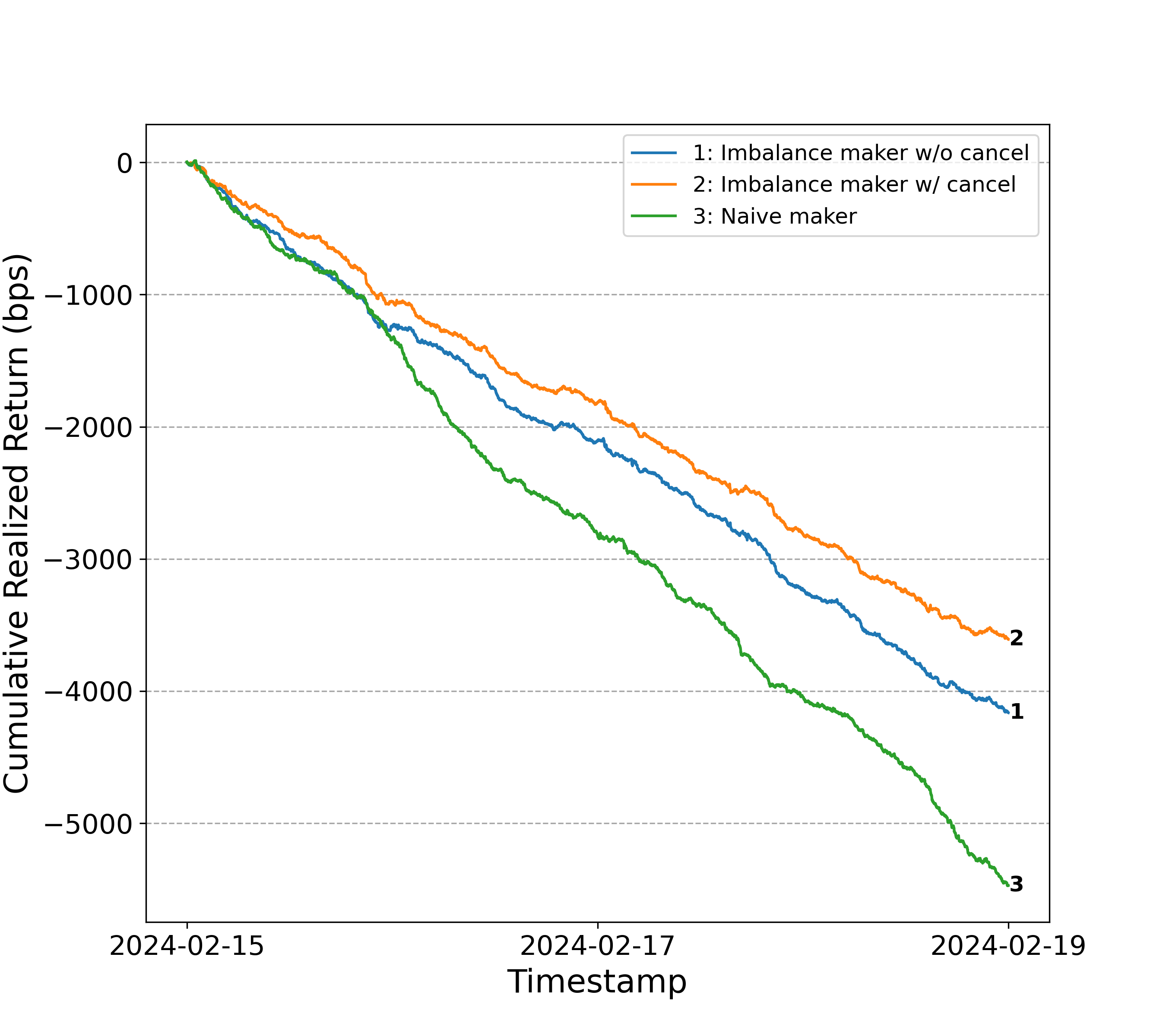}
        \caption{PnL plots of the maker strategies, net of fees.\\\\}
\label{fig:pnl_plots_maker_strategies}
    \end{minipage}%
    \hspace{0.1\textwidth}
    \begin{minipage}{0.4\textwidth}
        \centering
        \captionsetup{width=.8\linewidth}
        \includegraphics[scale=0.37]{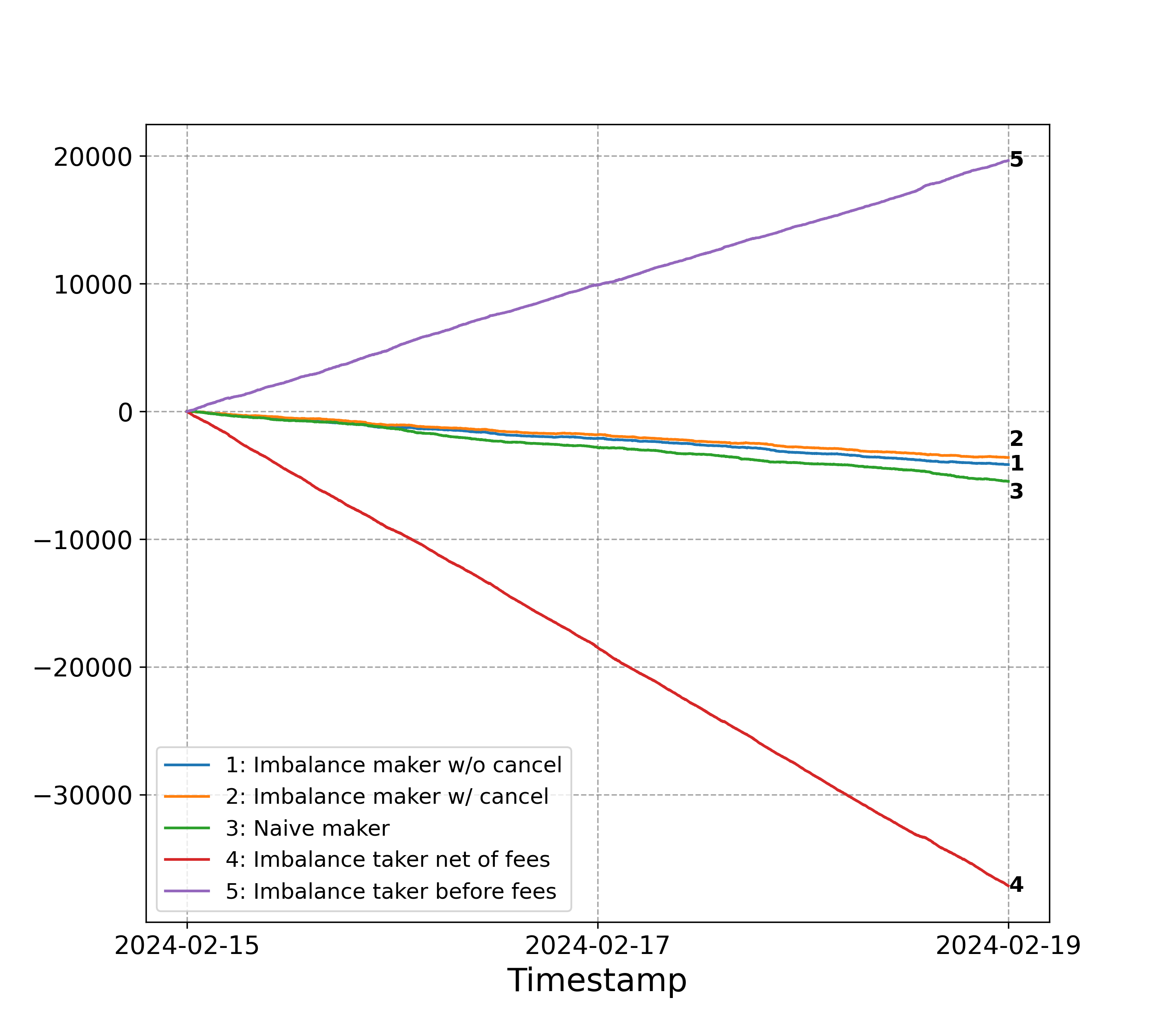}
        \caption{PnL plots of maker strategies (net of fees) and the taker strategy (both net of fees and before fees).}
        \label{fig:pnl_plots_maker_and_taker_strategies}
    \end{minipage}
    \hspace{0.15\textwidth}
\end{figure}

The cumulative PnL of the maker strategies, net of fees, is shown in Figure~\ref{fig:pnl_plots_maker_strategies}, while Figure~\ref{fig:pnl_plots_maker_and_taker_strategies} includes the PnL of all strategies, including the taker one. For the taker strategy, we provide two plots: one net of fees and the other before fees, to highlight the substantial impact of fees in high-frequency trading. Table~\ref{tab:maker_taker_strategies_statistics} summarizes key statistics for each strategy, including the number of trades, mean and standard deviation of post-fee returns, and average holding time.

The results are stark. All imbalance-based strategies—whether maker or taker—perform poorly, with negative returns across the board. The imbalance-based maker strategy with cancellations only slightly mitigates losses compared with its counterpart without cancellations. The taker strategy, which trades most frequently, suffers the worst performance, despite being highly profitable before fees.
These results underscore significant challenges of trading profitably using simple public signals such as the imbalance:
\begin{itemize}
    \item The taker strategy is \ja{rendered unprofitable} by the taker fee: while its pre-fee PnL is impressive (around +1 bp per roundtrip), paying the taker fee on each leg of the roundtrip erodes its profitability completely.
    \item The maker strategies are primarily undermined by fill probability: they almost always miss out on favorable price moves and are left with the smaller number of fills corresponding to reversals (situations where the imbalance incorrectly predicts the next price change). The cancellation rule only mitigates a subset of those adverse cases, namely the ones where imbalance slowly changes adversely, rather than sudden adverse price moves caused by, e.g., large taker orders.
\end{itemize}

Thus, we conclude that it is challenging to overcome the taker fee with taker strategies, even when the signal provides a  robust pre-fee edge.
Furthermore, maker strategies, despite not paying taker fees, do not offer an easy way around that, given the interplay of queue position, fill probability, and price drift.
More sophisticated approaches based on other signals are therefore necessary (if not necessarily sufficient). We describe one such in the remainder of this paper.

\begin{table}[h]
   \centering
   \caption{Statistics of trading strategies, net of fees.}
   \label{tab:maker_taker_strategies_statistics}
   \small
   \resizebox{\textwidth}{!}{%
   \begin{tabular}{c | c c c c}
       \toprule
       \textbf{Strategy} & \textbf{Trades} & \textbf{Mean Return} & \textbf{Std Dev Return} & \textbf{Avg Holding Time (sec)} \\
       \midrule
       Imbalance taker & 18927 & $-1.9621$ & 2.1653 & 15.04 \\
       Naive maker & 12623 & $-0.4307$ & 2.6486 & 22.56 \\
       Imbalance maker w/o cancel & 8851 & $-0.4705$ & 2.8529 & 32.17 \\
       Imbalance maker w/ cancel & 7333 & $-0.4921$ & 3.1143 & 38.83 \\
       \bottomrule
   \end{tabular}%
   }
\end{table}

\paragraph{Threading the needle.}
In light of the grim evidence we have presented so far in this section, what can market makers do to thread the needle between (1) having to absorb negative instantaneous price drift necessary to obtain a fill and (2) experiencing positive price drift at slightly longer horizons (several seconds, say)? That is, how can they be filled by a taker order whose direction is negatively correlated to that of subsequent orders within the markout time?

In Section \ref{sec:fill_outcome}, we saw that most configurations of the top-of-book queue sizes and near-side queue position result in mainly negative outcomes; the  exception is  for front-of-queue orders (\(\operatorname{QP}_T \approx 0\)) in a large near-side queue where
returns are notably positive when the opposite-side queue is small, i.e. with a favorable order book imbalance.
This suggests that trading in the direction of the imbalance with a maker order near the front of the queue has a high fill probability and is profitable if the order fills, but the fill becomes less likely, and the trade becomes increasingly unprofitable as the queue position moves to the back of the queue.
These points raise the question: How can one achieve an early queue position in a large near-side queue with a favorable order book imbalance, and thereby benefit from the well-documented positive correlation between imbalance and returns?

Posting at the end of a large queue when the imbalance is favorable is not the answer, as the order will be at the back of a large queue and subject to a poor expected return.
Instead, the order needs to be posted when the queue is still small but becomes large shortly thereafter.
This usually implies posting the order when the order book imbalance is unfavorable\footnote{This excludes cases where a recent price change temporarily widens the spread to more than one tick, creating an opportunity to post at the new top-of-book price—a scenario where being among the first to post is primarily a technical challenge, involving latency optimization and the strategic use of special order types.} (unless the opposite-side queue also happens to be small), again with an \emph{a-priori}  poor expected return.

That is, to position an order to trade profitably with the order book imbalance, two things need to happen: one first needs to trade against an unfavorable imbalance, and then the imbalance has to change favorably.
On the basis of the information we have presented thus far,  this  is not, in general, a profitable approach without further information in the form of a signal in favor of the required sequence of events.
We now discuss the construction of just such a signal.

\FloatBarrier

\section{Reversals}
\label{sec:reversal}

We noted in Section \ref{sec:fill_prob} that orders posted when the opposite-side queue is large, and the near-side queue is small (implying an adverse imbalance) are associated high fill probabilities.
While those conditions usually imply a poor expected return, as established in Section~\ref{sec:fill_outcome}, there is a subset of cases where the initially adverse imbalance shifts favorably after post-time, i.e. when the initial imbalances falsely predicts the next price change.
These scenarios, which we term \textit{reversals}, represent instances where the typical negative correlation between fill probability and post-fill returns can break down.
While not a fully-fledged trading strategy, identifying such reversals offers a potential resolution to the fundamental challenge facing maker orders. We present one such model not as a definitive solution, but to illustrate the mechanics involved and demonstrate that such opportunities can, in principle, be identified.

\begin{figure}[htbp]
\centering
\captionsetup{width=\textwidth}
\includegraphics[scale=0.22]{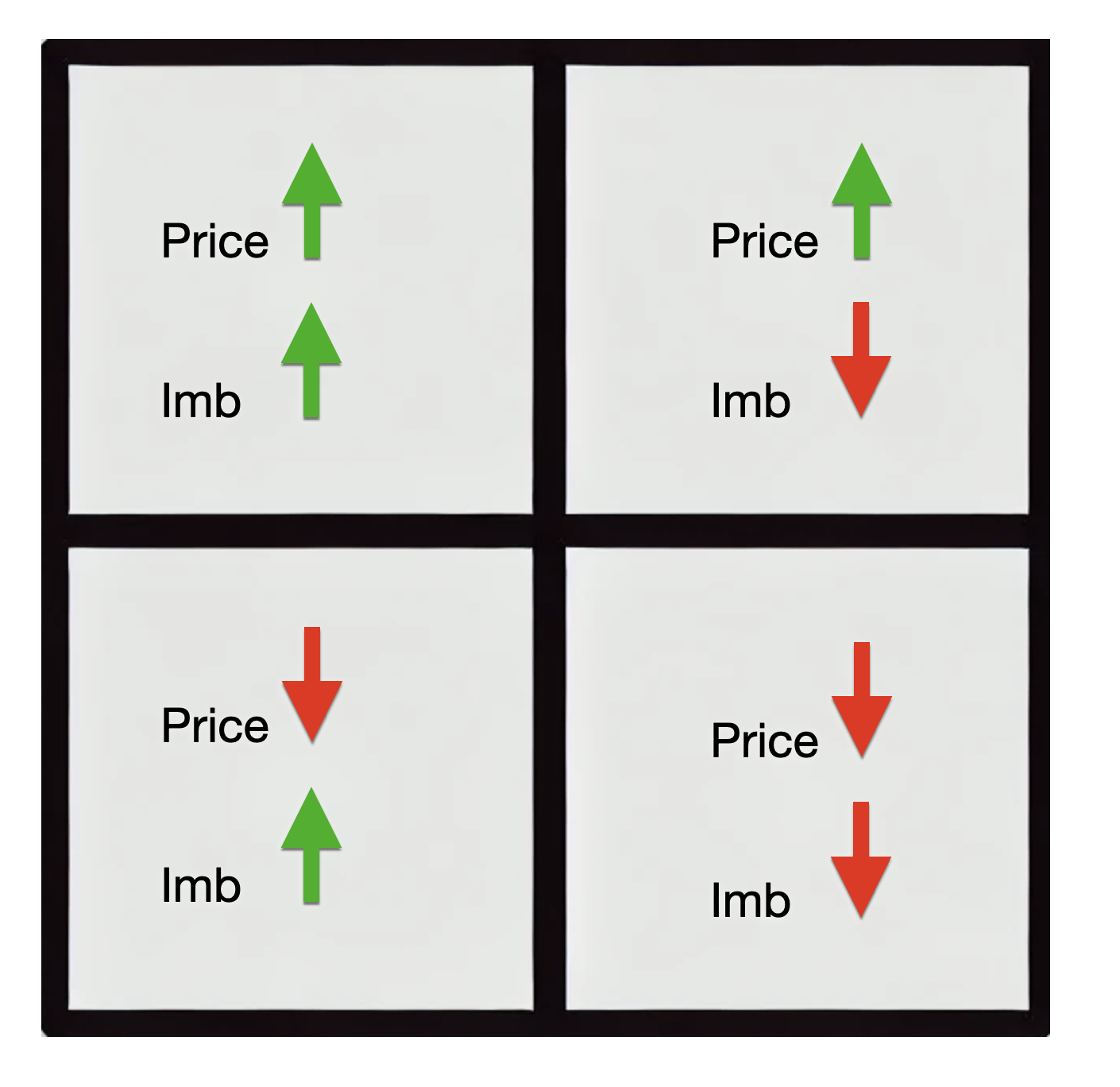}
\caption{Confusion matrix: imbalance as a predictor of next price change. The off-diagonal cells represent reversals: the top right (bottom left) cell is an opportunity to post a sell (buy) order with high fill probability and positive price drift.}
\label{fig:imbalance_confusion_matrix}
\end{figure}

Figure~\ref{fig:imbalance_confusion_matrix} shows a confusion matrix of the imbalance as a predictor of the next price change.
Orders trading against the imbalance in the off-diagonal cases (reversals), where the imbalance is a false predictor, have a high fill probability and yield a favorable subsequent return.

The challenge lies in distinguishing these off-diagonal cases (reversals) from the more common diagonal cases where the next price change aligns with the imbalance-implied direction.
The rest of this section is dedicated to detailing our modeling strategy for predicting reversals, developing a feature set, and evaluating the results.

\subsection{Modelling Approach}
\label{subsec:modelling_approach}

Let us now introduce an operational definition of the concept of a reversal, as introduced earlier, and outline our approach for predicting such events.

First, we extract the subset of orders with a high estimated fill probability, specifically those exceeding a chosen threshold of 80\% at submission time\footnote{The results remain consistent across different threshold probabilities,} -- our estimate is obtained by mapping the initial bid and ask queue sizes onto the fill probability surface illustrated in Figure~\ref{fig:mesh_buy}.
This subset of orders, denoted by \(\mathfrak{O}\), forms our data set, comprising 182,381 orders spanning a one-week period.
We utilize the first half of this data set for training and the second half for testing.
On average, the opposite-side queue in this data set is 40\% larger than the near-side queue, implying that these orders are submitted with an initially adverse imbalance.

We then set up a classification task aimed at distinguishing orders in the set \(\mathfrak{O}\) that experience the negative price drift implied by the high fill probability (adverse bid-ask skew) from those that fill and experience positive price drift.
That is, we define a binary label \(\varphi: \mathfrak{O} \rightarrow \{0,1\}\) such that \(\varphi(\mathfrak{o}) = 1\) for \(\mathfrak{o} \in \mathfrak{O}\) if it satisfies the following criteria:
\begin{enumerate}
    \item The order \(\mathfrak{o}\) fills, with some fill time \(T = T(\mathfrak{o})\).

    \item The order achieves a positive short-term return: let \(T' \geq T\) denote the timestamp of the first post-fill price change.
    The requirement is: \(\operatorname{ret}_{T, T'}(\mathfrak{o}) := \operatorname{sgn}(\mathfrak{o}) \left(p(\mathfrak{o}) / p_{T'} - 1\right) \cdot 10,000 > 0\). Here, \(p(\mathfrak{o})\) denotes the order's limit price, while \(p_{T'}\) is the top bid price at time \(T'\) for a buy order or the top ask price for a sell order, with the signed return accounting for order direction using \(\operatorname{sgn}(\mathfrak{o}) = 1\) for buys and \(-1\) for sells.
\end{enumerate}
Orders $\mathfrak o$ that do not satisfy the above criteria are labeled $\varphi(\mathfrak o)= 0$.
Here $\operatorname{ret}_{T, T'}(\mathfrak{o})$ is the return in basis points (bp) between the order's limit price and the price next midprice $p_{T'}$ at the next midprice change after the order's fill time.

To predict this output variable, we employ a logistic regression model using a set of features (discussed in the next subsection) known at the time of order submission. The model is defined as
\[
P(\varphi(\mathfrak{o}) = 1) = \frac{1}{1 + \exp\left(\alpha_0 + \sum_{i=1}^{n_f} \alpha_i x_i\right)},
\]
where \(\mathbf{x} = (x_1, \cdots, x_{n_f}) \in \mathbb R^{n_f}\) is the feature vector containing \(n_f \in \mathbb{N}\) distinct features.

As a benchmark for our logistic regression model, we also utilize random forests.
While these models extend beyond linear approaches, they lack the interpretability of logistic regression.

\subsection{Features}
\label{subsec:features}

In this subsection, we outline the features we use to predict reversals.  We expect this to be inherently challenging due to the ease with which accurate predictions can be parlayed into trading profits (cf. Unprofitability Principle, Section \ref{subsubsec:trade_offs}).
Although the literature has not directly addressed our specific problem, we can draw insights from related prediction tasks, such as short-term return predictions.
For instance, \cite{Ait_Sahalia_2022} constructs non-overlapping lookback windows to calculate various variables.
Similarly, \cite{Cohen_2023} develops a feature set designed to capture the current state of the LOB and recent LOB activity, which they use to forecast market participants' trading decisions.

Our approach encompasses four groups of features:
\begin{enumerate}
    \item \textbf{Price Dynamics}:
    This group includes two measures of volatility over relatively long lookback periods:
    \begin{itemize}
        \item \texttt{stdev\_100}, \texttt{stdev\_500}: Standard deviations of the 100 (500) most recent ten-second returns, calculated from the last traded price, spanning approximately 16.66 minutes and 83.33 minutes, respectively.
    \end{itemize}
    Additionally, we analyze price movements across multiple time scales—100 milliseconds, 1 second, 5 seconds, 30 seconds, and 5 minutes---denoted as \texttt{100ms}, \texttt{1s}, \texttt{5s}, \texttt{30s}, and \texttt{300s}. For each scale, we consider three consecutive non-overlapping windows (\texttt{w0}, \texttt{w1}, \texttt{w2}) and compute:
    \begin{itemize}
        \item \texttt{amplitude}: The range between maximum and minimum prices within the window.
        \item \texttt{ret\_vwap}: Log return of the volume-weighted average price (VWAP) within the window relative to the current top-of-book price.
    \end{itemize}

    \item \textbf{Trade Volume Patterns}: These features capture trading activity within the same windows:
    \begin{itemize}
        \item \texttt{max\_size}, \texttt{avg\_size}: Maximum and average trade sizes.
        \item \texttt{buy\_count}, \texttt{sell\_count}: Counts of buy and sell trades.
        \item \texttt{total\_buy}, \texttt{total\_sell}: Total volumes of buy and sell trades.
    \end{itemize}

    \item \textbf{Momentum}: This group reflects the persistence of price trends and trading activity, computed over the same lookback windows as the features in the preceding groups:
    \begin{itemize}
        \item \texttt{ret\_autocov}: Autocovariance of returns from consecutive trades.
        \item \texttt{ret\_sum}: Cumulative log returns of trades.
        \item \texttt{trade\_intensity}: Average time between consecutive trades.
    \end{itemize}

    \item \textbf{LOB State}: We assess the liquidity profile of the LOB:
    \begin{itemize}
        \item \texttt{top\_bid\_liq}, \texttt{top\_ask\_liq}: Liquidity at the top bid and ask levels.
        \item \texttt{ob\_bid\_half}, \texttt{ob\_ask\_half}: Measures of bid- and ask-side liquidity, defined as the price difference in basis points between the top bid (ask) price and the volume-weighted average price to execute \$500K on the bid (ask) side; lower values indicate high liquidity near the touch, while larger values indicate an illiquid order book side.
        \item \texttt{totb\_mean}: Average top-of-book survival time in the past ten minutes.
        \item \texttt{age}: Duration in seconds since the last top-of-book price change.
    \end{itemize}
\end{enumerate}

In total, our model uses 173 features, all of which are observable at the time of order submission, thus ensuring that the model's predictions are actionable from a trading standpoint: traders can compute the model's prediction in real-time and decide whether to proceed with posting the order or wait for a more opportune moment.

Furthermore, our features are pre-processed such that the model need not differentiate between the direction of the order (buy or sell).
Concretely this means that instead of using "buy" and "sell" we use "same-direction" and "opposite-direction", or instead of "bid" and "ask" we use "near-side" and "opposite-side", and for some other features such as returns it means multiplying them by a sign based on the direction of the order.
All of this allows the model to treat buys and sells equivalently during training and prediction.

While there is some correlation among these features---such as the high correlation between volatility and trading volumes---they are intentionally designed to minimize cross-correlation.
This is achieved by utilizing consecutive non-overlapping lookback windows for each fixed time scale.
When different time scales are considered, they differ by approximately an order of magnitude, ensuring that the overlap between lookback windows across these scales is small.
We have thus carefully considered cross-correlations in the construction of our feature set to ensure that even the largest values remain sufficiently low.
This allows us to train our models—logistic regression and random forest—directly on the original feature set without the need for additional pre-processing or dimensionality reduction, steps that could reduce the explainability of the results.

\subsection{Results}
\label{subsec:results}

In the realm of trading, traditional metrics for evaluating classification models, such as prediction accuracy, often fall short of capturing the true metric of interest: economic significance or trading profit.

Consider, for instance, that in our data set, only about 15\% of orders are reversals.
A model could achieve an ostensibly high prediction accuracy of 85\% simply by always predicting a non-reversal.
However, such a model would never trigger any trading activity, thereby failing to generate any profit or trading edge.

To further illustrate that the link between prediction accuracy and economic significance is tenuous, suppose the model's out-of-sample predictions of reversals are mostly false positives.
Even in that case, the model might still offer economic value: the post-fill returns of orders placed based on these predictions could still be substantially better than the post-fill returns of randomly placed orders, thus potentially providing traders with a valuable tool for improving the performance of their fills, a phenomenon that our subsequent analysis reveals in our model's results.

We evaluate the economic significance of our model's predictions via a
twofold approach, involving the assessment of two canonical trading strategies informed by our model's predictions:
\begin{enumerate}
    \item \textbf{Undirectional Strategy}: This strategy involves placing orders only when the model predicts a reversal with a probability exceeding a threshold $p$. For various values of $p$, we examine the fill probability of these orders and their subsequent returns. We impose no inventory constraints, allowing for arbitrary accumulation of inventory in either direction. Negative returns are interpreted as opportunity costs.

    \item \textbf{Balanced-Inventory Strategy}: This strategy operates similarly to the basic MM strategy described in Section~\ref{sec:new_sec}, but it restricts order submissions to situations where the model (logistic regression or random forest) predicts a reversal probability exceeding a threshold $p \in [0, 1]$. Specifically, it posts a buy order at the top bid price if a (downward) reversal is predicted with probability $>p$, and a sell order at the top ask price if an (upward) reversal is predicted with probability $>p$. If the inventory is non-zero (i.e., long or short one unit), the strategy only posts on the side that returns the position to zero, provided a reversal is predicted with probability $>p$.
    Since this strategy maintains a balanced inventory (similar to the basic MM strategy), we can meaningfully compute its realized PnL (unlike the Undirectional Strategy).
\end{enumerate}

\paragraph{Undirectional Strategy Results.}
Table~\ref{tab:comparison_logistic_rf} provides a comparative overview of the Undirectional Strategy results derived from both the logistic regression and random forest models, detailing the number of orders placed per day during the test period, the empirical fill probability of these orders, the average post-fill return over a 5-second horizon (measured from fill time), and the standard deviation of those returns.
The results are similar across different return horizons, although larger horizons introduce more noise due to increased standard deviation.

For both logistic regression and random forest models, the number of orders decreases as the threshold for predicted reversal probability increases, as one would expect.
Yet, the decrease in number of reversal predictions drops quite steeply with an increase in the probability threshold, likely due to the class imbalance in the training data, where reversals are relatively rare.
The fill probability remains largely consistently high across varying thresholds.

Most importantly, the post-fill performance significantly improves as the threshold increases.
This indicates that traders can capture economic benefit by placing orders only when there is a substantial predicted probability of a reversal.
For the logistic regression model, that economic benefit becomes significant at a reversal probability of approximately 24-26\%, where the post-fill returns are statistically significantly about 0.1-0.2 basis points (bp) better than those of randomly placed orders.

\begin{table}[h]
   \centering
   \captionsetup{width=\textwidth}
   \caption{Comparison of Undrectional Strategy across different thresholds.}
   \label{tab:comparison_logistic_rf}
   \small
 \adjustbox{width=\textwidth, center}{   \begin{tabular}{c | c c c c c | c c c c c}
       \toprule
       \textbf{Threshold} & \multicolumn{5}{c|}{\textbf{Logistic Regression}} & \multicolumn{5}{c}{\textbf{Random Forest}} \\
       \cmidrule(r){2-6} \cmidrule(l){7-11}
        & \# Orders / Day & Fill Prob & Return (5s) & Stdev & p-value & \# Orders / Day & Fill Prob & Return (5s) & Stdev & p-value \\
       \midrule
       0.00 & 21689 & 0.8824 & $-0.74$ & 1.55 & 0.5000 & 21657 & 0.8824 & $-0.74$ & 1.55 & 0.5000 \\
       0.02 & 19438 & 0.8809 & $-0.71$ & 1.62 & 0.0003 & 21402 & 0.8825 & $-0.71$ & 1.56 & 0.0001 \\
       0.04 & 16596 & 0.8791 & $-0.72$ & 1.72 & 0.0165 & 20780 & 0.8813 & $-0.67$ & 1.57 & 0.0000 \\
       0.06 & 13939 & 0.8773 & $-0.71$ & 1.82 & 0.0018 & 19733 & 0.8807 & $-0.67$ & 1.58 & 0.0000 \\
       0.08 & 11674 & 0.8767 & $-0.72$ & 1.92 & 0.0397 & 18375 & 0.8794 & $-0.68$ & 1.61 & 0.0000 \\
       0.10 & 9566 & 0.8770 & $-0.72$ & 2.05 & 0.0607 & 16633 & 0.8785 & $-0.68$ & 1.66 & 0.0000 \\
       0.12 & 7723 & 0.8778 & $-0.70$ & 2.17 & 0.0040 & 14790 & 0.8773 & $-0.69$ & 1.71 & 0.0000 \\
       0.14 & 5979 & 0.8829 & $-0.70$ & 2.28 & 0.0189 & 12848 & 0.8761 & $-0.71$ & 1.77 & 0.0008 \\
       0.16 & 4441 & 0.8855 & $-0.69$ & 2.43 & 0.0083 & 11006 & 0.8743 & $-0.71$ & 1.83 & 0.0034 \\
       0.18 & 3158 & 0.8880 & $-0.69$ & 2.57 & 0.0334 & 9304 & 0.8732 & $-0.69$ & 1.90 & 0.0000 \\
       0.20 & 2104 & 0.8887 & $-0.68$ & 2.77 & 0.0496 & 7732 & 0.8690 & $-0.70$ & 1.96 & 0.0017 \\
       0.22 & 1365 & 0.8814 & $-0.68$ & 3.00 & 0.0903 & 6265 & 0.8687 & $-0.69$ & 2.04 & 0.0006 \\
       0.24 & 882 & 0.8673 & $-0.62$ & 3.24 & 0.0278 & 5034 & 0.8642 & $-0.68$ & 2.13 & 0.0003 \\
       0.26 & 577 & 0.8612 & $-0.60$ & 3.55 & 0.0432 & 3965 & 0.8589 & $-0.68$ & 2.18 & 0.0018 \\
       0.28 & 349 & 0.8559 & $-0.45$ & 3.86 & 0.0055 & 3037 & 0.8572 & $-0.67$ & 2.27 & 0.0010 \\
       0.30 & 233 & 0.8545 & $-0.32$ & 3.78 & 0.0011 & 2271 & 0.8523 & $-0.63$ & 2.41 & 0.0000 \\
       0.32 & 162 & 0.8284 & $-0.40$ & 3.78 & 0.0187 & 1675 & 0.8548 & $-0.57$ & 2.52 & 0.0000 \\
       0.34 & 113 & 0.8128 & $-0.27$ & 3.82 & 0.0093 & 1219 & 0.8527 & $-0.50$ & 2.59 & 0.0000 \\
       0.36 & 69 & 0.7913 & $+0.27$ & 3.49 & 0.0000 & 837 & 0.8507 & $-0.41$ & 2.61 & 0.0000 \\
       0.38 & 45 & 0.7733 & $+1.34$ & 3.57 & 0.0000 & 582 & 0.8521 & $-0.32$ & 2.80 & 0.0000 \\
       0.40 & 33 & 0.7818 & $+2.11$ & 3.75 & 0.0000 & 375 & 0.8465 & $-0.33$ & 2.99 & 0.0000 \\
       0.42 & 24 & 0.8250 & $+2.65$ & 3.69 & 0.0000 & 248 & 0.8537 & $-0.30$ & 3.12 & 0.0000 \\
       0.44 & 19 & 0.8125 & $+2.66$ & 3.72 & 0.0000 & 159 & 0.8441 & $-0.13$ & 3.54 & 0.0000 \\
       0.46 & 13 & 0.7826 & $+3.45$ & 3.88 & 0.0000 & 96 & 0.8113 & $+0.05$ & 2.88 & 0.0000 \\
       0.48 & 11 & 0.8421 & $+3.49$ & 4.23 & 0.0000 & 52 & 0.8161 & $+0.76$ & 3.02 & 0.0000 \\
       0.50 & 9 & 0.8750 & $+3.02$ & 1.92 & 0.0000 & 23 & 0.7692 & $+1.68$ & 3.35 & 0.0000 \\
       \bottomrule
   \end{tabular}
   }
\end{table}

How can we ascertain if the improvements in post-fill returns shown in Table~\ref{tab:comparison_logistic_rf} genuinely surpass the performance of randomly placed orders, considering the possibility that these results might be mere coincidences?
This can be addressed through a two-sample t-test, which compares the average returns of two groups: the strategy for a given model and threshold versus random orders.
\ja{For instance, at the $0.30$ reversal probability threshold for the Logistic Regression model, where the average post-fill return is $-0.32$, we test the null hypothesis that there is no difference between the returns of this trading strategy and random orders.
The calculated t-statistic is $3.062602$ with a p-value of $0.0011$, allowing us to reject the null hypothesis and conclude that the observed return improvement is statistically significant at the 5\% level.
The table also reports p-values for all reversal probability thresholds, providing a view of the statistical significance across different levels.}

\FloatBarrier

\paragraph{Balanced-Inventory Strategy Results.}
Let us now examine Table~\ref{tab:comparison_logistic_rf_extended}, which presents the performance of the Balanced Inventory strategy based on a model and a specified reversal probability threshold. At a threshold of $0$, the strategy operates without restrictions, and is equivalent to the naive market-making strategy discussed in Section~\ref{sec:new_sec}.
As the threshold increases, the strategy restricts order submissions to scenarios where the model predicts a reversal with a certain probability, resulting in fewer orders being posted.

The returns are calculated as realized roundtrip returns, defined as the price difference (in bp) between the limit prices of two orders: the initial order that accumulates inventory (changing it from 0 to $\pm1$) and the subsequent order that returns the inventory to 0.\footnote{Returns are signed such that a positive return indicates a profit, while a negative return indicates a loss.}
Each leg of the roundtrip earns a maker rebate of $0.5$ bp, thus we add $1$ bp to each realized roundtrip return.
We also present the standard deviations of these returns, the average holding time (the duration between an inventory-acquiring trade and its subsequent hedge), and the annualized Sharpe ratio of the strategy.
The Sharpe ratio is calculated based on hourly periods, given that our test period spans approximately 3.5 trading days.

We observe a significant decline in the number of fills as the reversal probability threshold increases, implying longer intervals between consecutive trades in a roundtrip, i.e., increased holding time.
This makes interpretation challenging for higher thresholds, so we only display results where the holding time is less than 10 minutes (600 seconds).

\begin{figure}[htbp]
    \centering
    \begin{minipage}{0.5\textwidth}
        \centering
        \captionsetup{width=.8\linewidth}
        \includegraphics[scale=0.37]{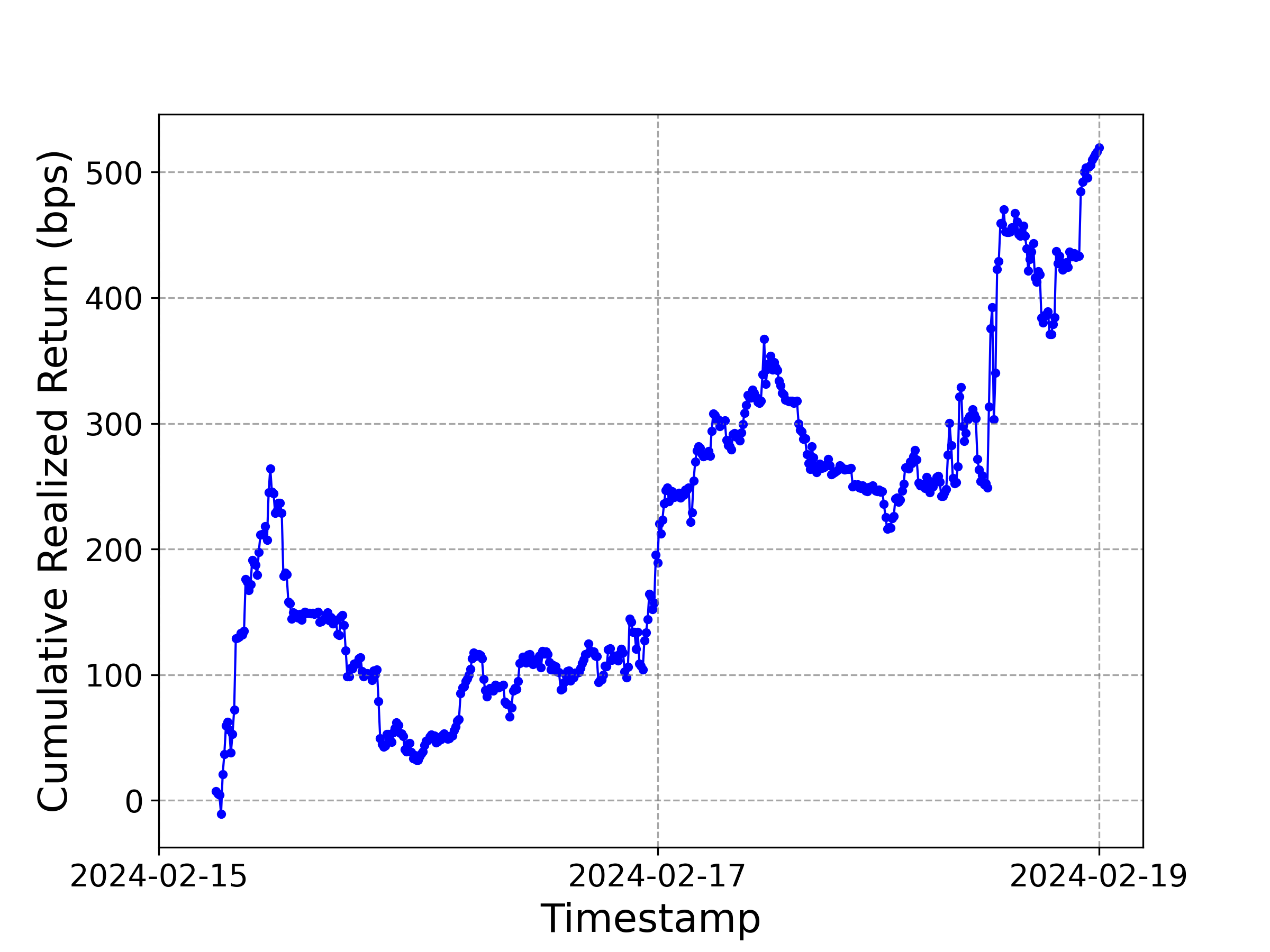}

        \caption{PnL plot of the model-based MM strategy, net of fees.}
        \label{fig:timeseries_realized_bp_diff_cumsum_w_model}
    \end{minipage}%
    \begin{minipage}{0.5\textwidth}
        \centering
        \captionsetup{width=.83\linewidth}
        \includegraphics[scale=0.37]{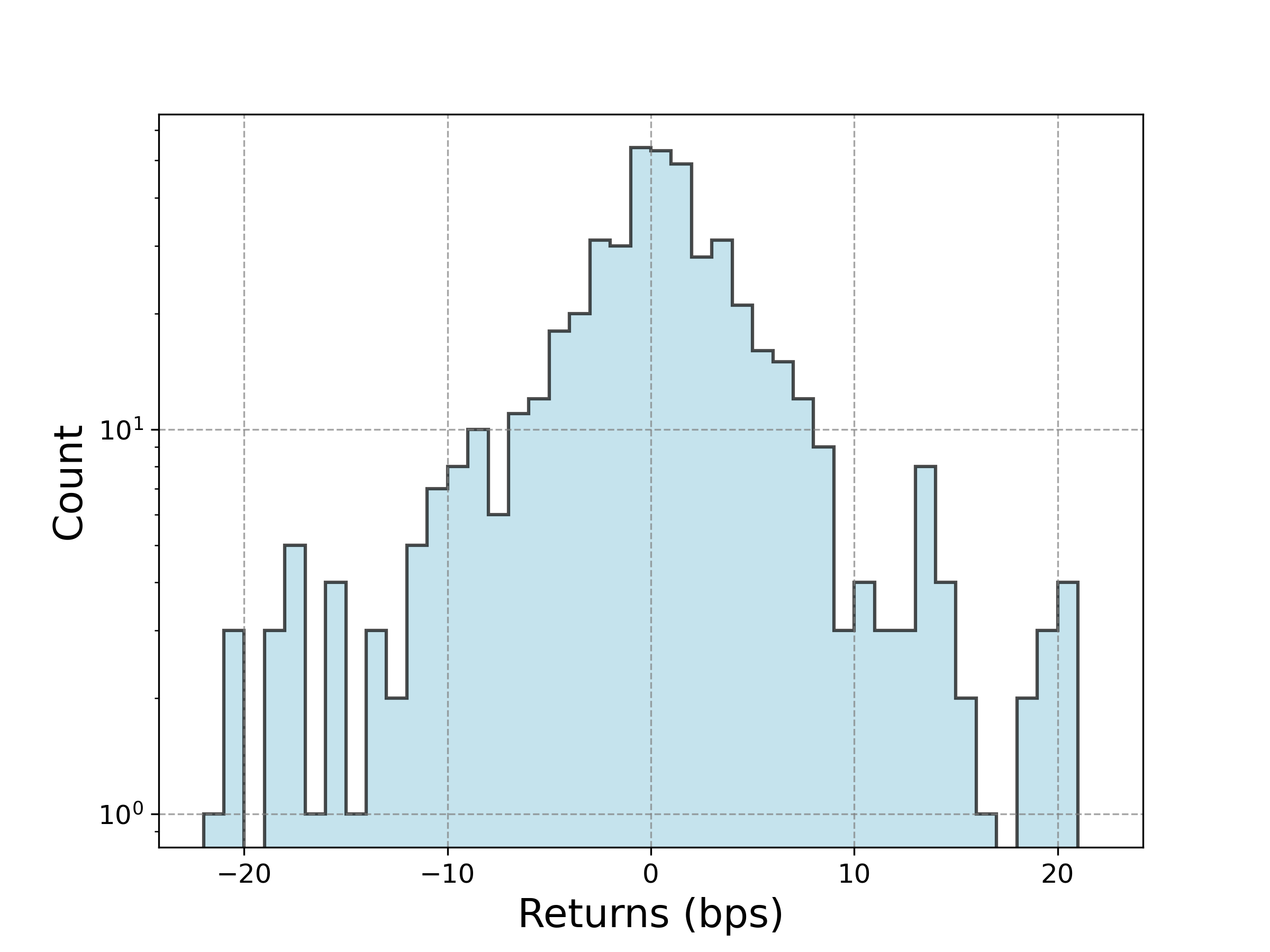}
        \caption{Histogram of pre-fee returns of roundtrip trades of the model-based MM strategy.}
        \label{fig:realized_bp_diff}
    \end{minipage}
\end{figure}

The improvement in realized roundtrip PnL compared with the naive strategy is striking.
The naive strategy incurs an average loss of 0.44 basis points per roundtrip (net of rebates), occurring at a high frequency, which explains its poor Sharpe ratio.
At a higher threshold, such as 0.24 for the Logistic Regression model, the number of roundtrips remains relatively large (327 per day), and the average return (0.71) constitutes a significant improvement over the naive strategy.
A hypothesis test similar to the one conducted earlier confirms that this difference is statistically significant and unlikely to be due to chance, despite the large standard deviation of returns.
\ja{The cumulative PnL plot and return histogram for this threshold are shown in Figures~\ref{fig:timeseries_realized_bp_diff_cumsum_w_model} and~\ref{fig:realized_bp_diff}, respectively, illustrating the strategy's improved performance and the distribution of realized returns; these are the analogues of the plots for the naive MM strategy shown in Figures~\ref{fig:timeseries_realized_bp_diff_cumsum} and~\ref{fig:realized_bp_diff_naive}.}
The Sharpe ratio improves from $-109$ for the naive strategy to 11.97 for the Logistic Regression-based strategy at the 0.24 threshold, demonstrating unequivocally that our model yields substantial economic benefit.
It is important to note that the impressive Sharpe ratio is unlikely to persist with larger-than-minimum-sized orders, and that our calculations assume the best possible maker rebate (0.5 bp).
Even minor changes to these assumptions could result in a significantly lower Sharpe ratio, yet it remains far superior to that of the naive strategy.

\begin{table}[h]
   \centering
   \captionsetup{width=\textwidth}
   \caption{Comparison of Balanced-Inventory Strategy across different thresholds.}
   \label{tab:comparison_logistic_rf_extended}
   \small
   \adjustbox{width=\textwidth, center}{
   \begin{tabular}{c | c c c c c c | c c c c c c}
       \toprule
       \textbf{Threshold} & \multicolumn{6}{c|}{\textbf{Logistic Regression}} & \multicolumn{6}{c}{\textbf{Random Forest}} \\
       \cmidrule(r){2-7} \cmidrule(l){8-13}
        & \# Fills / Day & Avg Ret & Stdev & Holding & Sharpe & p-value & \# Fills / Day & Avg Ret & Stdev & Holding & Sharpe & p-value \\
       \midrule
       0.00 & 15380 & $-0.44$ & $2.73$ & 11.39 & $-109.00$ & $0.5000$ & 15380 & $-0.44$ & $2.73$ & 11.39 & $-109.00$ & $0.5000$ \\
       0.02 & 14266 & $-0.45$ & $2.81$ & 12.28 & $-99.98$ & $0.6369$ & 15066 & $-0.43$ & $2.68$ & 12.54 & $-106.78$ & $0.3933$ \\
       0.04 & 12650 & $-0.46$ & $2.97$ & 13.85 & $-92.82$ & $0.8177$ & 14626 & $-0.43$ & $2.71$ & 12.92 & $-105.31$ & $0.4158$ \\
       0.06 & 10972 & $-0.45$ & $3.20$ & 15.97 & $-74.99$ & $0.6278$ & 13868 & $-0.44$ & $2.77$ & 13.62 & $-99.97$ & $0.5441$ \\
       0.08 & 9364 & $-0.44$ & $3.46$ & 18.71 & $-67.29$ & $0.5270$ & 12856 & $-0.45$ & $2.87$ & 14.69 & $-98.61$ & $0.6727$ \\
       0.10 & 7780 & $-0.44$ & $3.77$ & 22.52 & $-56.42$ & $0.5230$ & 11700 & $-0.44$ & $2.97$ & 16.14 & $-96.94$ & $0.5340$ \\
       0.12 & 6332 & $-0.44$ & $4.24$ & 27.67 & $-45.39$ & $0.5187$ & 10354 & $-0.44$ & $3.17$ & 18.24 & $-91.83$ & $0.5305$ \\
       0.14 & 4956 & $-0.43$ & $4.72$ & 35.35 & $-37.41$ & $0.4469$ & 8918 & $-0.45$ & $3.42$ & 21.18 & $-75.09$ & $0.5897$ \\
       0.16 & 3598 & $-0.43$ & $5.74$ & 48.68 & $-29.20$ & $0.4867$ & 7630 & $-0.41$ & $3.68$ & 24.75 & $-63.28$ & $0.2555$ \\
       0.18 & 2494 & $-0.40$ & $6.72$ & 70.24 & $-23.20$ & $0.3908$ & 6374 & $-0.41$ & $4.07$ & 29.63 & $-55.68$ & $0.3011$ \\
       0.20 & 1644 & $-0.39$ & $8.10$ & 106.53 & $-21.26$ & $0.3975$ & 5224 & $-0.36$ & $4.54$ & 36.15 & $-42.96$ & $0.1011$ \\
       0.22 & 1060 & $-0.22$ & $10.63$ & 165.32 & $-15.88$ & $0.2504$ & 4146 & $-0.39$ & $5.17$ & 45.54 & $-42.23$ & $0.2794$ \\
       0.24 & 654 & $+0.71$ & $13.69$ & 267.66 & $+11.97$ & $0.0160$ & 3222 & $-0.39$ & $5.90$ & 58.61 & $-34.28$ & $0.3252$ \\
       0.26 & 426 & $+1.21$ & $18.10$ & 409.26 & $+26.12$ & $0.0303$ & 2460 & $-0.32$ & $6.80$ & 76.73 & $-24.85$ & $0.1948$ \\
       0.28 & 270 & $+0.90$ & $22.52$ & 595.97 & $+12.32$ & $0.1650$ & 1826 & $-0.21$ & $7.82$ & 103.27 & $-18.78$ & $0.1076$ \\
       0.30 & - & - & - & - & - & - & 1330 & $+0.03$ & $8.90$ & 141.75 & $+1.30$ & $0.0287$ \\
       0.32 & - & - & - & - & - & - & 972 & $+0.31$ & $9.65$ & 194.22 & $+12.08$ & $0.0078$ \\
       0.34 & - & - & - & - & - & - & 692 & $+0.59$ & $12.20$ & 272.14 & $+13.48$ & $0.0133$ \\
       0.36 & - & - & - & - & - & - & 454 & $+0.02$ & $14.45$ & 414.03 & $+0.30$ & $0.2501$ \\
       \bottomrule
   \end{tabular}
   }
\end{table}

\subsection{Feature Importance}
\label{subsec:feature_importance}

The preceding subsection has shown that our reversal models have economically significant predictive power. Now, this final subsection considers interpretations of their parameters: What are the main economic factors driving the likelihood of reversals?
A convenient starting point to answering that question is an analysis of the coefficients of our logistic regression model.
To simplify our analysis, we focus on the case of a buy maker order, where a reversal occurs when a price-negative imbalance (often preceded by a price decline) is followed by a price increase; the term reversal is subsequently used only in the sense of such a buy-reversal.
One just needs to replace the words “bid” and “buy” with “ask” and “sell” in the sentences and plots below and the result would be a description from the perspective of a sell order.

\begin{figure}[htbp]
\centering

\begin{subfigure}[t]{0.4\textwidth}
    \centering
    \caption{Trade volume patterns}
    \includegraphics[width=\linewidth]{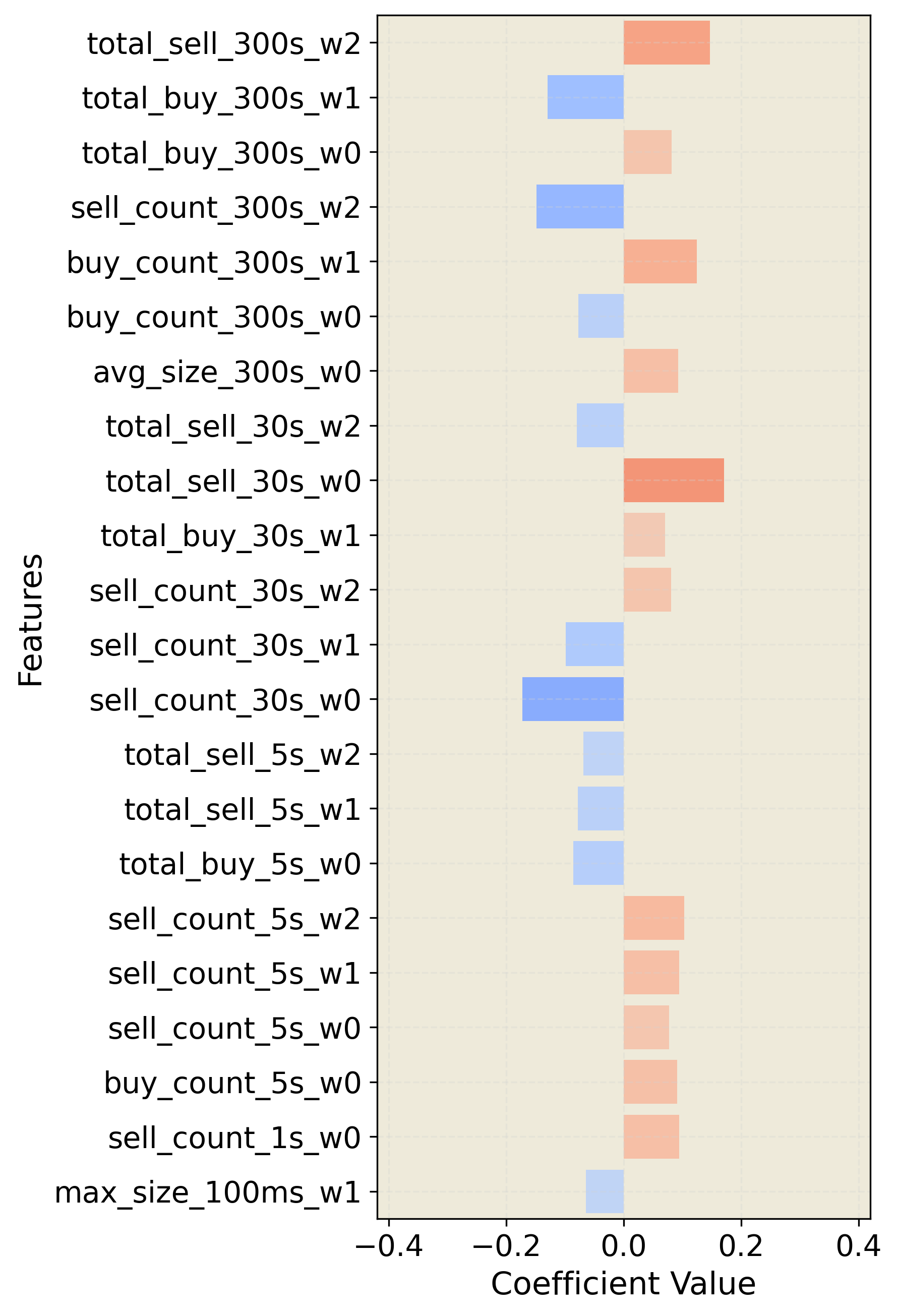}
    \label{fig:coeffs_dark_TRADE}
\end{subfigure}%
\hfill
\begin{subfigure}[t]{0.4\textwidth}
    \centering
    \caption{Momentum features}
    \includegraphics[width=\linewidth]{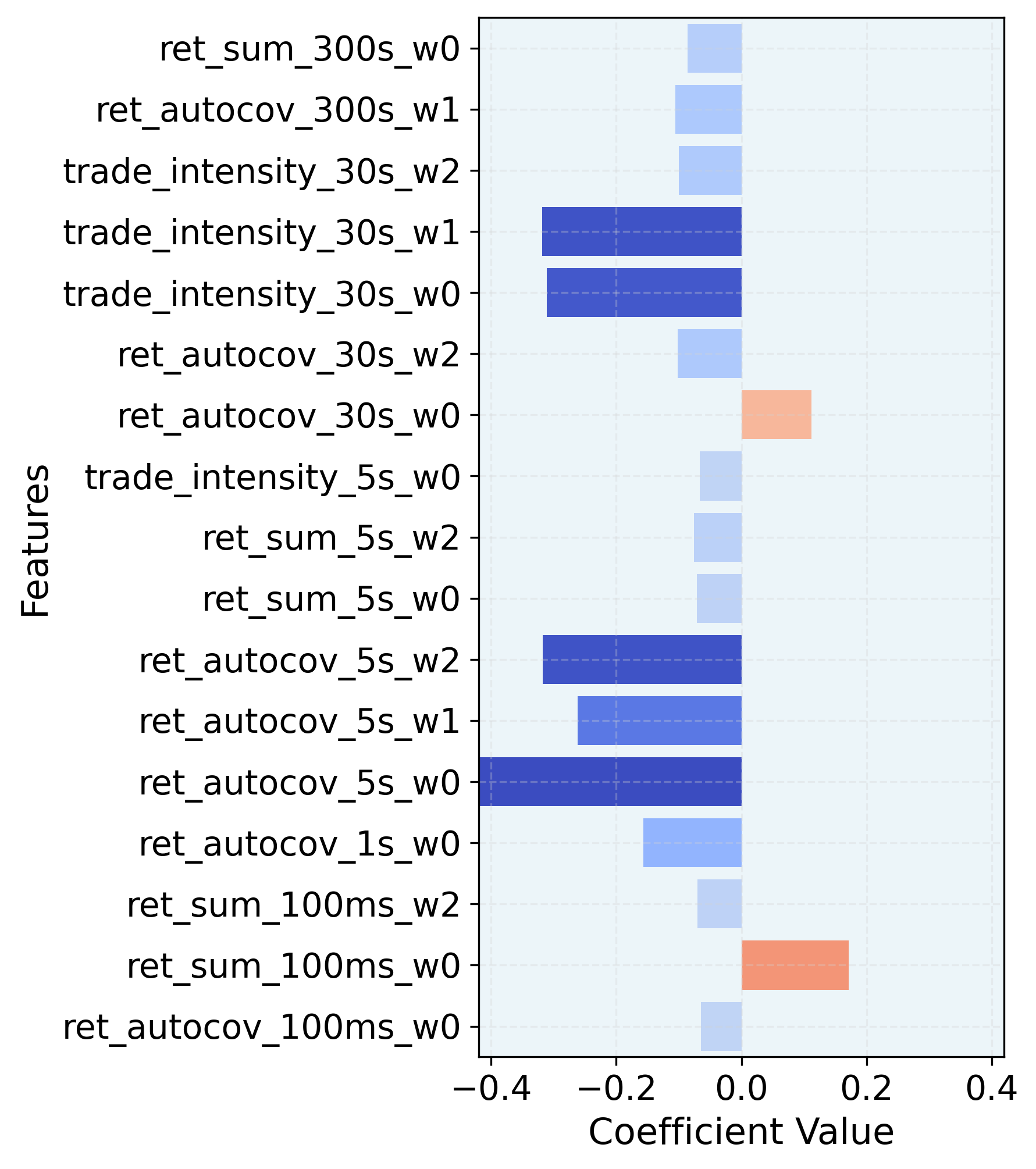}
    \label{fig:coeffs_dark_MOMENTUM}
\end{subfigure}

\begin{subfigure}[t]{0.4\textwidth}
    \centering
    \includegraphics[width=\linewidth]{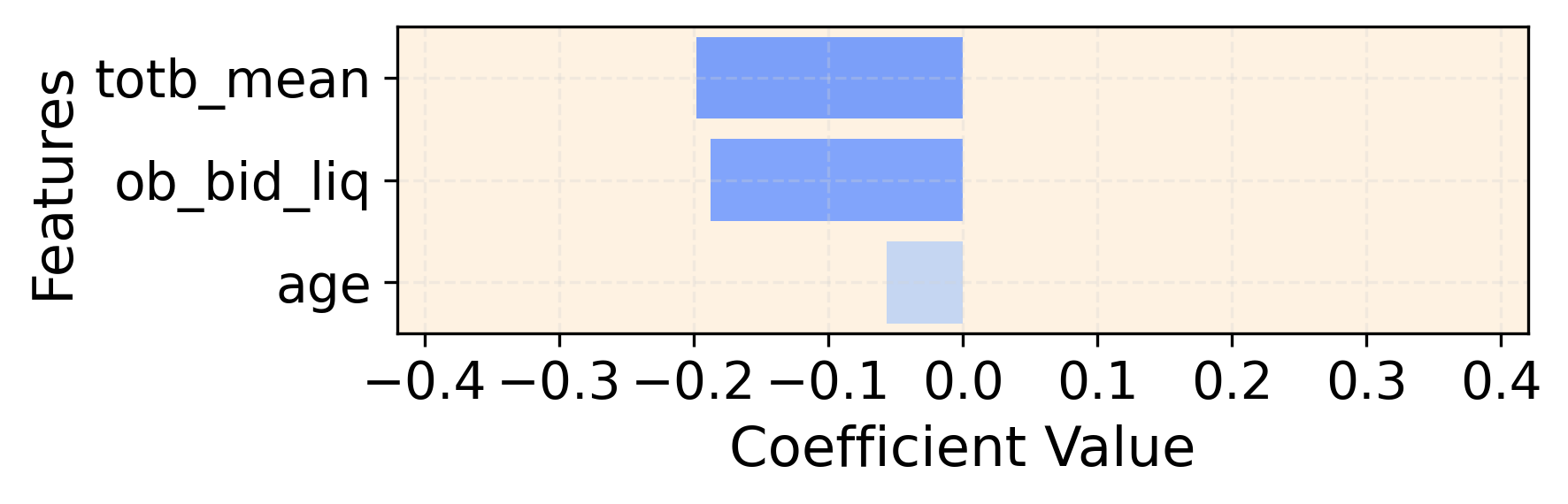}
    \caption{Order book features}
    \label{fig:coeffs_dark_DYNAMICS}
\end{subfigure}%
\hfill
\begin{subfigure}[t]{0.4\textwidth}
    \centering
    \includegraphics[width=\linewidth]{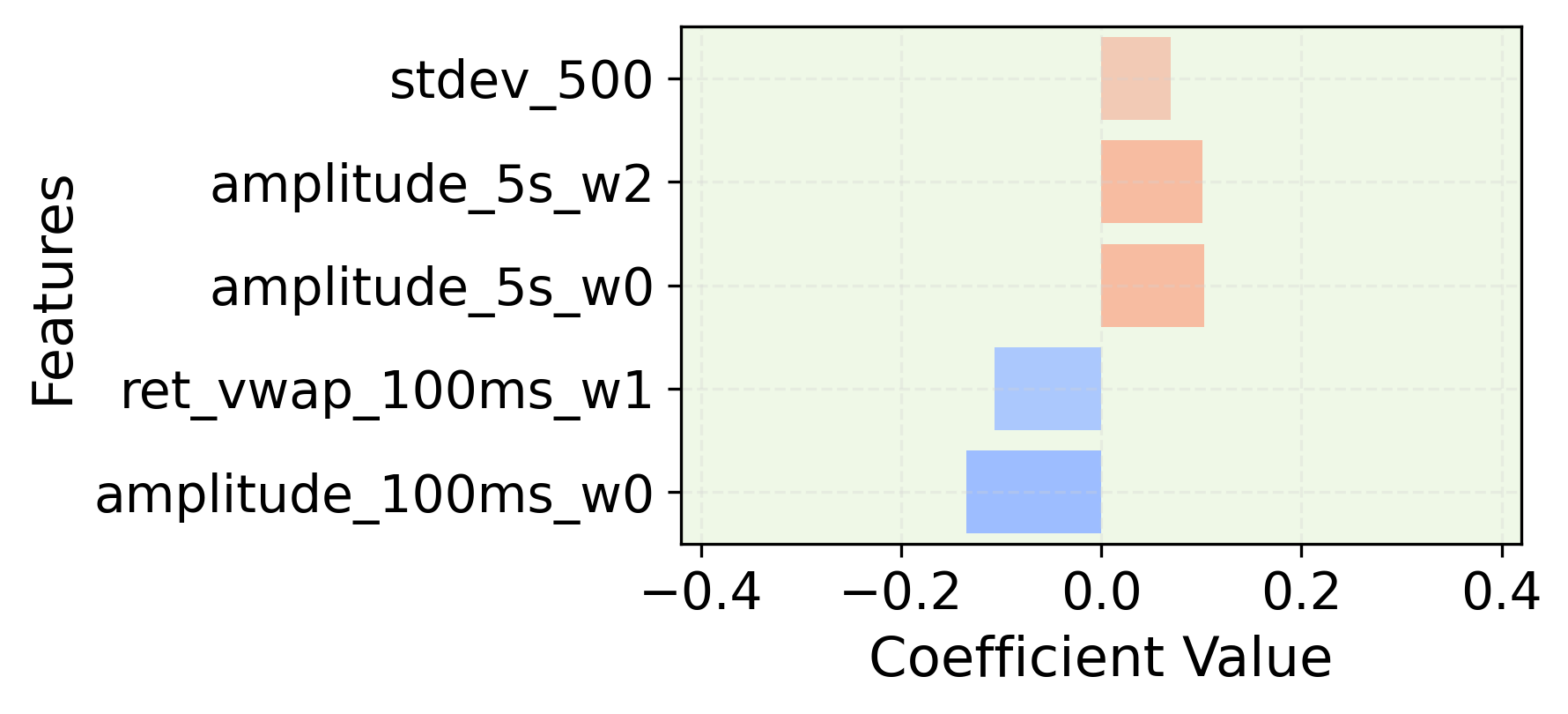}
    \caption{Price dynamics}
    \label{fig:coeffs_dark_OB}
\end{subfigure}

\captionsetup{width=\textwidth}
\caption{Coefficients of the logistic regression model grouped by feature categories, illustrating the relative importance and direction of each feature's impact on the likelihood of reversals. Only coefficients with absolute values greater than 0.05 are displayed.}

\label{fig:overall_label}

\end{figure}

\paragraph{Coefficient Analysis.}
The four bar plots in Figure~\ref{fig:overall_label} illustrate the model's coefficients across the four feature groups described in Section~\ref{subsec:features}, displaying only those with absolute values larger than $0.05$ to keep the visualization concise; the full set of coefficients can be found in the Appendix~\ref{sec:model_coefficients}.
We note that all variables are standardized prior to estimation, allowing for a comparison of coefficients across features.
We make a range of observations regarding those coefficients and possible economic interpretations, acknowledging that any such interpretations involve some degree of speculation, as they involve teasing apart the effect of many correlated influences:

\begin{itemize}

\item First, consider the momentum features in Figure~\ref{fig:coeffs_dark_MOMENTUM}.
Notably, the features \texttt{ret\_autocov\_5s\_w*} (where * denotes $\{0,1,2\}$) have large negative coefficients, suggesting that reversals are more likely when the last trade price oscillates (even if it is just between the top bid and top ask prices) rather than trend consistently in one direction across each of the last three 5-second windows, collectively amounting to the last 15 seconds.
Next, notice that the coefficient on \texttt{ret\_sum\_100ms\_w0} is significantly positive, implying that a recent sharp price drop (in the last 100ms) implies a higher probability of a reversal. Combining these insights, we arrive at the conclusion that a sudden price drop following a balanced up/down price action, is associated with a higher probability for a reversal.
This is economically sensible, as the new top bid price is likely to lack liquidity immediately following the price drop, and the absence of persistent downward pressure (as one would see from a high autocovariance) suggests that the trade causing the sudden price drop is more likely to be an isolated event rather than part of a series of orders driving the price downward in a sustained manner.

\item The coefficients \texttt{avg\_time\_tr\_30s\_w0} and \texttt{avg\_time\_tr\_30s\_w1},
are noticeably negative, indicating that a longer average time between trades over the past minute correlates with a significantly reduced likelihood of a reversal.
This aligns with the economic intuition that reversals occur more frequently during periods of heightened market activity,
as the frequency at which order book queues fluctuate is increased with order survival times lower than during quiet times.
Supporting this conclusion, the variable \texttt{totb\_mean} also has a significantly negative coefficient, suggesting that the model predicts a higher probability of reversal when recent top-of-book survival times are short.
Additionally, \texttt{stdev\_500} shows a marginally positive coefficient, further indicating that increased volatility (which is correlated with heightened market activity and decreased top-of-book survival times) slightly raises the probability of a reversal.

\item The model penalizes large pre-existing liquidity on the bid side, as shown by the negative coefficients for \texttt{ob\_bid\_liq} in Figure~\ref{fig:coeffs_dark_OB}.
This makes sense, as high liquidity at the touch reduces the likelihood of a fill, which was one of the requirements in our definition of a reversal.
Conversely, consider the explanatory variable \texttt{ob\_bid\_half} which is small (close to zero) if there is much liquidity at or near the touch, and large if the bid-side is very illiquid for various levels from the top bid price down: it has a marginally negative coefficient, suggesting that when the bid-side is illiquid for several book levels from the top bid downward, the model yields a slightly lower predicted probability of a reversal.
Combining those two observations, we arrive at the conclusion that the model prefers scenarios of an illiquid top bid price but lots of liquidity at deeper order book levels close to the top bid price.
This again makes economic sense: if other makers are willing to post significant liquidity only slightly deeper in the book than the top bid price, then it is likely not much worse to post an order at the top bid as well, given that this price is only slightly worse than those deeper levels.

\item In the trade volume patterns group (Figure~\ref{fig:coeffs_dark_TRADE}), the model assigns moderately positive coefficients to the three features \texttt{sell\_count\_5s\_w*} (where * denotes \{0,1,2\}) which quantify the number of filled sell maker orders in the past 15 seconds.
Conversely, the significantly negative coefficient for \texttt{sell\_count\_30s\_w0} suggests a reduced reversal probability in situations where sell activity was prevalent in the slightly more distant past.
This pattern implies recent sell pressure with a preceding period characterized by mostly buying activity.

\item The model's negative coefficients for the \texttt{total\_sell\_5s\_w*} (where * denotes \{0,1,2\}) features reveal that the reversal probability is higher if the aforementioned high count of sell trades during the same lookback windows collectively consume only little volume, suggesting they largely take small maker orders. This seems to suggest that large taker volume may be indicative of a further price drop.

\item The positive coefficient for \texttt{buy\_count\_5s\_w0} (and slightly negative ones for the preceding two lookback windows) implies that recent buy activity marginally increases the likelihood of a reversal, suggesting a shift from selling to buying and hinting at a potential reversal (despite there still being a price-negative imbalance).

\end{itemize}

Synthesizing our observations above, it appears that our model’s ideal criteria for a reversal are as follows:
\begin{enumerate}

    \item \textbf{Elevated Volatility:} The market should exhibit certain characteristics associated with heightened volatility: high trade-arrival intensity and short top-of-book survival times. In these conditions, reversals are more frequent and the time between order submission and its execution or cancellation is likely to be low (making it less likely for economic conditions to change during the order’s lifetime).

    \item \textbf{Price Movement Patterns:} In the recent past (15 seconds), the last trade returns should lack a strong autocorrelated trend, indicating that the price oscillates rather than trending strongly in one direction. Following such price action, there is a sudden drop in the top bid price (within the last 100 milliseconds) to a new local minimum, where there is minimal pre-existing liquidity.

    \item \textbf{Trade Volume Dynamics:} Over the past 15 seconds, there should be a large number of sell trades following a period of little selling activity; however, those sell trades should not amount to much total volume, i.e. it was executed predominantly against small maker orders. Accompanying this, there should be a significant price decline over the same period, and a recent emergence of buy trades in the past 5 seconds after a period of little or no buying activity.
\end{enumerate}
The price drop to a new local minimum with little existing liquidity suggests that the order has a high probability of filling, while the remaining criteria can be viewed as mitigating the risk of further price decline, despite a price-negative imbalance, thereby reducing the likelihood of a negative short-term price drift post-fill.

\paragraph{Permutation Importance.}
To complement our analysis of the logistic regression coefficients, we conducted a permutation importance analysis on the random forest model.
Permutation importance is a model-agnostic approach to assess the significance of each feature by randomly permuting its values and observing the impact on model performance. A substantial decrease in accuracy indicates a feature's importance.

The six most important features, as determined by the permutation importance analysis (with scores normalized by the maximum importance), are \texttt{ret\_autocorr\_sum\_30s\_w0} (1.00), \texttt{velocity\_tr\_30s\_w0} (0.93), \texttt{ob\_bid\_liq} (0.67), \texttt{totb\_mean} (0.57), \texttt{sell\_count\_30s\_w0} (0.27), and \texttt{amplitude\_100ms\_w0} (0.12).
The permutation importance drops off rather steeply beyond these six features, suggesting that they play a more minor role in the model’s performance.
Yet, the perspective offered by the permutation importance analysis is consistent with our interpretations of the coefficients from the logistic regression model:

\begin{itemize}

\item The high importance score of the top feature, \texttt{ret\_autocov\_sum\_30s\_w0}, stresses the importance of recent price autocovariance as an explanatory variable.
This corroborates the finding that negative autocovariance of recent returns (implying price oscillations up and down) are an important precursor of reversals, which we previously inferred from logistic regression's large negative coefficients for \texttt{ret\_autocov\_5s\_w*},  \( *=0,1,2 \).

\item The high (normalized) importance score of 0.93 for the feature \texttt{velocity\_tr\_30s\_w0} echoes the logistic regression's findings where a short average time between consecutive trades was found to be strongly positively correlated with a higher likelihood of a reversal.

\item The feature \texttt{ob\_bid\_liq}, with an importance score of 0.67, corroborates our earlier conclusion based on the coefficient analysis that large pre-existing liquidity at the bid side results in a significantly reduced reversal probability, which can be explained by the reduced fill probability associated with an already-large queue at post-time, coupled with the fact that our definition of a reversal included a requirement that the order be filled.

\item Furthermore, \texttt{totb\_mean} also appears prominently in both analyses, with a permutation importance score of 0.57 and a large negative coefficient, reinforcing the notion that low survival times of top-of-book prices are conducive to reversals.

\item The logistic regression model applied a large negative coefficient to the feature \texttt{sell\_count\_30s\_w0}, whose significance in predicting reversals is further corroborated by its high (normalized) permutation importance score of 0.27 in the random forest model.

\item Finally, the moderately high permutation importance score for \texttt{amplitude\_100ms\_w0} is consistent with its significantly positive coefficient in the logistic regression model, reflecting our earlier conclusion that reversals are more likely to occur following a very recent price drop.

\end{itemize}

A concluding afterword notes that all of the explanatory variables with a high permutation score for the random forest model also exhibit a large (in absolute terms) coefficient in our logistic regression model; and conversely, there is no feature with a very large coefficient that does not emerge as one with a high permutation importance score.
Thus, we conclude that, while the permutation importance analysis does not allow us to paint as detailed a picture as the coefficient analysis (since it only tells us how important a feature is without telling us how it influences the model’s predictions), its results are consistent with our earlier findings.

\FloatBarrier

\section{Conclusion}
\label{sec:conclusion}

Through granular empirical analysis, we have described the effect of (i) basic order book mechanics and (ii) the strong persistence of price changes from the immediate to short timescales, revealing interactions between returns, queue sizes, and orders' queue positions.
These interactions lead to negative correlation between an order's fill probability and its post-fill returns, creating a fundamental challenge in profitable maker trading at the touch. The practical consequences in high-frequency trading are immense, rendering several natural (and commonly-cited) trading strategies highly unprofitable in practice.

To overcome these challenges, a key to successful market making is to avoid being adversely selected. While this can in principle be achieved "defensively" by cancelling orders when adverse selection becomes likely (itself a complex task), our work explores a more proactive strategy: developing counter-signals to the order book imbalance. Such signals allow for the submission of orders during an apparently adverse imbalance to benefit from an early queue position when conditions shift favorably.
To show that this can be done, we develop such a signal and demonstrate its efficacy: it significantly outperforms benchmarks and gives an indication of the type of features that might be helpful in predicting when the order book imbalance falsely predicts the next price change;
the strategy is slightly profitable with min-sized orders and assuming low fees, albeit those profits are likely not scalable; however, we hope that our approach can give practitioners a pointer on what direction to focus their efforts.
The specific model we present is one of many possible constructions, intended to demonstrate a proof-of-concept rather than to serve as an optimized, production-ready strategy.

Our model's performance, which required a complex feature set to achieve even modest (and likely non-scalable) profits, affirms the "Unprofitability Principle" we informally introduced earlier as: forecasting ability and ease of exploitation are anti-correlated.
This principle helps explain why so many academic studies struggle to produce practically viable results—a tendency exemplified by a research focus on fill probabilities alone, which, by ignoring their negative correlation with returns, risks optimizing for a pyrrhic victory.
This flawed approach is symptomatic of two broader challenges that often bedevil academic research: (i) a failure to examine the problem at a sufficiently granular level to uncover the influence of order book mechanics on outcomes, and (ii) a lack of access to data that provides the full picture, as even L3 data show only what happened, not what did not happen (such as failed cancellations), which can be equally important.
By conducting live trading experiments, we address both these challenges, leveraging a methodology that provides insights difficult to attain with other approaches.

Ultimately, like all liquid LOB markets, the crypto exchange we used is highly competitive. The pervasive barrier to profitability for HFT (maker or taker) is partly a consequence of the mechanics of an LOB market, partly a consequence of competition between traders, and pervasively a consequence of the persistent wealth extraction via exchange fees.


\section*{Funding}
AYS acknowledges support from NSERC Discovery Grant RGPIN-2024-05996.

\section*{Data Availability}
The data that support the findings of this study were collected by the authors and are not publicly available.

\section*{Declaration of Competing Interests}
The authors declare no competing financial or personal interests in relation to the work described.

\bibliographystyle{abbrvnat}
\bibliography{backtest_bib}{}

\newpage

\appendix

\section{Histograms for Medium-Sized Queues}
\label{sec:more_histos}

\begin{figure}[htbp]
    \centering
    \begin{minipage}{0.5\textwidth}
        \centering
        \captionsetup{width=.8\linewidth}
        \includegraphics[scale=0.36]{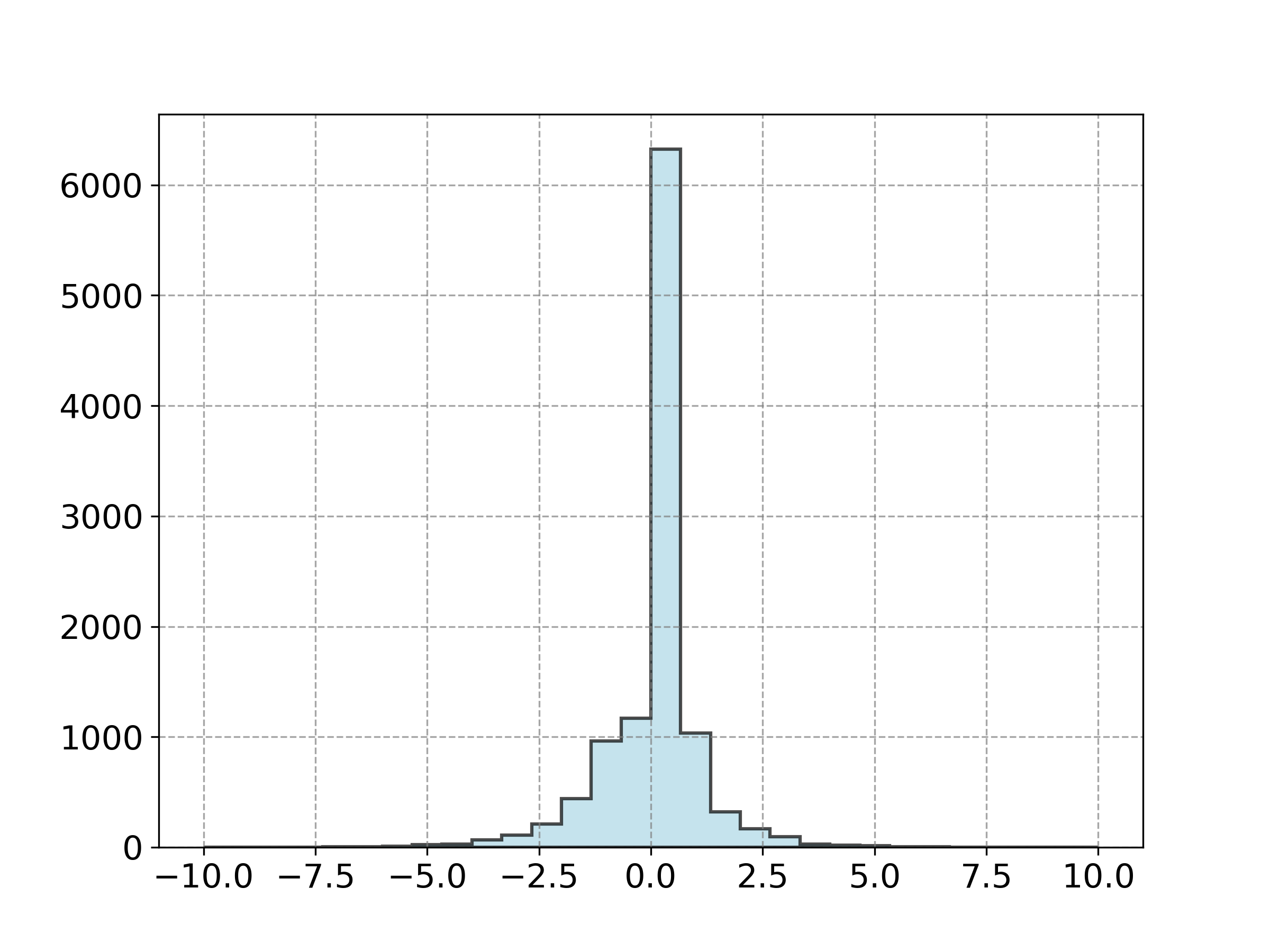}
        \caption{Queue position: 0--10\%; medium near-side queue; small opposite-side queue.}
    \end{minipage}%
    \begin{minipage}{0.5\textwidth}
        \centering
        \captionsetup{width=.8\linewidth}
        \includegraphics[scale=0.36]{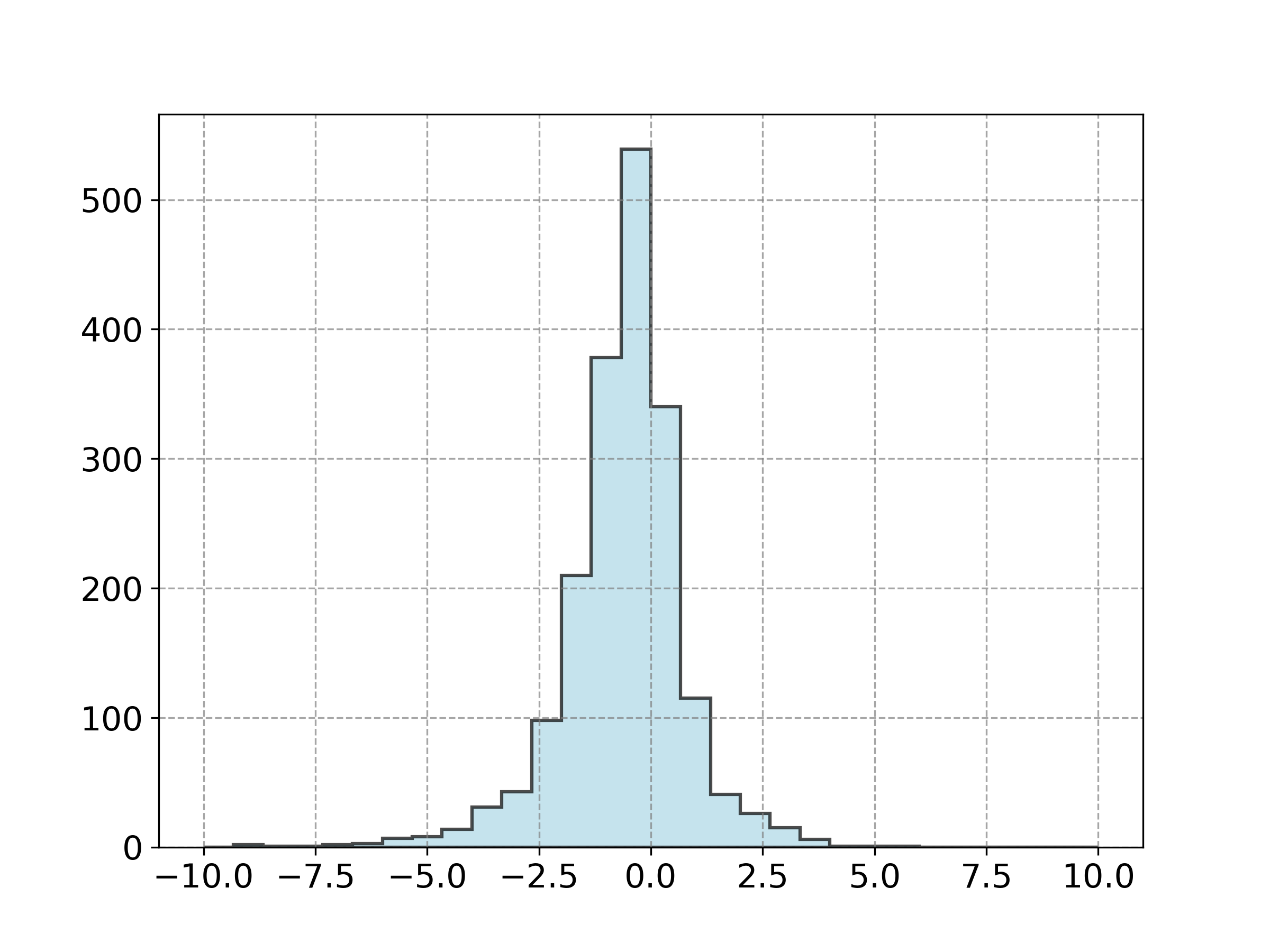}
        \caption{Queue position: 75--100\%; medium near-side queue; small opposite-side queue.}
    \end{minipage}
\end{figure}

\begin{figure}[htbp]
    \centering
    \begin{minipage}{0.5\textwidth}
        \centering
        \captionsetup{width=.8\linewidth}
        \includegraphics[scale=0.36]{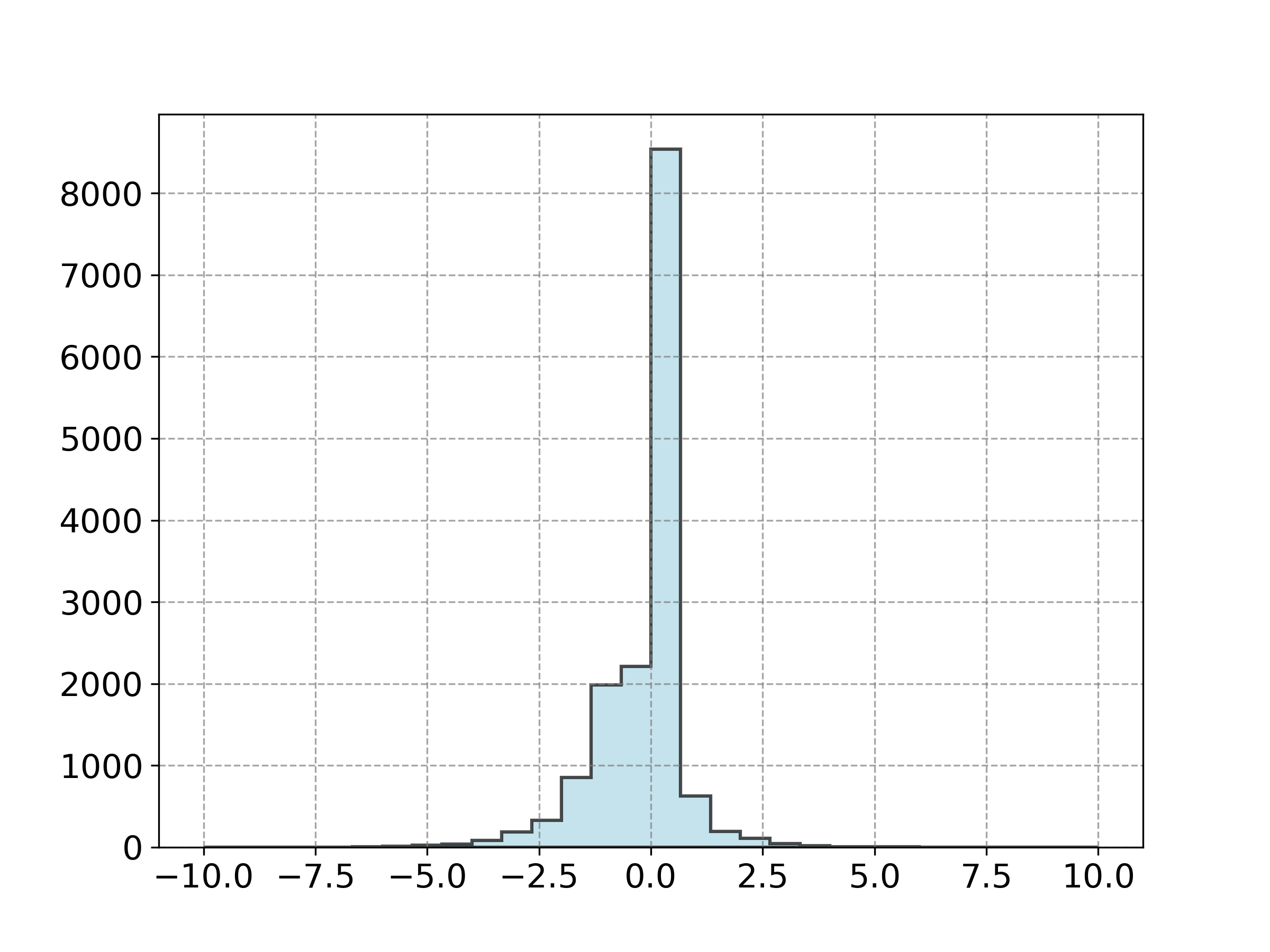}
        \caption{Queue position: 0--10\%; medium near-side queue; medium opposite-side queue.}
    \end{minipage}%
    \begin{minipage}{0.5\textwidth}
        \centering
        \captionsetup{width=.8\linewidth}
        \includegraphics[scale=0.36]{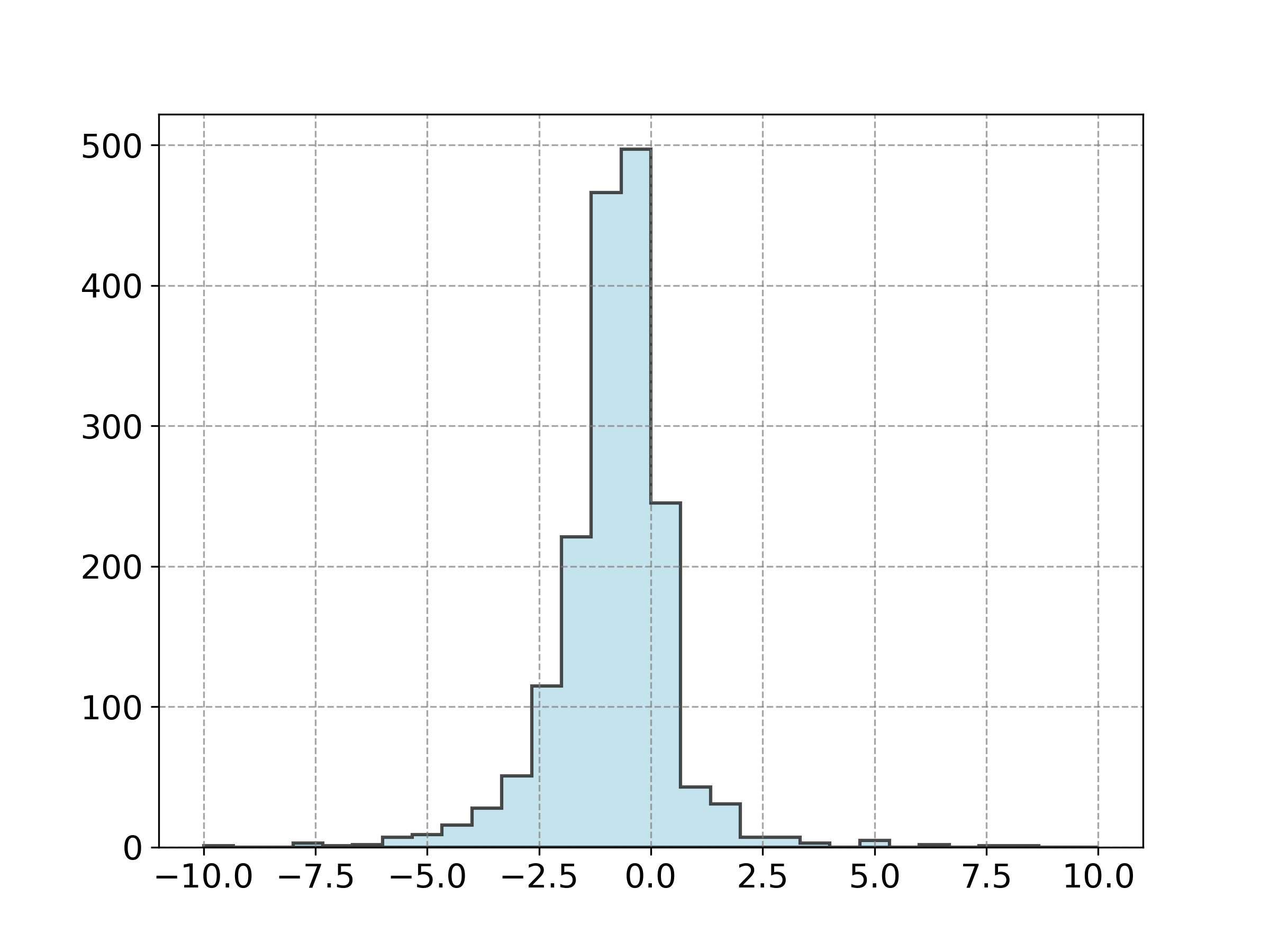}
        \caption{Queue position: 75--100\%; medium near-side queue; medium opposite-side queue.}
    \end{minipage}
\end{figure}

\begin{figure}[htbp]
    \centering
    \begin{minipage}{0.5\textwidth}
        \centering
        \captionsetup{width=.8\linewidth}
        \includegraphics[scale=0.36]{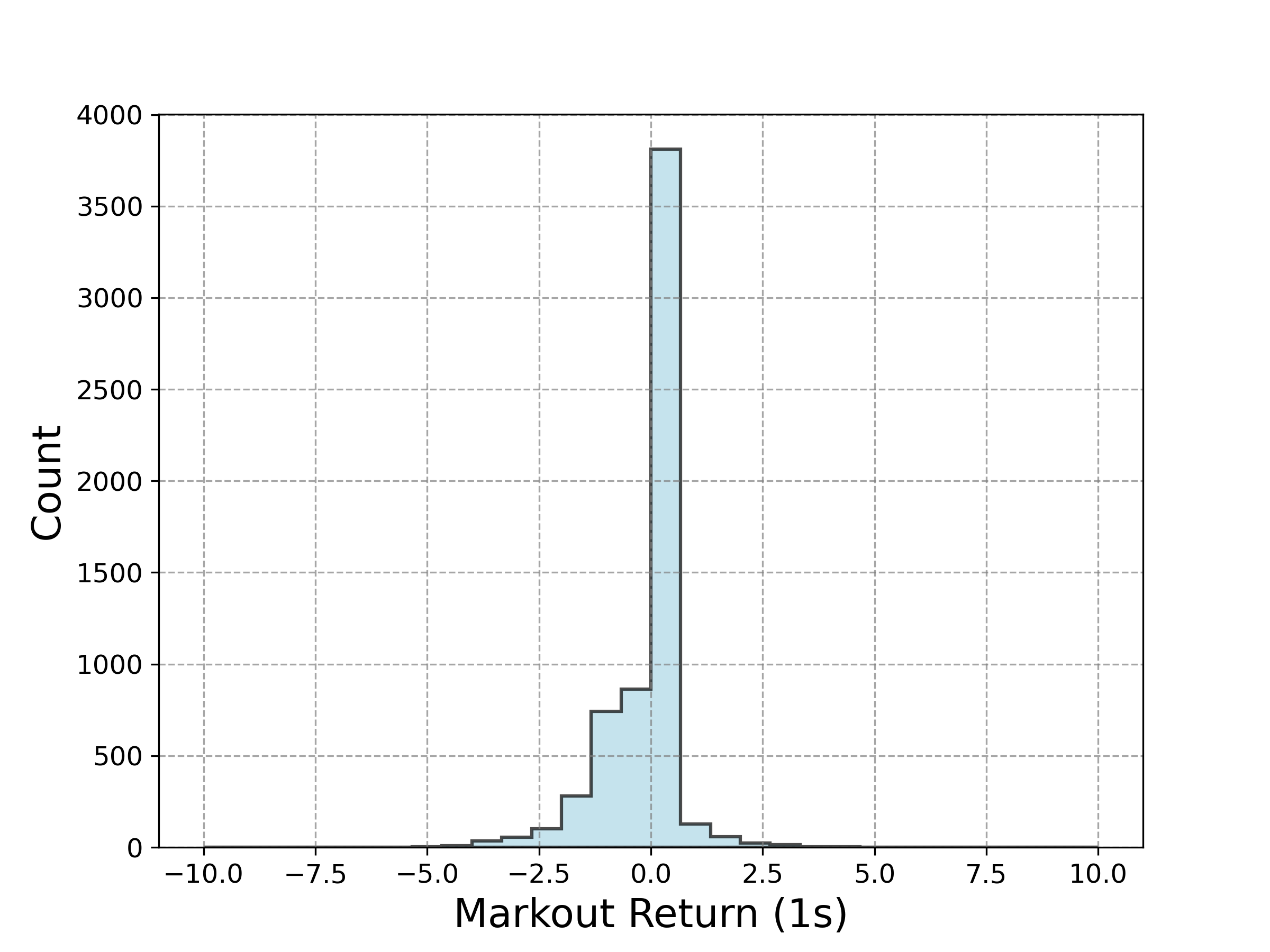}
        \caption{Queue position: 0--10\%; medium near-side queue; large opposite-side queue.}
    \end{minipage}%
    \begin{minipage}{0.5\textwidth}
        \centering
        \captionsetup{width=.8\linewidth}
        \includegraphics[scale=0.36]{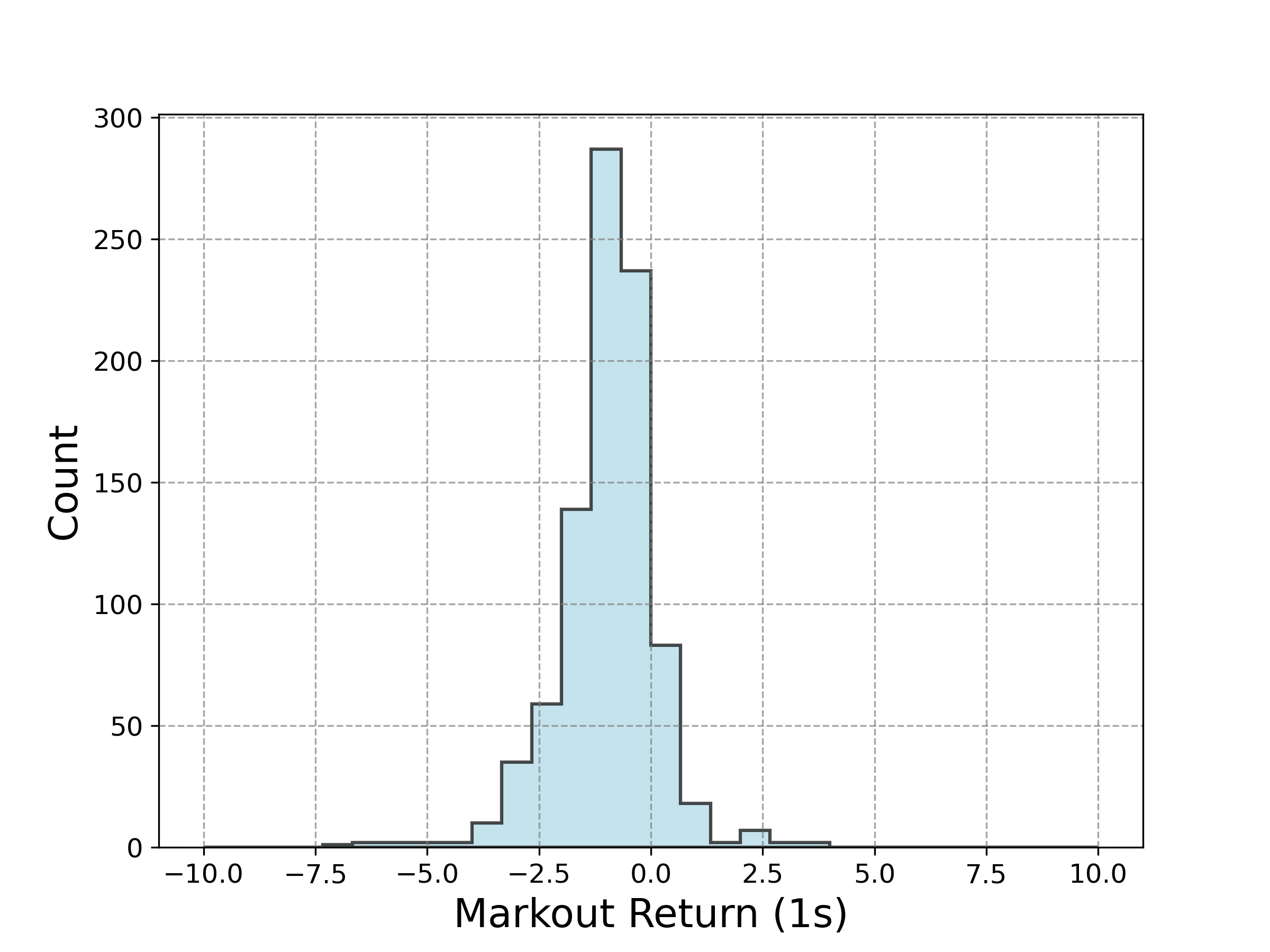}
        \caption{Queue position: 75--100\%; medium near-side queue; large opposite-side queue.}
    \end{minipage}
\end{figure}

\begin{figure}[htbp]
    \centering
    \begin{minipage}{0.5\textwidth}
        \centering
        \captionsetup{width=.8\linewidth}
        \includegraphics[scale=0.36]{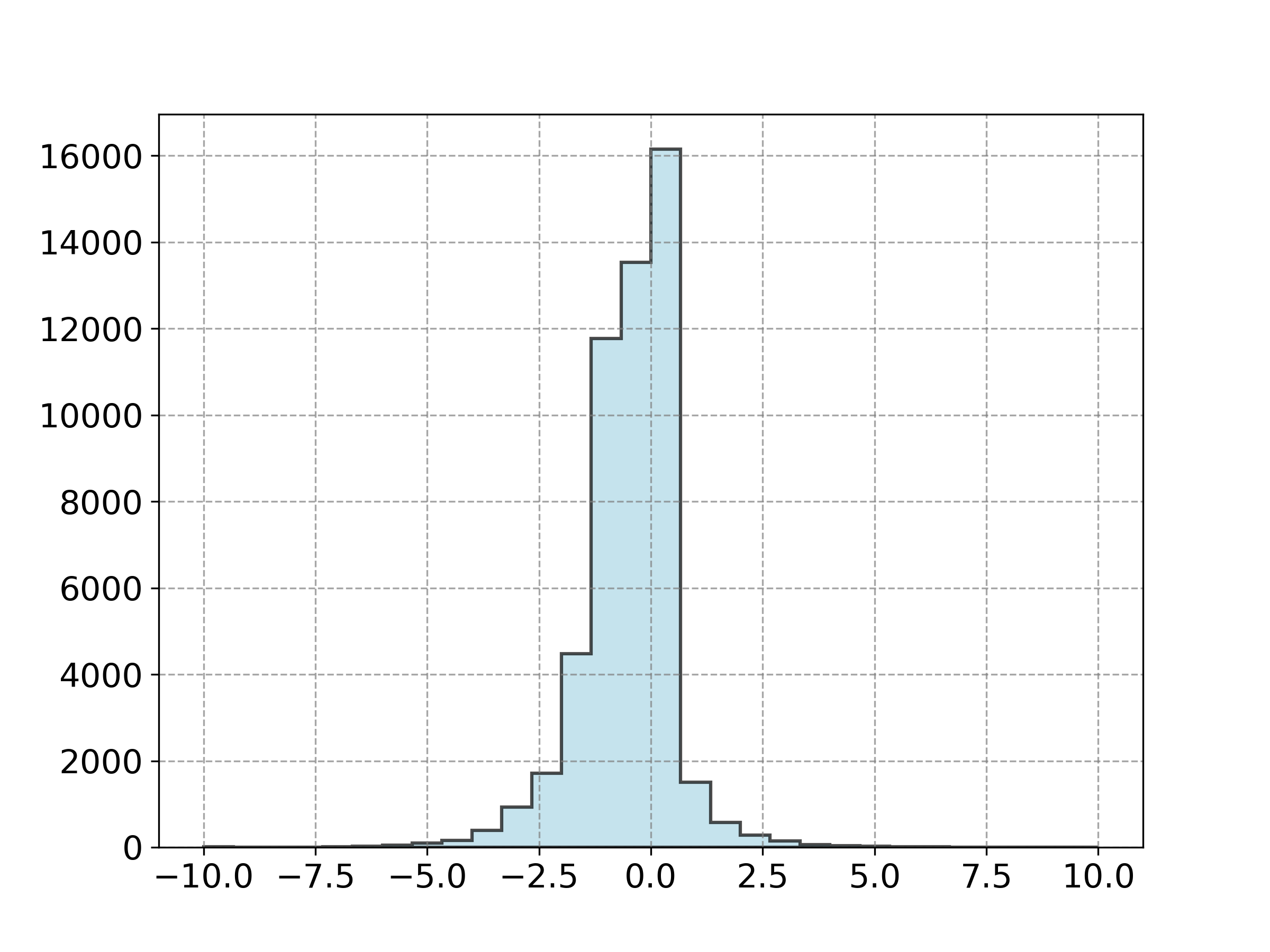}
        \caption{Queue position: 0--10\%; small near-side queue; medium opposite-side queue.}
    \end{minipage}%
    \begin{minipage}{0.5\textwidth}
        \centering
        \captionsetup{width=.8\linewidth}
        \includegraphics[scale=0.36]{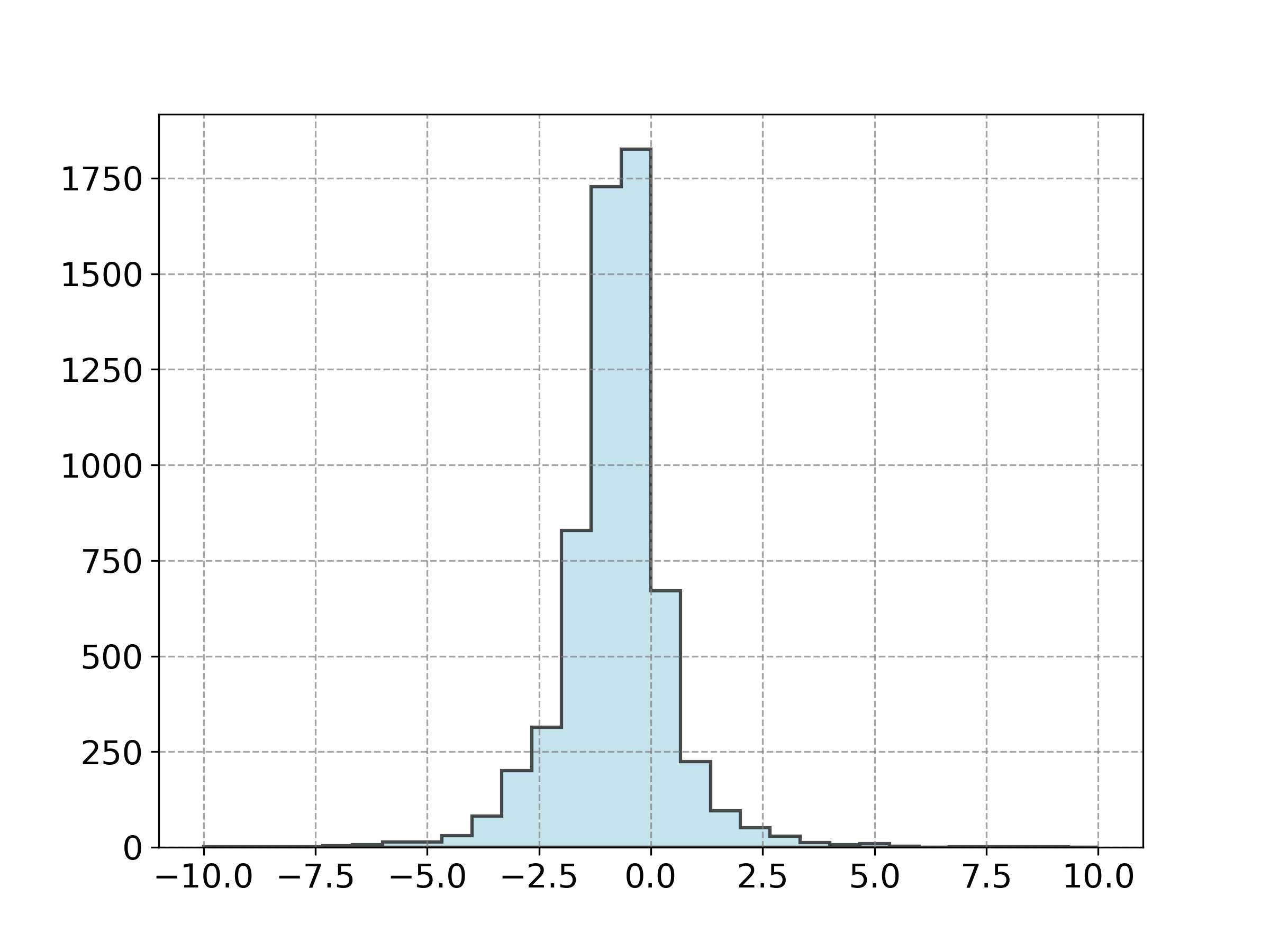}
        \caption{Queue position: 75--100\%; small near-side queue; medium opposite-side queue.}
    \end{minipage}
\end{figure}

\begin{figure}[htbp]
    \centering
    \begin{minipage}{0.5\textwidth}
        \centering
        \captionsetup{width=.8\linewidth}
        \includegraphics[scale=0.36]{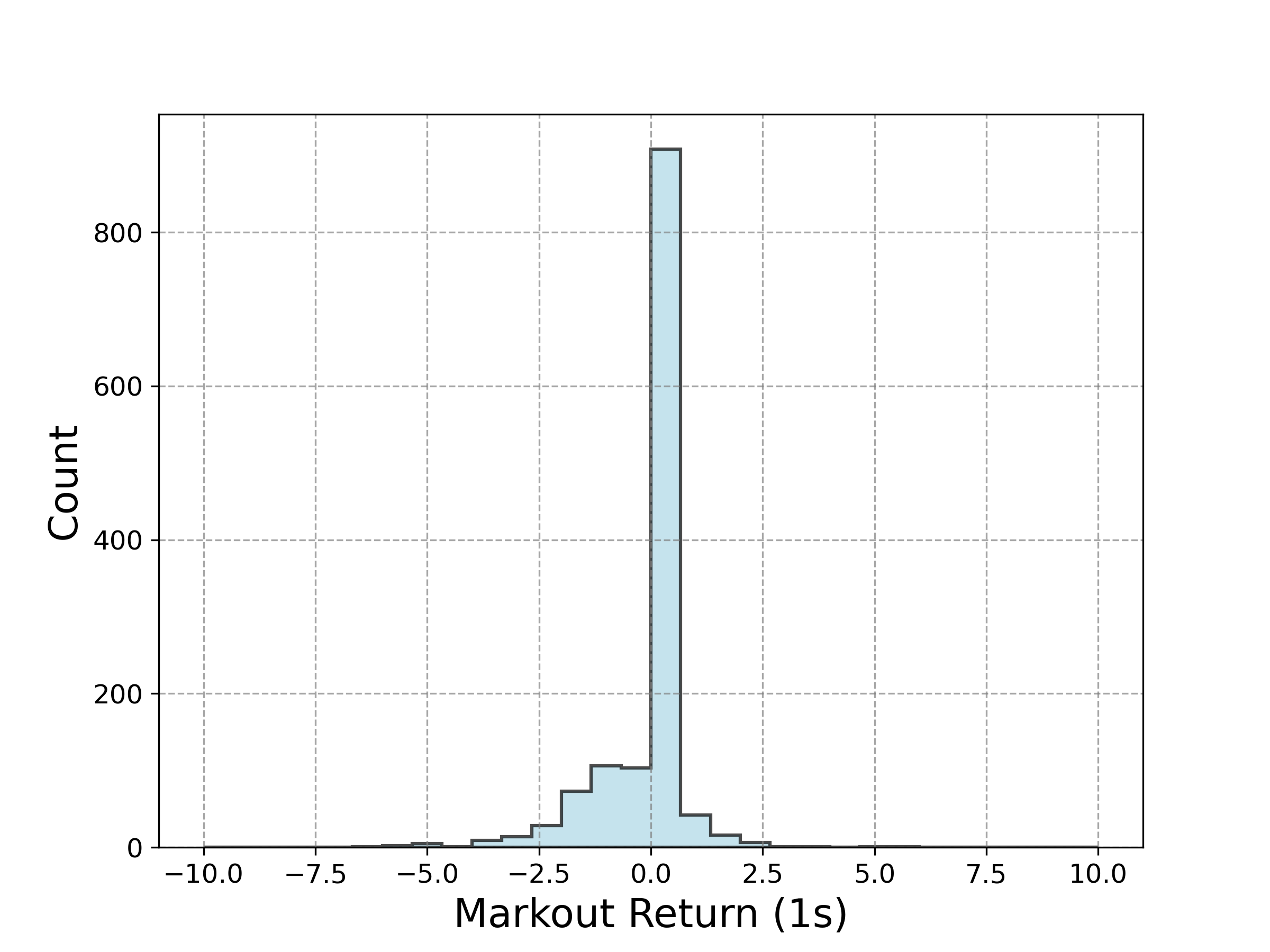}
        \caption{Queue position: 0--10\%; large near-side queue; medium opposite-side queue.}
    \end{minipage}%
    \begin{minipage}{0.5\textwidth}
        \centering
        \captionsetup{width=.8\linewidth}
        \includegraphics[scale=0.36]{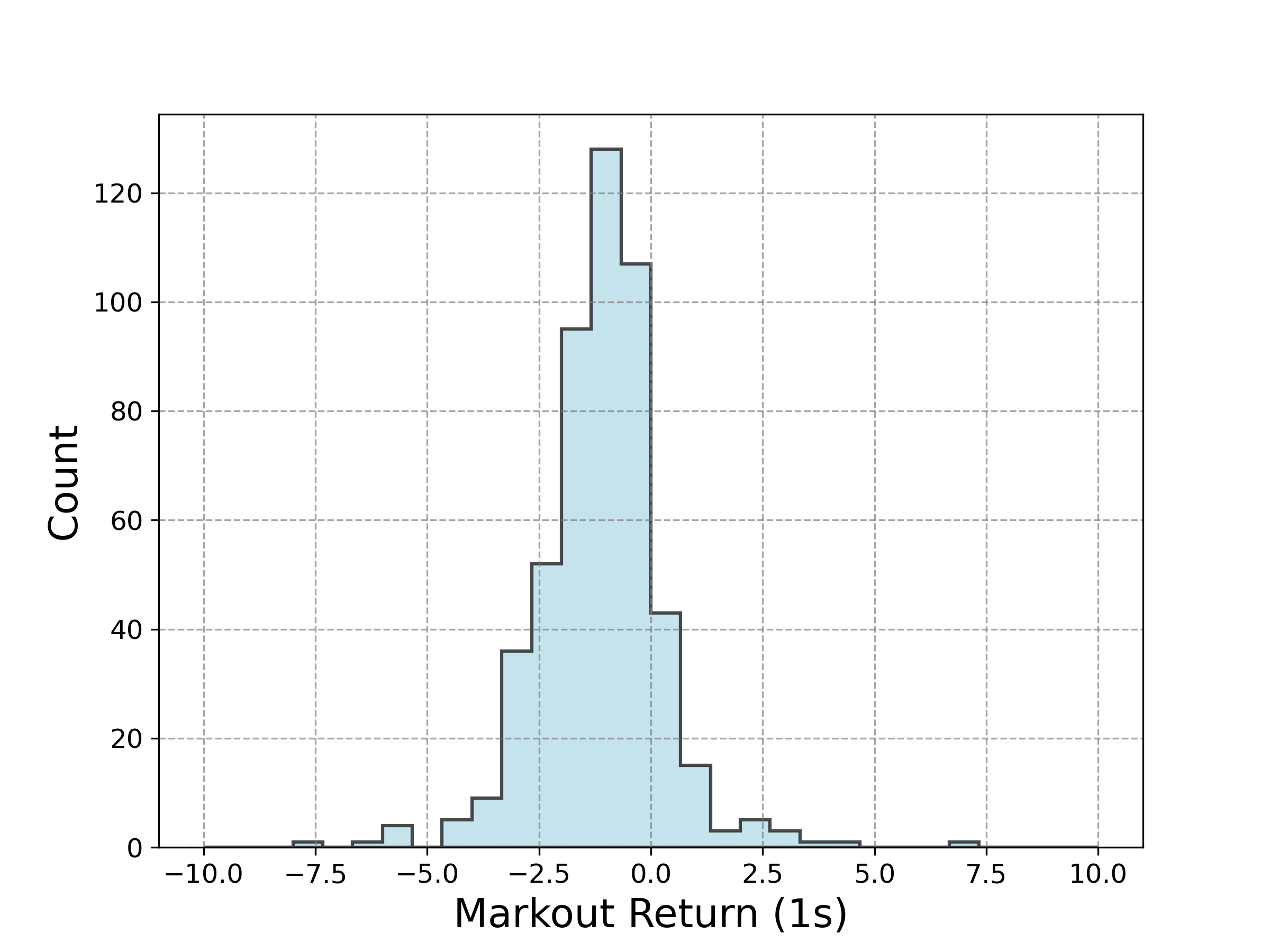}
        \caption{Queue position: 75--100\%; large near-side queue; medium opposite-side queue.}
    \end{minipage}
\end{figure}

\FloatBarrier

\section{Model Coefficients}
\label{sec:model_coefficients}

\begin{longtable}{l | l}
\caption{Coefficients of the logistic regression model grouped by feature categories.} \\
\toprule
\textbf{Feature} & \textbf{Coefficient value} \\
\midrule
\endfirsthead

\caption*{\textbf{Table \thetable{} (continued)}: Coefficients of the logistic regression model grouped by feature categories.} \\
\toprule
\textbf{Feature} & \textbf{Coefficient value} \\
\midrule
\endhead

\bottomrule
\endfoot

\multicolumn{2}{c}{\textbf{Momentum}} \\ \midrule
\texttt{ret\_autocov\_100ms\_w0} & -0.064670 \\
\texttt{ret\_autocov\_100ms\_w1} & -0.025752 \\
\texttt{ret\_autocov\_100ms\_w2} & -0.057328 \\
\texttt{ret\_sum\_100ms\_w0} & 0.171598 \\
\texttt{ret\_sum\_100ms\_w1} & 0.017349 \\
\texttt{ret\_sum\_100ms\_w2} & -0.070002 \\
\texttt{trade\_intensity\_100ms\_w0} & -0.028381 \\
\texttt{trade\_intensity\_100ms\_w1} & -0.048367 \\
\texttt{trade\_intensity\_100ms\_w2} & 0.025111 \\
\texttt{ret\_autocov\_1s\_w0} & -0.156862 \\
\texttt{ret\_autocov\_1s\_w1} & 0.002328 \\
\texttt{ret\_autocov\_1s\_w2} & -0.009059 \\
\texttt{ret\_sum\_1s\_w0} & -0.012417 \\
\texttt{ret\_sum\_1s\_w1} & 0.006796 \\
\texttt{ret\_sum\_1s\_w2} & 0.014250 \\
\texttt{trade\_intensity\_1s\_w0} & -0.027251 \\
\texttt{trade\_intensity\_1s\_w1} & -0.048310 \\
\texttt{trade\_intensity\_1s\_w2} & -0.039917 \\
\texttt{ret\_autocov\_5s\_w0} & -0.433133 \\
\texttt{ret\_autocov\_5s\_w1} & -0.261970 \\
\texttt{ret\_autocov\_5s\_w2} & -0.317830 \\
\texttt{ret\_sum\_5s\_w0} & -0.070847 \\
\texttt{ret\_sum\_5s\_w1} & 0.016905 \\
\texttt{ret\_sum\_5s\_w2} & -0.076009 \\
\texttt{trade\_intensity\_5s\_w0} & -0.066931 \\
\texttt{trade\_intensity\_5s\_w1} & -0.044181 \\
\texttt{trade\_intensity\_5s\_w2} & -0.056927 \\
\texttt{ret\_autocov\_30s\_w0} & 0.111506 \\
\texttt{ret\_autocov\_30s\_w1} & -0.044950 \\
\texttt{ret\_autocov\_30s\_w2} & -0.102125 \\
\texttt{ret\_sum\_30s\_w0} & -0.020848 \\
\texttt{ret\_sum\_30s\_w1} & 0.033951 \\
\texttt{ret\_sum\_30s\_w2} & -0.038579 \\
\texttt{trade\_intensity\_30s\_w0} & -0.310782 \\
\texttt{trade\_intensity\_30s\_w1} & -0.318268 \\
\texttt{trade\_intensity\_30s\_w2} & -0.100313 \\
\texttt{ret\_autocov\_300s\_w0} & 0.055847 \\
\texttt{ret\_autocov\_300s\_w1} & -0.105477 \\
\texttt{ret\_autocov\_300s\_w2} & -0.015074 \\
\texttt{ret\_sum\_300s\_w0} & -0.085791 \\
\texttt{ret\_sum\_300s\_w1} & 0.048547 \\
\texttt{ret\_sum\_300s\_w2} & 0.010486 \\
\texttt{trade\_intensity\_300s\_w0} & -0.013885 \\
\texttt{trade\_intensity\_300s\_w1} & -0.012627 \\
\texttt{trade\_intensity\_300s\_w2} & -0.012789 \\
\midrule

\multicolumn{2}{c}{\textbf{Order Book}} \\ \midrule
\texttt{age} & -0.056527 \\
\texttt{ob\_ask\_liq} & 0.001471 \\
\texttt{ob\_bid\_liq} & -0.187389 \\
\texttt{ob\_half} & -0.049546 \\
\texttt{ob\_other\_half} & -0.007837 \\
\texttt{totb\_mean} & -0.198196 \\
\midrule

\multicolumn{2}{c}{\textbf{Price Dynamics}} \\ \midrule
\texttt{amplitude\_100ms\_w0} & -0.134841 \\
\texttt{amplitude\_100ms\_w1} & -0.002490 \\
\texttt{amplitude\_100ms\_w2} & -0.015208 \\
\texttt{ret\_vwap\_100ms\_w0} & 0.054331 \\
\texttt{ret\_vwap\_100ms\_w1} & -0.106379 \\
\texttt{ret\_vwap\_100ms\_w2} & 0.035287 \\
\texttt{amplitude\_1s\_w0} & 0.009782 \\
\texttt{amplitude\_1s\_w1} & 0.007194 \\
\texttt{amplitude\_1s\_w2} & 0.019774 \\
\texttt{ret\_vwap\_1s\_w0} & -0.007336 \\
\texttt{ret\_vwap\_1s\_w1} & -0.019845 \\
\texttt{ret\_vwap\_1s\_w2} & 0.020953 \\
\texttt{amplitude\_5s\_w0} & 0.103106 \\
\texttt{amplitude\_5s\_w1} & 0.051163 \\
\texttt{amplitude\_5s\_w2} & 0.101474 \\
\texttt{ret\_vwap\_5s\_w0} & 0.041889 \\
\texttt{ret\_vwap\_5s\_w1} & 0.010773 \\
\texttt{ret\_vwap\_5s\_w2} & -0.009981 \\
\texttt{amplitude\_30s\_w0} & -0.025883 \\
\texttt{amplitude\_30s\_w1} & 0.009798 \\
\texttt{amplitude\_30s\_w2} & 0.049859 \\
\texttt{ret\_vwap\_30s\_w0} & -0.051273 \\
\texttt{ret\_vwap\_30s\_w1} & -0.058798 \\
\texttt{ret\_vwap\_30s\_w2} & 0.019591 \\
\texttt{amplitude\_300s\_w0} & 0.041433 \\
\texttt{amplitude\_300s\_w1} & 0.023701 \\
\texttt{amplitude\_300s\_w2} & -0.009184 \\
\texttt{ret\_vwap\_300s\_w0} & 0.033777 \\
\texttt{ret\_vwap\_300s\_w1} & -0.010073 \\
\texttt{ret\_vwap\_300s\_w2} & 0.017873 \\
\texttt{stdev\_100} & -0.014769 \\
\texttt{stdev\_500} & 0.069001 \\
\midrule

\multicolumn{2}{c}{\textbf{Trade Volume Patterns}} \\ \midrule
\texttt{max\_size\_100ms\_w0} & -0.015398 \\
\texttt{max\_size\_100ms\_w1} & -0.064591 \\
\texttt{max\_size\_100ms\_w2} & 0.001329 \\
\texttt{avg\_size\_100ms\_w0} & -0.012507 \\
\texttt{avg\_size\_100ms\_w1} & 0.031008 \\
\texttt{avg\_size\_100ms\_w2} & 0.004749 \\
\texttt{buy\_count\_100ms\_w0} & -0.002806 \\
\texttt{buy\_count\_100ms\_w1} & -0.028945 \\
\texttt{buy\_count\_100ms\_w2} & 0.044143 \\
\texttt{sell\_count\_100ms\_w0} & -0.042881 \\
\texttt{sell\_count\_100ms\_w1} & -0.031871 \\
\texttt{sell\_count\_100ms\_w2} & -0.010898 \\
\texttt{total\_buy\_100ms\_w0} & 0.017631 \\
\texttt{total\_buy\_100ms\_w1} & 0.031328 \\
\texttt{total\_buy\_100ms\_w2} & -0.025174 \\
\texttt{total\_sell\_100ms\_w0} & 0.027392 \\
\texttt{total\_sell\_100ms\_w1} & 0.037905 \\
\texttt{total\_sell\_100ms\_w2} & 0.019188 \\
\texttt{max\_size\_1s\_w0} & 0.009413 \\
\texttt{max\_size\_1s\_w1} & -0.024575 \\
\texttt{max\_size\_1s\_w2} & -0.006610 \\
\texttt{avg\_size\_1s\_w0} & 0.021593 \\
\texttt{avg\_size\_1s\_w1} & 0.015751 \\
\texttt{avg\_size\_1s\_w2} & 0.006351 \\
\texttt{buy\_count\_1s\_w0} & -0.034406 \\
\texttt{buy\_count\_1s\_w1} & -0.046010 \\
\texttt{buy\_count\_1s\_w2} & -0.023821 \\
\texttt{sell\_count\_1s\_w0} & 0.094811 \\
\texttt{sell\_count\_1s\_w1} & 0.024451 \\
\texttt{sell\_count\_1s\_w2} & -0.000842 \\
\texttt{total\_buy\_1s\_w0} & 0.002977 \\
\texttt{total\_buy\_1s\_w1} & -0.003422 \\
\texttt{total\_buy\_1s\_w2} & 0.004587 \\
\texttt{total\_sell\_1s\_w0} & -0.049719 \\
\texttt{total\_sell\_1s\_w1} & -0.012063 \\
\texttt{total\_sell\_1s\_w2} & -0.011760 \\
\texttt{max\_size\_5s\_w0} & 0.005391 \\
\texttt{max\_size\_5s\_w1} & 0.023364 \\
\texttt{max\_size\_5s\_w2} & 0.000746 \\
\texttt{avg\_size\_5s\_w0} & 0.000672 \\
\texttt{avg\_size\_5s\_w1} & -0.002132 \\
\texttt{avg\_size\_5s\_w2} & 0.019611 \\
\texttt{buy\_count\_5s\_w0} & 0.091383 \\
\texttt{buy\_count\_5s\_w1} & -0.026798 \\
\texttt{buy\_count\_5s\_w2} & -0.019723 \\
\texttt{sell\_count\_5s\_w0} & 0.077718 \\
\texttt{sell\_count\_5s\_w1} & 0.094188 \\
\texttt{sell\_count\_5s\_w2} & 0.103399 \\
\texttt{total\_buy\_5s\_w0} & -0.085354 \\
\texttt{total\_buy\_5s\_w1} & -0.041604 \\
\texttt{total\_buy\_5s\_w2} & 0.016422 \\
\texttt{total\_sell\_5s\_w0} & 0.023385 \\
\texttt{total\_sell\_5s\_w1} & -0.077652 \\
\texttt{total\_sell\_5s\_w2} & -0.068970 \\
\texttt{max\_size\_30s\_w0} & -0.053188 \\
\texttt{max\_size\_30s\_w1} & -0.004395 \\
\texttt{max\_size\_30s\_w2} & -0.009718 \\
\texttt{avg\_size\_30s\_w0} & 0.002545 \\
\texttt{avg\_size\_30s\_w1} & -0.010833 \\
\texttt{avg\_size\_30s\_w2} & -0.033334 \\
\texttt{buy\_count\_30s\_w0} & 0.021358 \\
\texttt{buy\_count\_30s\_w1} & -0.046941 \\
\texttt{buy\_count\_30s\_w2} & -0.019813 \\
\texttt{sell\_count\_30s\_w0} & -0.172218 \\
\texttt{sell\_count\_30s\_w1} & -0.098570 \\
\texttt{sell\_count\_30s\_w2} & 0.081075 \\
\texttt{total\_buy\_30s\_w0} & 0.023254 \\
\texttt{total\_buy\_30s\_w1} & 0.070232 \\
\texttt{total\_buy\_30s\_w2} & 0.021899 \\
\texttt{total\_sell\_30s\_w0} & 0.170826 \\
\texttt{total\_sell\_30s\_w1} & 0.001023 \\
\texttt{total\_sell\_30s\_w2} & -0.080049 \\
\texttt{max\_size\_300s\_w0} & -0.004522 \\
\texttt{max\_size\_300s\_w1} & -0.024765 \\
\texttt{max\_size\_300s\_w2} & -0.026390 \\
\texttt{avg\_size\_300s\_w0} & 0.092860 \\
\texttt{avg\_size\_300s\_w1} & -0.006021 \\
\texttt{avg\_size\_300s\_w2} & -0.019310 \\
\texttt{buy\_count\_300s\_w0} & -0.077479 \\
\texttt{buy\_count\_300s\_w1} & 0.124425 \\
\texttt{buy\_count\_300s\_w2} & 0.017996 \\
\texttt{sell\_count\_300s\_w0} & 0.021356 \\
\texttt{sell\_count\_300s\_w1} & -0.036636 \\
\texttt{sell\_count\_300s\_w2} & -0.148054 \\
\texttt{total\_buy\_300s\_w0} & 0.081260 \\
\texttt{total\_buy\_300s\_w1} & -0.129170 \\
\texttt{total\_buy\_300s\_w2} & -0.002067 \\
\texttt{total\_sell\_300s\_w0} & 0.047449 \\
\texttt{total\_sell\_300s\_w1} & -0.008577 \\
\texttt{total\_sell\_300s\_w2} & 0.147168 \\

\end{longtable}

\end{document}